\documentclass[prd,aps,
nofootinbib,
floatfix,
superscriptaddress]{revtex4}
\usepackage{graphicx}
\usepackage{epsfig}
\usepackage{rotating}
\usepackage{amssymb}
\usepackage{subfigure}
\usepackage{dsfont}
\usepackage{psfrag}
\usepackage{amsmath,euscript,array,mathrsfs}
\usepackage{bbold}
\usepackage{bm}
\usepackage{multirow}
\usepackage{dcolumn}
\usepackage{xcolor}

\newcommand{\beq}{\begin{equation}}
\newcommand{\eeq}{\end{equation}}
\newcommand{\beqs}{\begin{eqnarray}}
\newcommand{\eeqs}{\end{eqnarray}}
\newcommand{\lsim}{\mathrel{\raisebox{-
.6ex}{$\stackrel{\textstyle<}{\sim}$}}}
\newcommand{\gsim}{\mathrel{\raisebox{-
.6ex}{$\stackrel{\textstyle>}{\sim}$}}}
\newcommand{\Tr}{{\rm Tr}}
\newcommand{\STr}{{\rm STr}}

\def\hbar{\hspace{0pt}\raisebox{1pt}{$-$} \hspace{-7pt} h}

\def\r{\rho}

\newcommand{\be}{\begin{equation}}
\newcommand{\ee}{\end{equation}}

\newcommand{\bea}{\begin{eqnarray}}
\newcommand{\eea}{\end{eqnarray}}
\newcommand{\nn}{\nonumber}

\def\lbldef#1#2{\expandafter\gdef\csname #1\endcsname {#2}}

\def\href#1#2{#2}

\newcommand{\ber}{\begin{eqnarray}}
\newcommand{\eer}{\end{eqnarray}}

\newcommand{\beqar}{\begin{eqnarray}}

\newcommand{\eeqar}{\end{eqnarray}}

\def\calO{{\mathcal{O}}}

\newcommand{\dsl}
  {\kern.06em\hbox{\raise.15ex\hbox{$/$}\kern-.56em\hbox{$\partial$}}}

\newcommand{\eeqarr}{\end{eqnarray}}
\newcommand{\ZZ}{{\rm \kern 0.275em Z \kern -0.92em Z}\;}

\def\CC{{\mathchoice
{\rm C\mkern-8mu\vrule height1.45ex depth-.05ex
width.05em\mkern9mu\kern-.05em}
{\rm C\mkern-8mu\vrule height1.45ex depth-.05ex
width.05em\mkern9mu\kern-.05em}
{\rm C\mkern-8mu\vrule height1ex depth-.07ex
width.035em\mkern9mu\kern-.035em}
{\rm C\mkern-8mu\vrule height.65ex depth-.1ex
width.025em\mkern8mu\kern-.025em}}}

\def\RR{{\rm I\kern-1.6pt {\rm R}}}

\def\ZZ{{\rm Z}\kern-3.8pt {\rm Z} \kern2pt}
\def\IB{\relax{\rm I\kern-.18em B}}
\def\ID{\relax{\rm I\kern-.18em D}}
\def\II{\relax{\rm I\kern-.18em I}}
\def\IP{\relax{\rm I\kern-.18em P}}

\newcommand{\bear}{\begin{eqnarray}}
\newcommand{\eear}{\end{eqnarray}}

\def\r{\rho}                                     

\def\6{\partial}

\def\bea{\begin{eqnarray}}
\def\eea{\end{eqnarray}}

\newcommand\Sec[1]{Sec~\ref{Sec:#1}}
\newcommand\App[1]{Appendix~\ref{Sec:#1}}
\newcommand\Tab[1]{Table~\ref{tab:#1}}
\newcommand\Fig[1]{Fig.~\ref{fig:#1}}
\newcommand\Eq[1]{Eq.~(\ref{eq:#1})}

\def\beqx{\begin{displaymath}}
\def\eeqx{\end{displaymath}}

\newcommand{\bmat}{\left(\begin{array}}
\newcommand{\emat}{\end{array}\right)}

\def\r{\rho}

\def\bo{{\raise-.3ex\hbox{\large$\Box$}}}               
\def\face{{\raise.2ex\hbox{$\displaystyle \bigodot$}\mskip-2.2mu \llap {$\ddot
        \smile$}}}                                   
\def\>{\rangle}                                      
\def\<{\langle}                                      

\def\leftrightarrowfill{$\mathsurround=0pt \mathord\leftarrow \mkern-6mu
        \cleaders\hbox{$\mkern-2mu \mathord- \mkern-2mu$}\hfill
        \mkern-6mu \mathord\rightarrow$}        
\def\dvec#1{\vbox{\ialign{##\crcr
        \leftrightarrowfill\crcr\noalign{\kern-1pt\nointerlineskip}
        $\hfil\displaystyle{#1}\hfil$\crcr}}}           
\def\Tr{{\rm Tr \,}}                                    

\def\-{\hphantom{-}}

\begin{document}

\preprint{PNUTP-19/A03}

\title{$Sp(4)$ gauge theories on the lattice: Quenched fundamental and antisymmetric fermions}

\begin{abstract}
We perform lattice studies of meson mass spectra and
decay constants of the $Sp(4)$ gauge theory in the quenched approximation. 
We consider two species of (Dirac) fermions as matter field content,
transforming in the 2-index antisymmetric and the fundamental 
representation of the gauge group, respectively.
All matter fields are formulated as Wilson fermions. 
We extrapolate to the continuum and massless limits, 
and compare to each other the results obtained for the two species of mesons.
In the case of two fundamental and three antisymmetric fermions, 
the long-distance dynamics is relevant for composite Higgs models. 
This is the first lattice study of this class of theories. 
The global $SU(4)\times SU(6)$ symmetry is broken to 
the $Sp(4)\times SO(6)$ subgroup, and the condensates align with the explicit mass terms  
present in the lattice formulation of the theory. 

The main results of our quenched calculations are that, with 
 fermions in the 2-index antisymmetric representation of the group, 
 the masses squared and decay constant squared of all the mesons we 
 considered are larger than the corresponding quantities for the fundamental 
 representation, by factors that vary between $\sim 1.2$ and $\sim 2.7$. 
We also present technical results that will be useful for future lattice investigations 
of dynamical simulations,  
 of composite chimera baryons,
and of the approach to large $N$ in the $Sp(2N)$ theories considered. 
We briefly discuss their high-temperature behaviour, where symmetry
 restoration and enhancement are expected.
\end{abstract}

\author{Ed Bennett}
\email{e.j.bennett@swansea.ac.uk}
\affiliation{Swansea Academy of Advanced Computing, Swansea University,
Bay Campus, SA1 8EN, Swansea, Wales, United Kingdom}

\author{Deog Ki Hong}
\email{dkhong@pusan.ac.kr}
\affiliation{Department of Physics, Pusan National University, Busan 46241, Korea}

\author{Jong-Wan Lee}
\email{jwlee823@pusan.ac.kr}
\affiliation{Department of Physics, Pusan National University, Busan 46241, Korea}
\affiliation{Extreme Physics Institute, Pusan National University, Busan 46241, Korea}

\author{C.-J. David Lin}
\email{dlin@mail.nctu.edu.tw}
\affiliation{Institute of Physics, National Chiao-Tung University, 
1001 Ta-Hsueh Road, Hsinchu 30010, Taiwan}
\affiliation{Centre for High Energy Physics, Chung-Yuan Christian University,
Chung-Li 32023, Taiwan}

\author{Biagio Lucini}
\email{b.lucini@swansea.ac.uk}
\affiliation{Department of Mathematics, College of Science, Swansea university,
Bay campus, SA1 8EN, Swansea, Wales, United Kingdom}

\author{Michele Mesiti}
\email{michele.mesiti@swansea.ac.uk}
\affiliation{Swansea Academy of Advanced Computing, Swansea University,
Bay Campus, SA1 8EN, Swansea, Wales, United Kingdom}

\author{Maurizio Piai}
\email{m.piai@swansea.ac.uk}
\affiliation{Department of Physics, College of Science, Swansea University,
Singleton Park, SA2 8PP, Swansea, Wales, UK}

\author{Jarno Rantaharju}
\email{jarno.rantaharju@helsinki.fi}
\affiliation{Swansea Academy of Advanced Computing, Swansea University,
Bay Campus, SA1 8EN, Swansea, Wales, UK}
\affiliation{Department of Physics \& Helsinki Institute of Physics, 
P.O. Box 64, FI-00014 University of Helsinki, Helsinki, Finland}

\author{Davide Vadacchino}
\email{davide.vadacchino@pi.infn.it}
\affiliation{INFN, Sezione di Pisa, Largo Pontecorvo 3, 56127 Pisa, Italy}

\date{\today}

\maketitle
\flushbottom


\section{Introduction}
\label{Sec:Introduction}

In composite Higgs models
 (CHMs)~\cite{Kaplan:1983fs,Georgi:1984af,Dugan:1984hq},
 the Higgs fields, responsible for electroweak symmetry breaking, 
arise as  pseudo-Nambu-Goldstone bosons (pNGBs) in a more fundamental theory,
hence addressing the little hierarchy problem of generic extensions
of the Standard Model (SM)  of particle physics.
In comparison to the other SM fermions, the top quark has a large mass, 
making it heavier than the $W$, the $Z$, and even 
the recently discovered Higgs boson~\cite{Aad:2012tfa,Chatrchyan:2012xdj}.
It is then natural to complete the CHM scenario by postulating
that also the top quark has composite nature, at least partially, at the fundamental level.
The additional model-building dimension added to this framework by (partial) top compositeness 
yields a richness of potential implications that has been explored in the literature on the subject
in a range of possible directions, and motivates us to study its dynamical
 origin with nonperturbative techniques.
The literature on composite Higgs models is indeed vast
(see for instance Refs.~\cite{Agashe:2004rs, Contino:2006qr,
Barbieri:2007bh,Lodone:2008yy,Marzocca:2012zn,Grojean:2013qca,Ferretti:2013kya,
Cacciapaglia:2014uja,Arbey:2015exa,Vecchi:2015fma,Panico:2015jxa,Ferretti:2016upr,
Agugliaro:2016clv,Alanne:2017rrs,Feruglio:2016zvt,DeGrand:2016pgq,Fichet:2016xvs,
Galloway:2016fuo,Csaki:2017cep,
Chala:2017sjk,Csaki:2017jby,Ayyar:2017qdf,Alanne:2017ymh,Ayyar:2018zuk,Ayyar:2018ppa,
Cai:2018tet,
Agugliaro:2018vsu,Ayyar:2018glg,
Cacciapaglia:2018avr,Witzel:2019jbe,Cacciapaglia:2019bqz,
Ayyar:2019exp,
Cossu:2019hse,
Cacciapaglia:2019ixa,BuarqueFranzosi:2019eee}), 
especially in connection with dynamical theories characterised by the $SU(4)/Sp(4)$ coset 
(see for instance Refs.~\cite{Katz:2005au,Gripaios:2009pe,Barnard:2013zea,Lewis:2011zb,
Hietanen:2014xca,Arthur:2016dir,Arthur:2016ozw,
Pica:2016zst,Detmold:2014kba,Lee:2017uvl,Cacciapaglia:2015eqa,
Bizot:2016zyu,Hong:2017smd,Golterman:2017vdj,Drach:2017btk,
Sannino:2017utc,Alanne:2018wtp,Bizot:2018tds,BuarqueFranzosi:2018eaj,
Gertov:2019yqo,Cacciapaglia:2019dsq}).

In Ref.~\cite{Bennett:2017kga} (see also Refs.~\cite{Bennett:2017tum,Bennett:2017ttu} and the more recent 
Refs.~\cite{Bennett:2017kbp,Lee:2018ztv,Bennett:2019jzz,Bennett:2019ckt}), some of us proposed 
a systematic programme of exploration of the
lattice dynamics of  $Sp(2N)$ gauge theories. Our main scientific motivation is the application 
of the results of such  studies to the CHM context. 
In order to realise also top compositeness,
it is necessary to implement on the lattice matter fields with mixed representations.
For example, the model discussed in Refs.~\cite{Barnard:2013zea,Ferretti:2013kya}
requires that the matter content consists of $N_f=2$ Dirac 
fields transforming in the fundamental representation of $Sp(2N)$, 
supplemented by $n_f=3$ Dirac fields transforming in the antisymmetric representation of $Sp(2N)$.
This dynamical system is expected to  yield the spontaneous breaking of the
 $SU(4)\times SU(6)$ global symmetry to its $Sp(4)\times SO(6)$ subgroup. The introduction of 
diagonal mass terms for the fermions is compatible ({\it aligned}) with the vacuum structure, 
and provides a degenerate nonvanishing mass for the resulting $5+20$ pNGBs.
The lattice treatment of such a system with multiple dynamical  fermion representations
is a novel arena for lattice gauge theories,
and only  recently have calculations of this type been published,
in the specific context of theories with 
$SU(4)$ gauge group~\cite{DeGrand:2016pgq,Ayyar:2017qdf,Ayyar:2018zuk,
Ayyar:2018glg,Cossu:2019hse}.

In this paper, we take a first step in this direction for $Sp(2N)$ gauge theories.
We consider the $Sp(4)$ gauge theory, and  treat 
the two species of  fermions in the quenched approximation;
only the gluon dynamics is  captured by the lattice numerical study,
but the operators used to compute the relevant correlation functions involve both types of matter fields.
We compute the mass spectra and decay constants of the mesons 
built both with fundamental and antisymmetric fermions,
and perform their continuum extrapolation.
We compare the properties of mesonic observables
obtained with the two representations, 
which, in the dynamical theory, is important for CHM phenomenology. 
Since very little is known about the $Sp(2N)$ gauge theories, 
our quenched study is a first benchmark of these theories and 
would serve as a starting point for a more extensive and detailed investigation of such models. 

We treat  the 
relevant degrees of freedom with a low-energy effective field theory (EFT)
that we employ to analyse the 
numerical data extrapolated to the continuum limit.
The EFT  proposed in Ref.~\cite{Bennett:2017kga} for the theory with
$SU(4)/Sp(4)$  coset is based on the ideas of  hidden local symmetry, adapted from Refs.~\cite{
Bando:1984ej,Casalbuoni:1985kq,Bando:1987br,Casalbuoni:1988xm,Harada:2003jx}
(and~\cite{Georgi:1989xy,Appelquist:1999dq,Piai:2004yb,Franzosi:2016aoo}), and supplemented by
some simplifying working assumptions. 
Here we return to the EFT to improve it 
and to generalise it to the case of the $SU(6)/SO(6)$ coset.

The paper is organised as follows. 
In Sec.~\ref{Sec:model}, we define the $Sp(4)$ theory 
with field content we are interested in,
by writing both the Lagrangian 
density of the microscopic continuum theory 
as well as its low-energy EFT description.
We devote Sec.~\ref{Sec:lattice} to describing the lattice action we adopt, 
the Monte Carlo algorithm we employ,
and other important aspects of the lattice study we perform, such as scale setting
 and topology.
In Sec.~\ref{Sec:mesons} we present our results for the calculation of the masses 
and the (renormalised) decay constants
of the lightest mesons 
in the quenched approximation. We compare the results for 
quenched fundamental and antisymmetric fermions.
We also discuss in Sec.~\ref{Sec:fits} a first attempt at matching the 
results to the low-energy EFT description
applicable to pseudoscalar (PS), vector (V) and axial-vector (AV) states.
We conclude by summarising and discussing our main findings 
and by outlining future avenues for investigation
in Sec.~\ref{Sec:conclusions}.

The presentation is complemented by a rather generous set of
 Appendixes, 
intended to be of use also beyond the specific aims of this paper,
  for the research programme we are carrying out as a whole.
 We expose some details and conventions in the treatment of
  spinors in Appendix~\ref{Sec:AppendixA} and some technical points about
  the treatment of massive spin-1 particles in Appendix~\ref{Sec:AppendixB}.
Technical points about the 
  embedding of the SM gauge group in the context of CHMs are 
  highlighted in Appendix~\ref{Sec:AppendixC}.
  Appendix~\ref{Sec:AppendixF} contains some numerical tests
of the topological charge history and of its effect on spectral observables, 
in the illustrative case of a numerical ensemble that has a fine lattice spacing.
  In Appendix~\ref{Sec:AppendixD}, besides briefly summarising some properties of QCD light flavoured 
  mesons, we discuss general symmetry properties of the mesons in theories with symmetric cosets,
  that are important for spectroscopy. 
  We also touch upon possible high-temperature symmetry restoration and enhancement 
  in Appendix~\ref{Sec:AppendixD1}.
  We explicitly write the operators relevant as sources of all the mesons 
  in Appendix~\ref{Sec:AppendixE}, and in Appendix~\ref{Sec:AppendixE1}
  we specify the sources of PS, V and AV mesons in the $SU(4)/Sp(4)$ case,
by adopting  a specific choice of $SU(4)$ generators and normalisations.

\section{The model}
\label{Sec:model}

In this section, we describe the specific model of interest, 
borrowing  ideas from Refs.~\cite{Ferretti:2013kya,Barnard:2013zea}, 
 and we describe the basic properties of the long-distance EFT description(s) we use later.

\subsection{Continuum microscopic theory}
\label{Sec:microscopic}

\begin{table}
\begin{center}
\begin{tabular}{|c|c|c|c|c|}
\hline\hline
{\rm ~~~Fields~~~} &$Sp(4)$  &  $SU(4)$ & $SU(6)$ \cr
\hline
$V_{\mu}$ & $10$ & $1$ & $1$  \cr
$q$ & $4$ & $4$ & $1$ \cr
$\psi$ & $5$ & $1$ & $6$\cr
\hline
$\Sigma_6$ & $1$ & $6$ & $1$ \cr
$M_6$ & $1$ & $\bar{6}\sim6$ & $1$\cr 
$\Sigma_{21}$ & $1$ & $1$ & $21$ \cr
$M_{21}$ & $1$ & $1$ & $\overline{21}$\cr 
\hline\hline
\end{tabular}
\end{center}
\caption{Field content of the microscopic theory ($V_{\mu}$, $q$, $\psi$)
 and of the low-energy EFT describing  the pNGBs ($\Sigma_{6,21}$, $M_{6,21}$).
$Sp(4)$ is the gauge group, while $SU(4)$ and $SU(6)$ are the global symmetries. 
The elementary fields $V_{\mu}$ are gauge bosons, while  $q$ and $\psi$ are two-component spinors.
$\Sigma_{6}$ and $\Sigma_{21}$ are composite scalar fields. They capture the long-distance dynamics of operators 
that are bilinear in $q$ and $\psi$, 
 the VEVs of which are responsible
 for the breaking $SU(4)\rightarrow Sp(4)$ and $SU(6)\rightarrow SO(6)$, respectively.
 The mass matrices $M_6$ and  $M_{21}$ are treated  as scalar spurions, 
 formally transforming as $\sim \bar{6}\sim6$ of $SU(4)$,
 and $\sim \overline{21}$ of $SU(6)$, respectively. 
}
\label{Fig:fields}
\end{table}

The  $Sp(4)$ gauge theory we started to study in 
Ref.~\cite{Bennett:2017kga} has matter content consisting of
two Dirac fermions $Q^{i\,a}$, where $a=1\,,\,\cdots\,,\,4$ is the colour index
and $i=1,2$ the flavour index, or equivalently  four two-component spinors $q^{j\,a}$ with $j=1\,,\,\cdots\,,\,4$.
Following~\cite{Ferretti:2013kya,Barnard:2013zea}, we
supplement it by three Dirac fermions $\Psi^{i\,ab}$ transforming in the antisymmetric 
2-index representation of $Sp(4)$,
or equivalently by six two-component spinors $\psi^{j\,ab}$, with $j=1\,,\,\cdots\,,\,6$. 
The field content is summarised in Table~\ref{Fig:fields}.
The Lagrangian density is 
\beqs
{\cal L}&=& -\frac{1}{2} \Tr V_{\mu\nu} V^{\mu\nu}
\,+\,\frac{1}{2}\left(i\overline{Q^{i}}_a \gamma^{\mu}\left(D_{\mu} Q^i\right)^a
\,-\,i\overline{D_{\mu}Q^{i}}_a \gamma^{\mu}Q^{i\,a}\right)\,-\,M\overline{Q^i}_a Q^{i\,a}+\nonumber\\
&&
\,+\,\frac{1}{2}\left(i\overline{\Psi^{k}}_{ab} \gamma^{\mu}\left(D_{\mu} \Psi^k\right)^{ab}
\,-\,i\overline{D_{\mu}\Psi^{k}}_{ab} \gamma^{\mu}\Psi^{k\,ab}\right)\,-\,m\overline{\Psi^k}_{ab} \Psi^{k\,ab}\,.
\label{eq:lagrangian}
\eeqs
The covariant derivatives are defined by making use of the transformation 
properties under the action of an element $U$ of the 
$Sp(4)$ gauge group---$Q\rightarrow U Q$ and $\Psi \rightarrow U \Psi U^{\mathrm{T}}$---so that
\beqs
V_{\mu\nu}&\equiv& \partial_{\mu}V_{\nu}-\partial_{\nu}V_{\mu} + i g \left[V_{\mu}\,,\,V_{\nu}\right]\,,\\
D_{\mu} Q^i&=& \partial_{\mu} Q^i \,+\,i g V_{\mu} Q^{i}\,,\\ 
D_{\mu} \Psi^j&=& \partial_{\mu} \Psi^j \,+\,i g V_{\mu} \Psi^{j}\,+\,i g \Psi^{j} V_{\mu}^{\mathrm{T}}\,,
\eeqs
where $g$ is the gauge coupling.

The Lagrangian density possesses a global $SU(4)$ symmetry acting on the fundamental fermions $Q$
and a global $SU(6)$ acting on 
the antisymmetric-representation fermions $\Psi$. The mass terms break 
them to the $Sp(4)$ and $SO(6)$ subgroups, respectively.
The unbroken subgroups consist of the  transformations
that leave invariant
the symplectic matrix $\Omega$ and the symmetric matrix $\omega$, respectively, that are defined by
\beqs
\Omega&=&\Omega_{jk}\,=\,\Omega^{jk}\,\equiv\,
\left(\begin{array}{cccc}
0 & 0 & 1 & 0\cr
0 & 0 & 0 & 1\cr
-1 & 0 & 0 & 0\cr
0 & -1 & 0 & 0\cr
\end{array}\right)\,,
\label{Eq:symplectic}
~~~~
\omega\,=\,\omega_{jk}\,=\,\omega^{jk}\,\equiv\,
\left(\begin{array}{cccccc}
0 & 0 & 0 & 1 & 0 & 0 \cr
0 & 0 & 0 & 0 & 1 & 0 \cr
0 & 0 & 0 & 0 & 0 & 1  \cr
1 & 0 & 0 & 0 & 0 & 0 \cr
0 & 1 & 0 & 0 & 0 & 0 \cr
0 & 0 & 1 & 0 & 0 & 0 \cr
\end{array}\right)\,.
\eeqs
By rewriting explicitly the fermion contributions to the Lagrangian density in two-component notation as follows
 (see Appendix~\ref{Sec:AppendixA} for the list of  conventions about spinors)
\beqs
Q^{i\,a}&=&\left(
\begin{array}{c}
q^{i\,a} \cr \Omega^{ab}(-\tilde{C}q^{i+2\,\ast})_b
\end{array}
\right)\,,
~~~~
\Psi^{i\,ab}\,=\,\left(
\begin{array}{c}
\psi^{i\,ab} \cr
\Omega^{ac}\Omega^{bd} (-\tilde{C}\psi^{i+3\,\ast})_{cd}
\end{array}
\right)\,,
\eeqs
 the global symmetries become manifest:
\beqs
{\cal L}&=& -\frac{1}{2} \Tr V_{\mu\nu} V^{\mu\nu}
\,+\,\frac{1}{2}\left(i(q^{j})^{\dagger}_{\,\,\,a} \bar{\sigma}^{\mu}\left(D_{\mu} q^j\right)^a
\,-\,i(D_{\mu}q^{j})^{\dagger}_{\,\,\,a} \bar{\sigma}^{\mu}q^{j\,a}\right)\,+\,\nonumber\\
&&
\,-\,\frac{1}{2} M \Omega_{jk}\left( q^{j\,a\, T} \Omega_{ab} \tilde{C} q^{k\,b} - (q^{j})^{\dagger}_{\,\,\,a}\Omega^{ab} \tilde{C} (q^{k\,\ast})_b\right)+\nonumber\\
&&
\,+\,\frac{1}{2}\left(i(\psi^{k})^{\dagger}_{\,\,\,ab} \bar{\sigma}^{\mu}\left(D_{\mu} \psi^k\right)^{ab}
\,-\,i(D_{\mu}\psi^{k})^{\dagger}_{\,\,\,ab} \bar{\sigma}^{\mu}\psi^{k\,ab}\right)\,+\,\nonumber\\
&&
\,-\,\frac{1}{2}m\omega_{jk}\left(\psi^{j\,ab\,\mathrm{T}} \Omega_{ac}\Omega_{bd} \tilde{C}\psi^{k\,cd}\,
-\,(\psi^j)^{\dagger}_{\,\,\,ab}\Omega^{ac}\Omega^{bd} \tilde{C}\psi^{k\,\ast})_{cd}\right)\,.
\eeqs

Of the 15 generators $T^A$ of the global $SU(4)$, and 35 generators $t^B$ of $SU(6)$, 
we denote with $A=1\,,\,\cdots\,,\,5$ and  with $B=1\,,\,\cdots\,,\,20$ the broken ones,
which obey
\beqs
\Omega T^A-T^{A\,\mathrm{T}} \Omega &=&0\,,~~~~\omega t^B-t^{B\,\mathrm{T}} \omega \,=\,0\,,
\eeqs
while the unbroken generators with $A=6\,,\,\cdots\,,\,15$ and  with $B=21\,,\,\cdots\,,\,35$ satisfy
\beqs
\Omega T^A+T^{A\,\mathrm{T}} \Omega &=&0\,,~~~~\omega t^B+t^{B\,\mathrm{T}} \omega \,=\,0\,.
\eeqs

As described in Appendix~\ref{Sec:AppendixC},
the Higgs potential in the SM has a global symmetry with
group  $SU(2)_L\times SU(2)_R\sim SO(4)$, which in the present case
is a subgroup of the unbroken global $Sp(4)$. The $SU(3)_c$ gauge group characterising QCD  is
a subgroup of the unbroken global $SO(6)$.
And finally the generator $Y$ of the hypercharge $U(1)_Y$ group  is a linear combination of
one of the generators of $SU(2)_R$ and of the generator of the 
additional $U(1)_X$ unbroken subgroup of $SO(6)$ that commutes with $SU(3)_c$.

\subsection{The pNGB fields}
\label{Sec:pions}

At low energies, the gauge theory with $Sp(4)$ group is best described by an 
EFT that contains only the fields corresponding to the pNGBs parametrising the
$\frac{SU(4)\times SU(6)}{Sp(4)\times SO(6)}$ coset.
We define the  fields $\Sigma_6$  and $\Sigma_{21}$ 
 in terms of the transformation properties of the 
operators that are responsible for spontaneous symmetry breaking, hence identifying 
\beqs
\Sigma_{6}^{\,\,nm}&\sim&\Omega_{ab} q^{n\,a\,\mathrm{T}} \tilde{C} q^{m\,b}\,,\\
\label{Eq:composite}
\Sigma_{21}^{\,\,nm}&\sim&-\Omega_{ab}\Omega_{cd} \psi^{n\,ac\,\mathrm{T}} \tilde{C} \psi^{m\,bd}\,.
\eeqs
$\Sigma_6$ transforms as the antisymmetric representation of $SU(4)$, and $\Sigma_{21}$ as the symmetric representation
of $SU(6)$. 
We parameterise them in terms of  fields $\pi_6$ and $\pi_{21}$ as
\beqs
\Sigma_6&\equiv&e^{\frac{2i \pi_6}{f_6}}\Omega=\Omega e^{\frac{2i \pi_6^{\mathrm{T}}}{f_6}}\,,
~~~~
\Sigma_{21}\,\equiv\,e^{\frac{2i \pi_{21}}{f_{21}}}\omega=\omega e^{\frac{2i \pi_{21}^{\mathrm{T}}}{f_{21}}}\,,
\label{Eq:Sigmas}
\eeqs
where $\pi_6\equiv \pi_6^AT^A$ with $A=1\,,\cdots\,,5$
and $\pi_{21}\equiv \pi_{21}^Bt^B$ with $B=1\,,\cdots\,,20$ are Hermitian matrix-valued fields, 
and the generators $T^A$ are normalised 
by the relation $\Tr T^AT^B=\frac{1}{2}\delta^{AB}=\Tr t^At^B$. 
The decay constants of the pNGBs are denoted by $f_{6}$ and $f_{21}$,
and the normalisation conventions we adopt correspond to those 
in which the decay constant in the chiral Lagrangian of QCD
is $f_{\pi}\simeq 93$ MeV.

In order to identify the operators to be included in the Lagrangian density
 describing the mass-deformed theory,
one treats the (diagonal) mass matrices as (nondynamical) spurions
$M_6 \equiv M\, \Omega$ and $M_{21} \equiv - m \,\omega$ (see Table~\ref{Fig:fields}).
The vacuum expectation value (VEV) of the operators $\Sigma_i$ yields
the symmetry breaking pattern $SU(4)\times SU(6)\rightarrow Sp(4)\times SO(6)$,
aligned with the explicit breaking terms  controlled by $M_6$ and $M_{21}$,
and hence in the vacuum of the theory we have $\langle \pi_i \rangle=0$.

At the leading order in both the derivative expansion and the expansion in small masses, 
the Lagrangian densities of the EFT describing the dynamics of the pNGBs 
of both the $SU(4)/Sp(4)$ and $SU(6)/SO(6)$ cosets are given by
\beqs
\label{Eq:Lpi}
{\cal L}_i&=&\frac{f_i^2}{4}\Tr\left\{\frac{}{}\partial_{\mu}\Sigma_i (\partial^{\mu}\Sigma_i )^{\dagger}\frac{}{}\right\}
\,-\,\frac{v_i^3}{4}\Tr \left\{\frac{}{}M_i \Sigma_i \frac{}{}\right\}\,+\,{\rm h.c.}\\
&=&\Tr\left\{\frac{}{}\partial_{\mu}\pi_i\partial^{\mu}\pi_i\frac{}{}\right\}\,+\,\frac{1}{3f_i^2}\Tr\left\{\frac{}{}
\left[\partial_{\mu}\pi_i\,,\,\pi_i\right]\left[\partial^{\mu}\pi_i\,,\,\pi_i\right]\frac{}{}\right\}\,+\,\cdots\nonumber\,+\\
&&\,+\,\frac{1}{2}\,m_i v_i^3\,\Tr (\Sigma_i\Sigma_i^{\dagger})  \,-\, \frac{m_i v_i^3}{f_i^2}\Tr \pi_i^2 \,+\,\frac{m_i v_i^3}{3 f_i^4}\Tr \pi_i^4 \,+\,\cdots\,,
\eeqs
for $i=6,21$, and with $m_6=M$ and $m_{21}=m$.\footnote{In order to make 
 the expansion for the $SU(6)/SO(6)$ 
formally identical to the $SU(4)/Sp(4)$ case, 
 we chose opposite signs in the definition of the mass matrices
and condensing operators. The origin for this technical annoyance is the fact
 that $\Omega^2=-\mathbb{1}_4$, while
$\omega^2=\mathbb{1}_6$. We also note that one has to exercise caution 
with the trace of the identity matrix, which may introduce numerical factors that differ in the expansions 
when traces are taken in products that do not include the group generators.
\label{Foot1}} The condensates are parameterised by $v_6$ and $v_{21}$,
which  have dimension of a mass. In the $SU(4)/Sp(4)$ case 
$\Tr \Sigma_6\Sigma_6^{\dagger}=4$, 
and in the $SU(6)/SO(6)$ case
$\Tr \Sigma_{21}\Sigma_{21}^{\dagger}=6$.

In order to describe the coupling to the Standard Model, one  chooses 
appropriate embeddings for the
relevant $SU(2)_L\times SU(2)_R$ and $SU(3)_c\times U(1)_X$ 
groups and promotes the ordinary derivatives to covariant derivatives.
By doing so, the irreducible representations of the unbroken 
$Sp(4)\times SO(6)$ can be decomposed in representations of the 
SM groups (see Appendix~\ref{Sec:AppendixC}).

Starting from the $SU(4)/Sp(4)$ coset, the five pNGBs transform as the 
fundamental representation of $SO(5)\sim Sp(4)$.
Because $SO(4)\sim SU(2)_L\times SU(2)_R$ is a natural subgroup of $SO(5)$,
one finds the decomposition $5=1+4$, and hence four of the pNGBs are identified with the SM Higgs doublet, 
while the one additional degree of freedom is a real singlet of $SU(2)_L\times SU(2)_R$.
In the conventions of~\cite{Lee:2017uvl,Bennett:2017kga}, the latter 
is denoted by $\pi^3$---or  $\pi^3_6$ if one needs to avoid ambiguity with the
set of pNGBs from the $SU(6)/SO(6)$ coset (see also Appendix~\ref{Sec:AppendixGW}).

A similar exercise can be performed for the $SU(6)/SO(6)$ coset.
By remembering that $SO(6)\sim SU(4)$, the $20$ pNGBs transform as 
 the $20^{\prime}$ irreducible representation of this  $SU(4)$
(the only self-conjugate among the three representations of $SU(4)$ that
 has 20 real elements).~\footnote{In the rest of the paper, we will always denote 
 this representation as $20^{\prime}$, 
for the purpose of avoiding confusion with the representations of the unrelated broken global $SU(4)$.}
The decomposition of $SU(4)$ in its maximal $SU(3)_c\times U(1)_X$ subgroup dictates that 
$20^{\prime}=8+6_{\mathbb{C}}$ (see also Appendix~\ref{Sec:AppendixGW}).

\subsection{EFT: Hidden local symmetry}
\label{Sec:HLS}

This subsection is devoted to the treatment of spin-1 composite states.
All  irreducible representations coming from the $SU(4)/Sp(4)$ theory can be decomposed following 
the same principles illustrated by the pNGBs, into representations of the groups relevant to SM physics. 
For example the $10$ of $SO(5)$ decomposes 
as $10=4+6$ of $SO(4)$, so that the composite vector mesons V
of the $SU(4)/Sp(4)$ theory  (corresponding to the $\rho$ mesons of QCD) decompose into a 
complex doublet and a complex triplet of $SU(2)_L\times SU(2)_R$. The axial vectors 
AV (corresponding to the
$a_1$ mesons in QCD) transform with the same internal quantum numbers as the pNGBs 
and hence give rise to a complex doublet and a real singlet.
In the $SU(6)/SO(6)$ coset, 
 the composite vector mesons V transform as the $15$ of $SO(6)\sim SU(4)$, which decomposes
as $15=1+3_{\mathbb{C}}+8$ of $SU(3)_c$, and the axial-vector mesons AV 
transform as the $20^{\prime}$ of $SO(6)$, which decomposes as $20^{\prime}=8+6_{\mathbb{C}}$ of $SU(3)_C$.

 \begin{figure}
\begin{center}
\begin{picture}(330,100)
\put(0,0){\includegraphics[width=.375\textwidth]{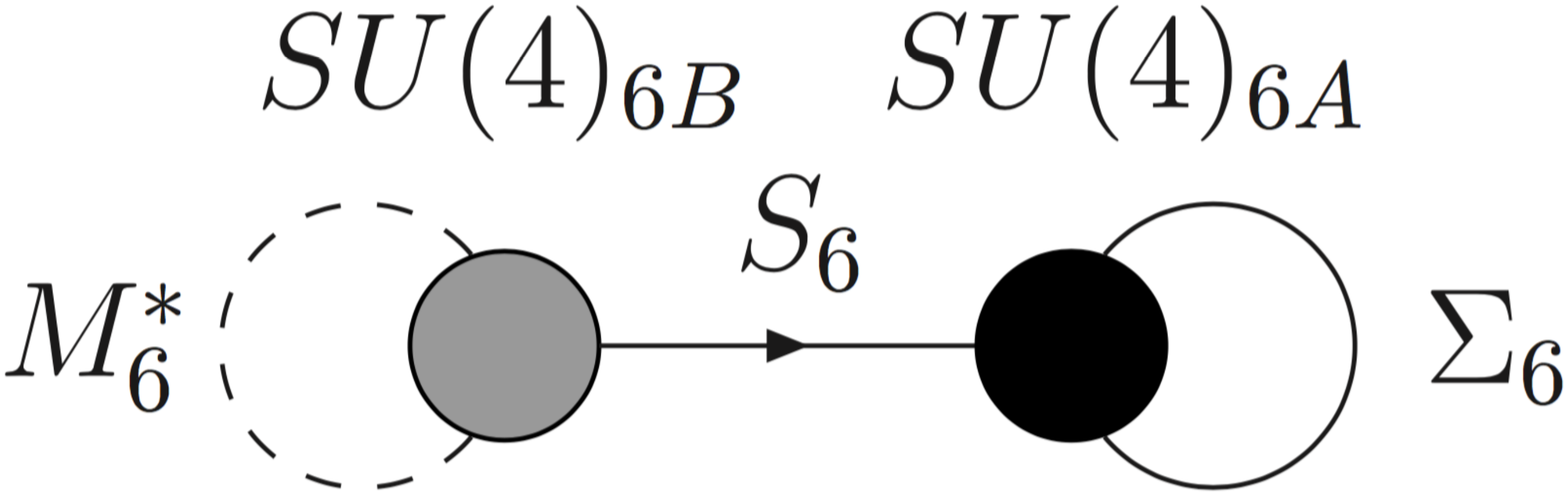}}
\put(165,0){\includegraphics[width=.375\textwidth]{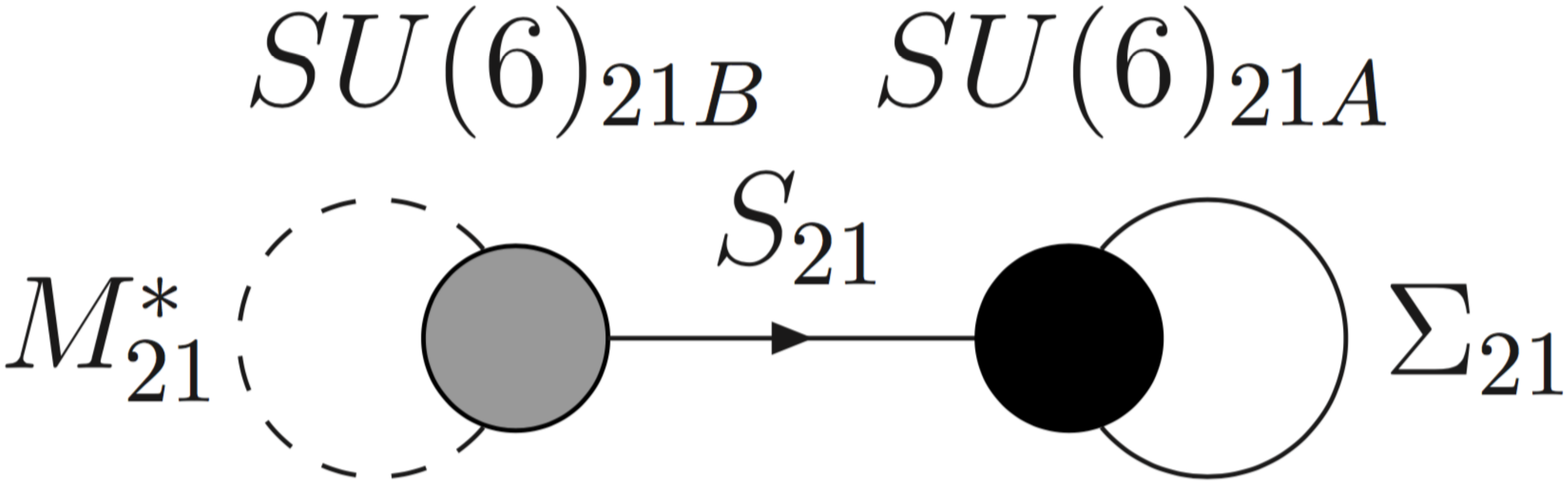}}
\end{picture}
\caption{The moose diagrams representing the low-energy 
EFT descriptions. On the left $SU(4)_{6A}$ is gauged, while $SU(4)_{6B}$ 
is a global symmetry (including the $SU(2)_L\times SU(2)_R$), and the combination
of the nontrivial VEVs of $S_6$ and $\Sigma_6$ breaks the symmetry 
to $Sp(4)$, giving mass to all the vector mesons and 
leaving a set of five light pions. On the right, the same 
principles are applied to $SU(6)_{21B}\times SU(6)_{21 A}$ 
and to its breaking to the $SO(6)$ subgroup.}
\label{Fig:EFT}
\end{center}
\end{figure}

We study a reformulation of the 
low-energy EFT description
of the model, that is intended to capture also the behaviour of the lightest 
vector and axial-vector states, in addition to the pNGBs (as in the chiral Lagrangian).
It is based on hidden local symmetry ~\cite{
Bando:1984ej,Casalbuoni:1985kq,Bando:1987br,Casalbuoni:1988xm,Harada:2003jx} 
(see also~\cite{Georgi:1989xy,Appelquist:1999dq,Piai:2004yb,Franzosi:2016aoo}) 
and illustrated by the diagram in  Fig.~\ref{Fig:EFT}.
There are well-known limitations to the applicability of this type of EFT treatment, 
which we will discuss in due time.

We consider the two moose diagrams as completely independent from one another.
We follow closely the notation of Ref.~\cite{Bennett:2017kga} in describing the $SU(4)/Sp(4)$ 
coset, except for the fact that we include only single-trace operators in the Lagrangian density. 
Because the breaking is due to the condensate of the operator transforming in the $6$ of $SU(4)$, 
we label all the fields of relevance to the low-energy EFT with a 
subscript, as in $S_6$.
The scalar fields  $S_6$ transform as a bifundamental of $SU(4)_{6B}\times SU(4)_{6A}$, 
while $\Sigma_6$  transform as the antisymmetric
representations of $SU(4)_{6A}$.
Hence the transformation rules are as follows:
\beqs
S_6 &\rightarrow& U_{6B} S_6 U_{6A}^{\dagger} \,,~~~~~~~~
\Sigma_6 \,\rightarrow\, U_{6A} \Sigma_{6} U_{6A}^{\mathrm{T}}
\,,
\eeqs
where $U_{6A}$ and $U_{6B}$ are group elements  of $SU(4)_{6A}$ and $SU(4)_{6B}$, respectively.

The EFT is built by imposing the nonlinear constraints
$\Sigma_6\,\Sigma_6^{\dagger} = \mathbb{1}_4 \,=\, S_6\,S_6^{\dagger}$,
which are solved by parameterising $S_6 =e^{\frac{2i \sigma_6}{F}}$ 
and $\Sigma_6=e^{\frac{2i \pi_6}{f}}\Omega=\Omega e^{\frac{2i \pi_6^{\mathrm{T}}}{f}}$.
$M_6=M \Omega$ is a constant matrix, introducing explicit symmetry breaking. One can think
of it as a spurion in the antisymmetric representation of $SU(4)_{6B}$, so that 
as a field it would transform according to
  $M_6^{\ast} \,\rightarrow\, U_{6B} M_{6}^{\ast} U_{6B}^{\mathrm{T}}$.
The 15 real Nambu-Goldstone fields $\sigma_6=\sigma_6^A T^A$ and five real $\pi_6=\pi_6^AT^A$
are in part gauged
into providing the longitudinal components for the $15$ gauge bosons of $SU(4)_{6A}$, 
so that only five linear combinations 
remain in the spectrum as physical pseudoscalars.
One then uses $\Sigma_i$ and its derivatives,  as well as $M_i$, to build 
all possible operators allowed by the symmetries, organises them
as an expansion in derivatives (momenta $p^2$) and explicit mass terms ($M$), 
and writes a Lagrangian density that includes all
such operators up to a given order in the expansion.
We also restrict attention to operators that can be written as single traces, as anticipated.

Truncated at the next-to-leading order, the Lagrangian density takes the following form, which we borrow 
from Ref.~\cite{Bennett:2017kga}\footnote{The very last term of the Lagrangian density
differs from Ref.~\cite{Bennett:2017kga}, as we
rewrite the subleading  correction to the pion mass in terms of a 
single-trace operator. The 
equations giving the masses and decay constants are independent 
of the dimensionality of the matrices used.
We notice also an inconsequential typo in Eq.~(2.16) 
of~\cite{Bennett:2017kga}, in which the last term should have a 
$+$ sign rather than a $-$ sign, in order to be consistent with Eqs.~(2.30)
and (2.31) of~\cite{Bennett:2017kga} itself.}:
\beqs
{\cal L}_6&=& 
-\frac{1}{2}\Tr\,A_{\,\mu\nu}A^{\mu\nu}
-\frac{\kappa}{2}\Tr\left\{ A_{\mu\nu} \Sigma (A^{\mu\nu})^{\mathrm{T}} \Sigma^{\ast}\right\}\nonumber+\\
&&+\frac{f^2}{4}\Tr\left\{\frac{}{}D_{\mu}\Sigma\,(D^{\mu}\Sigma)^{\dagger}\right\}\,+\,\frac{F^2}{4}\Tr\left\{\frac{}{}D_{\mu}S\,(D^{\mu}S)^{\dagger}\right\}\nonumber+\\
&&+b \frac{f^2}{4}\Tr\left\{D_{\mu}(S\Sigma )\left(D^{\mu}(S\Sigma)\right)^{\dagger}\right\}\,+\,
c\frac{f^2}{4} \Tr\left\{D_{\mu}(S\Sigma S^{\mathrm{T}})\left(D^{\mu}(S\Sigma S^{\mathrm{T}})\right)^{\dagger}\right\}\,+\nonumber\\
&&-\frac{v^3}{8}\Tr\left\{\frac{}{}M\, S\, \Sigma\,S^{\mathrm{T}} \right\}\,+\,{\rm h.c.} +\label{Eq:L6}\\
&&- \frac{v_1}{4} \Tr\left\{\frac{}{} M\, (D_{\mu} S)\, \Sigma \, (D^{\mu} S)^{\mathrm{T}} \right\}\,
 -\frac{v_2}{4} \Tr\left\{\frac{}{} M\, S\,(D_{\mu}  \Sigma) \, (D^{\mu} S)^{\mathrm{T}} \right\}\,+\,{\rm h.c.}\,+\nonumber\\
 &&
 -\frac{y_3}{8}\Tr\left\{A_{\mu\nu}\Sigma\left[(A^{\mu\nu})^{\mathrm{T}}S^{\mathrm{T}} M S-S^{\mathrm{T}} M S A^{\mu\nu}\right]\right\}\,+\,{\rm h.c.}\nonumber+\\
  &&
 -\frac{y_4}{8}\Tr\left\{A_{\mu\nu}\Sigma\left[(A^{\mu\nu})^{\mathrm{T}}S^{\mathrm{T}} M S+S^{\mathrm{T}} M S A^{\mu\nu}\right]\right\}\,+\,{\rm h.c.}\,\nonumber+
 \\
&&
+\frac{v_5^2}{32}\Tr \left\{\frac{}{}M S \Sigma S^{\mathrm{T}} M S \Sigma S^{\mathrm{T}}\frac{}{}\right\}\,+\, {\rm h.c.}\,.\nonumber
\eeqs
We omitted, for notational simplicity, the subscript ``$6$" on all fields and all the parameters.
We should stress that we made some simplifications, and omitted some operators, as discussed in~\cite{Bennett:2017kga}.
The covariant derivatives introduce the parameter $g_{\rm V}$, controlling the coupling of the spin-1 states.
They can be written as follows: 
\beqs
D_\mu S = \partial_\mu S - i S g_{\rm V} A_\mu,
\eeqs
and
\beqs
D_\mu \Sigma = \partial_\mu \Sigma + i \left[
(g_{\rm V} A_\mu) \Sigma + \Sigma (g_{\rm V} A_\mu)^T
\right].
\eeqs

The analogue of Eq.~(\ref{Eq:L6}) in the $SU(6)/SO(6)$ case is obtained in the same way. 
The only changes are 
the replacement of $\Sigma_6$ by $\Sigma_{21}$, that now depends on $20$ $\pi_{21}^i$ 
fields, of $S_6$ by $S_{21}$, that depends on
35 $\sigma_{21}^i$ fields, of $M_6$ by 
$M_{21}\equiv -m \omega$, and of $A^A_{6\,\mu}$ by the 35 vector
 bosons $A^A_{21\,\mu}$ of $SU(6)_{21 A}$.
 Finally, one must also require the change of sign
  $\kappa_{21} \leftrightarrow -{\kappa}_{6}$ in the second term  of the Lagrangian,
 for the same reason explained in footnote~\ref{Foot1}.

With the conventions outlined above, 
masses and decay constants are given by the same relations
as in Ref.~\cite{Bennett:2017kga}, both for the mesons sourced by fundamental and 
antisymmetric fermion bilinears:
\beqs
\label{eq:gfit_mv}
M_{\rm V}^2&=&\frac{1}{4(1+\kappa+m\,y_3)} {g_{\rm V}}^2 \left(b f^2+F^2+2 m v_1\right)\,,\\
M_{\rm AV}^2&=&\frac{1}{4(1-\kappa-m \,y_4)} {g_{\rm V}}^2 \left(b f^2+F^2+2 m v_1\right)+\\
&&\nonumber+\frac{g_{V}^2}{1-\kappa-m \,y_4} \left(f^2+m
   (v_2-v_1)\right)\,,\\
f_{\rm V}^2&=&
   \frac{1}{2} \left(b f^2+F^2+2 m v_1\right)\,,\\
f_{\rm AV}^2&=&
   \frac{\left(b f^2-F^2+2 m (v_1-v_2)\right)^2}{2 \left((b+4) f^2+F^2-2 m
   v_1+4 m v_2\right)}\,,\\
f_0^2&=&\frac{}{}F^2+(b+2c)f^2\,.
\eeqs
The pNGB decay constants obey the following 
relation:\footnote{In Ref.~\cite{Bennett:2017kga} we denoted the decay constant 
of the PS mesons
as $f_{\pi}(0)$, to explicitly highlight that this is not the constant that naturally 
appears in the $\pi\pi\rightarrow\pi\pi$ scattering amplitude.}
\beqs
f_{\rm PS}^2&=&
f_0^2-f_{\rm V}^2-f_{\rm AV}^2\,.
\label{eq:gfit_fps}
\eeqs
It was observed   in Ref.~\cite{Bennett:2017kga} that
 $f_0^2=f_{\rm PS}^2+f_{\rm V}^2+f_{\rm AV}^2$ is independent of $m$ 
as the accidental consequence of the truncations and of the omission of some operators.
It was also shown that some of the couplings parameterise the violation of the saturation
of the Weinberg sum rules,
when truncated at this level---retaining only the lightest excitations sourced by the V and AV operators 
rather than the whole infinite tower of states.

In both the $SU(4)/Sp(4)$ as well as $SU(6)/Sp(6)$ cosets, truncated at this level
the Lagrangian implies that 
the  mass of the pions satisfies a generalised Gell-Mann-Oakes-Renner relation, which reads as follows:
\beqs
m_{\rm PS}^2f_{\rm PS}^2 &=&m\left(v^3+\frac{}{}m v_5^2\right)\,,
\eeqs
which implies a dependence of the condensate on $m$.
We notice the presence of the constant $g_{\rm V}$, which enters the $g_{\rm VPP}$ coupling
between V and two PS states and has an important role in controlling the EFT expansion.

\section{Lattice Model}
\label{Sec:lattice}

The lattice action and its numerical treatment via Monte Carlo methods
are the main topics of this section. 
Most of the material covered here
 is based upon well-established processes,  and we discussed its application to our
 programme elsewhere~\cite{Bennett:2017kga,Bennett:2019jzz},
 hence we summarise it briefly, mostly for the purpose of defining the notation
 and language  we adopt later in the paper.

\subsection{Lattice definitions}
\label{Sec:definitions}

In the numerical (lattice) studies, we should adopt a discretised  four-dimensional Euclidean-space
version of \Eq{lagrangian}.
But as we perform our numerical work in the quenched approximation, 
we only need the pure gauge part of the Lagrangian density,
as in pioneering studies of $Sp(2N)$ 
Yang-Mills theories in
Ref.~\cite{Holland:2003kg}.
We employ the standard Wilson action
\beq
S_g\equiv\beta \sum_x \sum_{\mu<\nu} \left(1-\frac{1}{4} {\rm Re}\, {\rm Tr}\, \mathcal{P}_{\mu\nu}(x)\right),
\label{eq:action}
\eeq
where $\beta=8/g^2$ is the bare lattice coupling and the trace is over colour indices. 
 The elementary  plaquette $\mathcal{P}_{\mu\nu}$ 
is a path-ordered product of (fundamental) link variables $U_\mu(x)$, 
the group elements of $Sp(4)$, and reads as follows:
\beq
\mathcal{P}_{\mu\nu}(x)\equiv
U_\mu (x) U_\nu (x+\hat{\mu}) U_\mu^\dagger(x+\hat{\nu}) U_\nu^\dagger(x).
\eeq

Given the action in \Eq{action}, we generate the gauge configurations by implementing 
a heat bath (HB) algorithm with microcanonical overrelaxation updates.
Technical  details, including the modified Cabbibo-Marinari procedure~\cite{Cabibbo:1982zn} 
 and the resymplectisation process we adopted,
can be found in Refs.~\cite{Bennett:2017kga,Bennett:2017kbp}. 
 The HiRep code~\cite{DelDebbio:2008zf}, appropriately adapted to the
 requirements of this project, is used for the numerical 
calculations.

\begin{table}
\begin{center}
\begin{tabular}{|c|c|c|c|c|}
\hline\hline
Ensemble & $\beta$ & 
$N_t\times N_s^3$ & $\langle P \rangle$  & $w_0/a$ \\
\hline
QB1 & $7.62$ & $48\times 24^3$ & $0.60192$ & $1.448(3)$\\
QB2 & $7.7$ & $60\times 48^3$ & $0.608795$ & $1.6070(19)$\\
QB3 & $7.85$ & $60\times 48^3$ & $0.620381$ & $1.944(3)$\\
QB4 & $8.0$ & $60\times 48^3$ & $0.630740$ & $2.3149(12)$\\
QB5 & $8.2$ & $60\times 48^3$ & $0.643228$ & $2.8812(21)$\\
\hline\hline
\end{tabular}
\caption{%
\label{tab:ensemble_Q}%
List of ensembles used for quenched calculations. For each ensemble, we report the bare coupling
$\beta$, the lattice size $N_t\times N_s^3$, the average plaquette $\langle P \rangle$
and the gradient-flow scale $w_0/a=1/\hat{a}$.
}
\end{center}
\end{table}

The pure $Sp(4)$ Yang-Mills  lattice theory at any values of $\beta$ 
can in principle be  connected smoothly to the continuum, as
 no evidence of  bulk transitions has been found~\cite{Holland:2003kg}. 
In this study, we work in the regime with $\beta > 7.5$. 
In a previous publication~\cite{Bennett:2017kga}, some of us used 
two values of the coupling ($\beta=7.62$ and $\beta=8.0$), 
and performed preliminary studies of the meson spectrum 
with fermions in the fundamental representation, in the quenched limit. 
In order to carry out the continuum extrapolation,
here we extend those studies by including three additional
 values of the bare lattice coupling, 
$\beta=7.7,\,7.85,\,8.2$. 
The four-dimensional Euclidean lattice has size $N_t\times N_s^3$, 
with $N_t$ and $N_s$ the temporal and spatial extents, respectively. 
We impose periodic boundary conditions in all directions for the gauge fields. 
While for the ensemble at $\beta=7.62$ we reuse the configurations generated on a $48\times24^3$ lattice 
already employed in the quenched calculations in Ref.~\cite{Bennett:2017kga}, 
for all the other values of the coupling we  generate new configurations 
with $60\times 48^3$ lattice points. 
For each lattice coupling we generate $200$ gauge 
configurations, separated by $12$ trajectories\footnote{
Conventionally, for heat bath simulations like those used in this work, 
a full update of the lattice link variables is called a {\em sweep} rather than a {\em trajectory}. 
However, to match the terminology of our dynamical simulations \cite{Bennett:2017tum,Bennett:2017ttu,Bennett:2017kga,Bennett:2019jzz,Bennett:2019ckt}, 
we use the term {\em trajectory} for a full lattice gauge field update also in the present context. 
} between adjacent configurations. 
To ensure thermalisation, we  discard the first $600$ trajectories. 
In \Tab{ensemble_Q} we summarise the ensembles. 
In addition to the ensemble name, the lattice coupling and the lattice size, we also present 
two measured quantities: the average plaquette $\langle P \rangle$ and the gradient-flow scale 
$w_0/a$ in lattice units. 
The former is defined by $\langle P \rangle \equiv {\rm Re}
 \sum_x \sum_{\mu<\nu} {\rm Tr}\, \mathcal{P}_{\mu\nu}(x) / (24\times N_t\times N_s^3)$, 
while the latter will be defined and discussed in the next subsection. 
The statistical uncertainties are estimated by using a standard bootstrapping technique for resampling, 
which will also be applied to the rest of this work.

\subsection{Scale setting and topology}
\label{Sec:scale}

\begin{figure}
  \center
  \includegraphics[width=.80\textwidth]{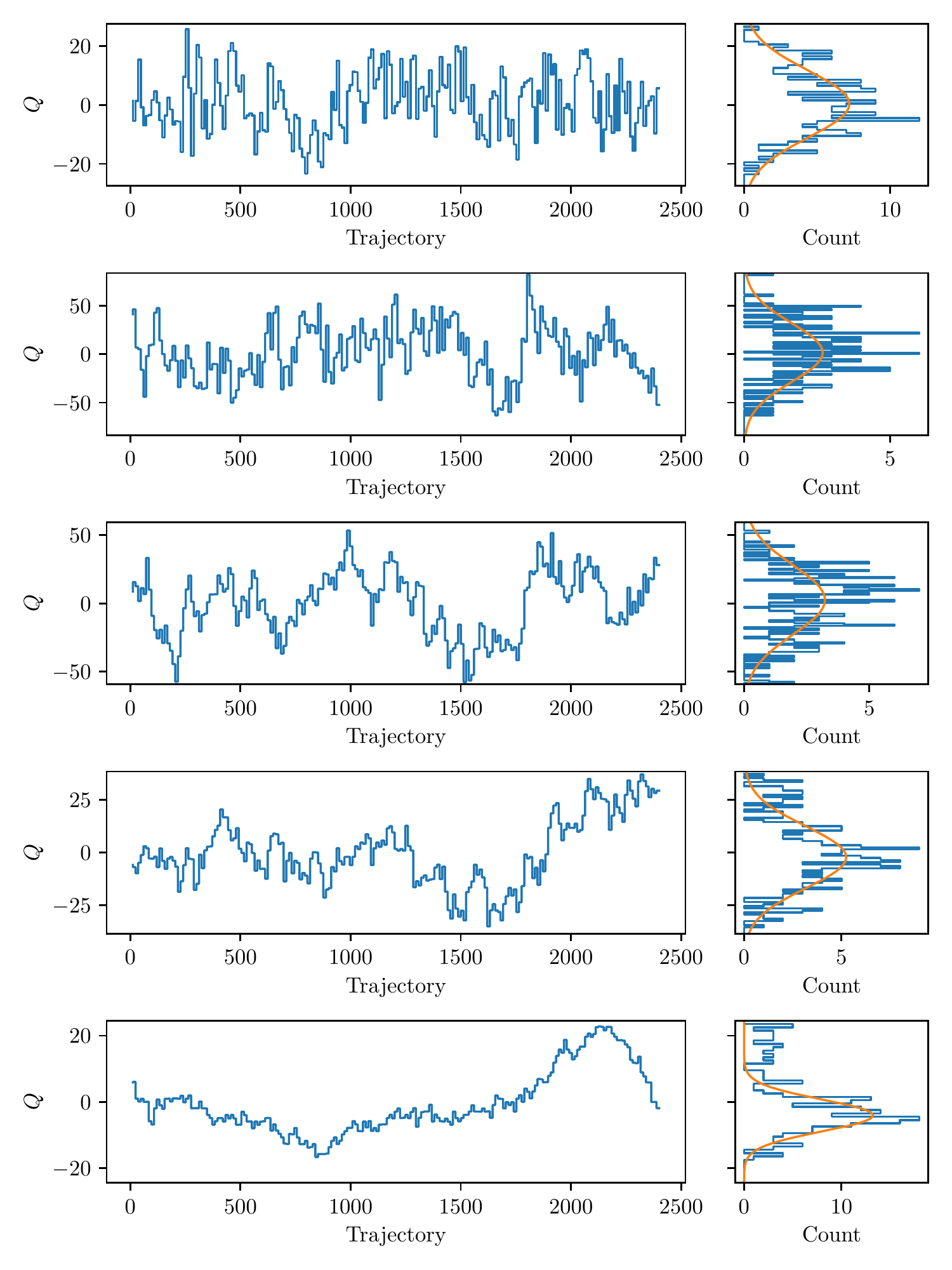}
  \caption{
Topological charge histories (left panels), and histograms (right panels), for the five ensembles
 QB1--QB5,  listed from top to bottom. Fitted parameters are
  given in Table~\ref{Fig:tabtop}. In the plots we show $Q$ evaluated only on the configurations used
  in the data analysis.
}
  \label{fig:topcharge}
\end{figure}

\begin{table}
\begin{center}
\begin{tabular}{|c|c|c|c|}
 \hline \hline
     & $Q_0$ & $\sigma$ & $\tau_{\exp}$ \\
    \hline
    QB1 & $0.29(92)$ & $11.43(94)$ & $1.35(21)$ \\
    QB2 & $1.6(2.3)$ & $30.2(2.3)$ & $2.95(24)$ \\
    QB3 & $2.5(2.3)$ & $25.4(2.3)$ & $7.73(12)$ \\
    QB4 & $-2.2(1.1)$ & $14.7(1.1)$ & $15.79(65)$ \\
    QB5 & $-4.12(46)$ & $4.81(46)$ & $34.1(1.7)$\\
     \hline \hline
\end{tabular}
\caption{Fitted parameters from topological charge histories; see also Fig.~\ref{fig:topcharge}.
The autocorrelation time $\tau_{\exp}$ is expressed in units of consecutive configurations.}
\label{Fig:tabtop}
\end{center}
\end{table}

In numerical lattice calculations,
 all dimensional quantities can be written in terms of  the lattice spacing $a$,
for example by defining a dimensionless mass as $m^{\rm latt}=m a$.
But in taking the continuum limit,
the lattice spacing vanishes, $a\rightarrow 0$.
Hence, in order to connect the lattice observables to continuum ones,
we have to set a common physical scale that  allows the comparison.
We adopt as our scale-setting method  L\"{u}scher's gradient-flow (GF) scheme,
using the definition of Wilson flow in Ref.~\cite{Luscher:2010iy} 
(see also Refs.~\cite{Luscher:2009eq,
Luscher:2011bx,Fujikawa:2016qis}).
This method is particularly suitable for the purpose of this work,
since it  relies on  theoretically defined quantities that do not require 
direct experimental input.

The scale-setting procedure with the GF scheme in  $Sp(4)$ theories
has been first discussed  in Ref.~\cite{Bennett:2017kga},
both for the pure Yang-Mills and for 
the theory with two fundamental Dirac fermions (see also~\cite{Bennett:2017tum,Bennett:2019jzz}).
We follow the  same procedure throughout this work:
we define the flow scale $w_0$ by $\mathcal{W}|_{t=w_0^2}=\mathcal{W}_0$ \cite{Borsanyi:2012zs},
where $\mathcal{W}(t)$ is the derivative of the action density built from gauge fields
at nonzero fictitious flow time $t$.
The reference value $\mathcal{W}_0=0.35$ has been chosen
to minimise both discretisation and finite-volume effects~\cite{Bennett:2017kga}
(though with the caveats discussed in Refs.~\cite{Fodor:2014cpa,Ramos:2015baa,Lin:2015zpa}).
We also choose a four-plaquette clover for the definition of the  field-strength tensors~\cite{Luscher:2010iy}.
The resulting values of the flow scale in lattice units $w_0/a$ are shown in \Tab{ensemble_Q}.

We measure the history of the topological charge $Q$ to monitor the possible emergence of topological freezing, 
which might affect spectral measurements~\cite{Luscher:2011kk,Galletly:2006hq} (see also Refs.~\cite{Brower:2003yx,Bernard:2017npd}). 
Since $Q\equiv\sum_x \frac{1}{32\pi^2} \epsilon_{\mu\nu\rho\sigma} \Tr \{U_{\mu\nu}(x) U_{\rho\sigma}(x)\}$ is dominated by 
ultraviolet (UV) fluctuations when calculated directly on the configurations in ensembles QB1--5, 
configurations that have been smoothed by the gradient flow are instead used. 
$Q$ is measured at the point in the flow such that the smoothing radius $\sqrt{8t}=L/2$.

In Fig.~\ref{fig:topcharge} we present the histories and histograms of $Q$ along the Markov chain for all ensembles in Table~\ref{tab:ensemble_Q}, 
the latter of which is fitted with the Gaussian fit form $n(Q) \propto \exp\left(-\frac{(Q-Q_0)^2}{2\sigma^2}\right)$. 
In Table \ref{Fig:tabtop} we present the results of this fit, and the exponential autocorrelation time $\tau_{\mathrm{exp}}$ 
calculated via a fit to the autocorrelation function of $Q$. 
In the five ensembles, there is no clear evidence of a freeze-out of the topology; 
the histograms clearly show sampling from multiple topological sectors, 
and the distributions are peaked within $1\sigma$ of $Q=0$.

However, as we move to finer lattice spacing, we observe that the autocorrelation time of the topological charge grows significantly; 
in the case of QB5, this has grown to around 34 configurations. 
In this case specifically $Q_0=-4.12$ is also marginal compared to $\sigma=4.81$. 
This effect may be due to the fact that a change of the discrete global quantity $Q$ by local updates becomes 
disfavoured in the approach to the continuum limit.

To verify that this increasing $\tau_{\mathrm{exp}}$ and marginal $Q_0$ do not affect the spectroscopic results 
we obtain from these ensembles, we generate an additional ensemble QB$\overline{5}$ of 2400 trajectories starting 
from the last configuration in QB5. We repeated the measurements of meson masses and decay constants, 
and of the topological charge history, on 200 configurations sampled from QB$\overline{5}$. 
While the value of $Q_0$ differs between the two ensembles, the meson masses and decay constants do not show significant deviations 
(beyond the statistical fluctuations). 
We report these tests in detail in Appendix~\ref{Sec:AppendixF}. This suggests that any systematic effect associated with the long autocorrelation 
time of the topological charge and the marginal $Q_0$ on the spectroscopy is comfortably smaller than the statistical error 
for the ensembles and observables we study, and we use ensembles QB1--5 for the remainder of the analysis.

\section{Of quenched mesons}
\label{Sec:mesons}

In this section, we present the main numerical results of our study.
We start by defining the mesonic two-point correlation functions that are computed numerically,
and the observables we extract from them, namely the meson masses 
and decay constants. We provide some technical details
about the otherwise standard procedure we follow,  
in order to clarify how different representations of the gauge group are implemented.
Perturbative renormalisation of the
decay constants is summarised towards the end of Sec.~\ref{Sec:correlators}. 
We perform  continuum extrapolations with
the use of Wilson chiral perturbation theory (W$\chi$PT) in Sec.~\ref{Sec:continuum}.
We devote Secs.~\ref{Sec:quenched} and~\ref{Sec:antisymmetric} to present the numerical results
for the mesons made of fermions transforming in the fundamental and 
2-index antisymmetric 
representations, respectively, and conclude with a comparison of the two 
representations in Sec.~\ref{Sec:comparison}.
For practical reasons, in this section we specify our results to the theory with $N_f=2$ 
fermions on the fundamental  representation and $n_f=3$ on the antisymmetric, 
though the results of the quenched calculations apply for generic $N_f$ and $n_f$.

\begin{table}
\begin{center}
\begin{tabular}{|c|c|c|c|c|c|}
\hline\hline
{\rm Label} & {\rm ~Interpolating operator~} & {\rm ~~~Meson~~~} 
& {\rm ~~~$J^{P}$~~~}
& $Sp(4)$ & $SO(6)$  \cr
 $M$ & $\mathcal{O}_M$& {\rm ~~~in QCD~~~} 
& 
& &   \cr
\hline
PS & $\overline{Q^i}\gamma_5 Q^j$ & $\pi$ & $0^{-}$ & $5 (+1)$ & $1$\cr
S & $\overline{Q^i} Q^j$ & $a_0$ & $0^{+}$ & $5 (+1)$ & $1$\cr
V & $\overline{Q^i}\gamma_\mu Q^j$ & $\rho$ & $1^{-}$ & $10$ & $1$\cr
T & $\overline{Q^i}\gamma_0\gamma_\mu Q^j$ & $\rho$ & $1^{-}$ & $10 (+5+1)$ & $1$ \cr
AV & $\overline{Q^i}\gamma_5\gamma_\mu Q^j$ & $a_1$ & $1^{+}$ & $5 (+1)$ & $1$\cr
AT & $\overline{Q^i}\gamma_5\gamma_0\gamma_\mu Q^j$ & $b_1$ & $1^{+}$ & $10 (+5+1)$ & $1$\cr
\hline
ps & $\overline{\Psi^k}\gamma_5 \Psi^m$ & $\pi$ & $0^{-}$ & $1$ & $20^{\prime} (+1)$\cr
s & $\overline{\Psi^k} \Psi^m$ & $a_0$ & $0^{+}$ & $1$ & $20^{\prime}  (+1)$\cr
v & $\overline{\Psi^k}\gamma_\mu \Psi^m$ & $\rho$ & $1^{-}$ & $1$ & $15$ \cr
t & $\overline{\Psi^k}\gamma_0\gamma_\mu \Psi^m$ & $\rho$ & $1^{-}$ & $1$  & $15 (+20^{\prime} +1)$\cr
av & $\overline{\Psi^k}\gamma_5\gamma_\mu \Psi^m$ & $a_1$ & $1^{+}$ & $1$ & $20^{\prime}  (+1)$\cr
at & $\overline{\Psi^k}\gamma_5\gamma_0\gamma_\mu \Psi^m$ & $b_1$ & $1^{+}$ & $1$ & 
$15 (+20^{\prime} +1)$\cr
\hline\hline
\end{tabular}
\end{center}
\caption{
Interpolating operators $\mathcal{O}_M$ built of Dirac fermions on the fundamental $Q^{i\,a}$
and antisymmetric $\Psi^{k\,ab}$.  We show explicitly the
 flavour indices $i,j=1\,,\,2$  and $k,m=1\,,\,2\,,\,3$,
 while colour and spinor indices are implicit and summed over.
We also show the $J^P$ quantum numbers, the  corresponding QCD 
mesons sourced by the analogous operator, and
 the irreducible representation  of  the unbroken global 
$Sp(4)\times SO(6)$ spanned by the meson (see also~\cite{Lewis:2011zb}).
We indicate in parentheses other nontrivial representations 
that are obtained with the
same operator structure but that we do not study in this paper as they source heavier states.
The singlets ($1$ of both $Sp(4)$ and $SO(6)$) 
are ignored, as we choose to analyse only the operators with
$i\neq j$ or $k\neq m$.
More details about the symmetries can be found in Appendix~\ref{Sec:AppendixD},
and the details of a specific choice of basis for the global $SU(4)$ are presented in
Appendix~\ref{Sec:AppendixE}.
}
\label{tab:mesons}
\end{table}

\subsection{Correlation functions}
\label{Sec:correlators}

We extract  masses and  decay constants of the lightest flavoured spin-0 and spin-1 mesons 
from the corresponding Euclidean two-point correlation functions of operators ${\cal O}_M$
involving Dirac fermions $Q$ transforming  in the fundamental and $\Psi$ in the 2-index
antisymmetric representation, as listed in Table~\ref{tab:mesons}.
In the table, colour and spin indices are implicitly summed over, while the flavour indices 
 $i\neq j$ ($k\neq m$) are chosen.
The operators of the form $\overline{Q^i} \Gamma_M Q^j$
are gauge invariant
and they source the meson states $M$. Spin and parity $J^P$ are determined by
the choice of  $\Gamma_M$. 
The operators built with $\Gamma=\gamma_5,\,\gamma_\mu,\,\gamma_5\gamma_\mu$,
 with $\mu=$1, 2, 3,
 correspond to the  pseudoscalar (PS), vector (V), and axial-vectors (AV) mesons, respectively.
They appeared in the EFT  discussion in \Sec{HLS}.
For all of them, we measure both the masses and the decay constants
of the particles that they source.
For  completeness, we also calculate the correlation functions
 built with $\Gamma=\mathbb{1}_4,\,\gamma_0\gamma_\mu,\,\gamma_5\gamma_0\gamma_\mu$, 
which refer to scalar (S), (antisymmetric) tensor (T), and axial tensor (AT),
but we extract only the masses of the lightest states sourced by these operators. 
The operators $\overline{\Psi^k} \Gamma_M \Psi^m$ are defined and classified in the
same way, except that we denote them with lowercase letters as ps, v, av, s, t, and at, respectively.
In \Tab{mesons}, we also show the irreducible representation of the
 unbroken global symmetry $Sp(4)\times SO(6)$, 
as well as the corresponding mesons in QCD, to provide intuitive guidance to the reader. 
We also recall that because of the (pseudo)real nature of the 
representations we use, there is no  difference between meson and diquark operators.
More details about the classification of the mesons and the relation between 
 four-component  and two-component spinors can be found 
in Appendixes \ref{Sec:AppendixD} and \ref{Sec:AppendixE}.

The two-point correlation functions 
at positive Euclidean time $t$ and vanishing momentum $\overrightarrow{p}$
can be written as
\beq
C_{M,\,M'}(t)\equiv\sum_{\vec{x}}\langle 0 | \mathcal{O}_M(\vec{x},t) 
\mathcal{O}^\dagger_{M'}(\vec{0},0) |0 \rangle. 
\eeq
We extract physical observables from these objects.
In most of our calculations we set $M=M'$, with the exception of the extraction of the
 pseudoscalar decay constant, 
which involves both ${\cal O}_{\rm PS}$ and ${\cal O}_{\rm AV}$ ( 
${\cal O}_{\rm ps}$ and ${\cal O}_{\rm av}$ in the case of  fermions $\Psi$).  
The standard procedure requires rewriting $C(t)$ 
in terms of fermion propagators 
$S^{a \alpha}_{\,\,\,\,\,\,\,\,\,b \beta }(x)\equiv
\langle Q^{a \alpha}(x)\overline{Q}_{\,b \beta}(0)\rangle$ (and analogous expressions for the 
propagators involving $\Psi$),
 to yield 
\beq
C_{M,M'}(t)=-\sum_{\vec{x}}{\rm Tr}
\left[\Gamma_M S(x) \Gamma_{M'} \gamma_5 S^\dagger(x) \gamma_5\right], 
\eeq
where the trace is over both spinor indices $\alpha,\beta$ and 
gauge indices $a,b$. 

In the simplest case of a point source, the fermion propagator $D^R$ 
($R$ labeling the fermion representation)
 is calculated by solving the Dirac equation 
\beq
D^R_{a \alpha ,  b \beta }(x,y) S^{b \beta}_{\,\,\,\,\,\,\,\, c \gamma }(y)
=\delta_{x,0}\delta_{\alpha\gamma}\delta_{ac}.
\label{Eq:DiracEquation}
\eeq
In order to improve the signal,
in our numerical studies throughout this work 
we use the $Z_2\times Z_2$ single time slice stochastic wall 
sources~\cite{Boyle:2008rh} with three different
sources considered individually for each configuration, 
instead of the point sources, on the right-hand side of Eq.~(\ref{Eq:DiracEquation}).

In all the spectroscopic measurements using quenched ensembles,
we use the (unimproved) Wilson action for the fermions.
The corresponding massive Wilson-Dirac operator in the fundamental representation $D^F$ is defined by
its action on the fermions $Q$, that takes the form
\beqs
D^{F} Q(x) \equiv (4/a+m_0)Q(x)-\frac{1}{2a}\sum_\mu
&&\left\{\frac{}{}(1-\gamma_\mu)U_\mu(x)Q(x+\hat{\mu})\right.+ \nn\\
&&\left.+(1+\gamma_\mu)U_\mu(x-\hat{\mu})Q(x-\hat{\mu})\frac{}{}\right\},
\label{Eq:DiracF}
\eeqs
where $U_\mu(x)$ are the link variables in the fundamental representation of $Sp(4)$, 
$a$ is the lattice spacing, and $\hat{\mu}$ is the unit vector in the spacelike direction $\mu$.

In order to  construct the Dirac operator $D^{AS}$ for  fermion fields $\Psi^{ab}$ in the
 2-index antisymmetric representation, we follow 
 the prescription in~\cite{DelDebbio:2008zf}. 
For $Sp(2N)$, we define an orthonormal basis $e_{AS}^{(ab)}$ (with
the multi-index $(ab)$ running over ordered  pairs with   $1 \leq a < b \leq 2N$)
for the appropriate vector space of 
$2N\times 2N$ antisymmetric matrices.
The $N(2N-1)-1$ such matrices have the following nonvanishing entries.
For $b=N+a$ and $2\leq a\leq N$
\beqs
(e_{AS}^{(ab)})_{c,N+c}\equiv -(e_{AS}^{(ab)})_{N+c,c}\equiv 
\left\{\begin{matrix}
&\frac{1}{\sqrt{2\,a\,(a-1)}},~~~\textrm{for}~c<a,\\
&\frac{-(a-1)}{\sqrt{2\,a\,(a-1)}},~~~\textrm{for}~c=a,\\
\end{matrix}\right.
\eeqs
and for $b\neq N+a$ 
\beqs
(e_{AS}^{(ab)})_{cd}\equiv \frac{1}{\sqrt{2}}(\delta_{ac}\delta_{bd}
-\delta_{bc}\delta_{ad})\,. 
\eeqs
The main difference compared to the case of $SU(N)$ is that the base $e_{AS}$ is $\Omega$-traceless, 
satisfying $ \Omega^{dc} \left(e^{(ab)}_{AS}\right)_{cd}=0$.
In the $Sp(4)$ case, one can verify that the resulting five nonvanishing matrices satisfy the 
orthonormalisation condition $\Tr e_{AS}^{(ab)}e_{AS}^{(cd)}=-\delta^{(ab)(cd)}$, while the
matrix $e_{AS}^{(13)}$ vanishes identically.
The explicit form
 of the antisymmetric link variables $U^{AS}_\mu(x)$
 descends from the fundamental link variables $U_\mu(x)$,  as
\beq
\left(U^{AS}_\mu\right)_{(ab)(cd)}(x)\equiv
{\rm Tr}\left[
(e_{AS}^{(ab)})^\dagger U_\mu(x) e_{AS}^{(cd)} U^{\mathrm{T}}_\mu(x)\right],~~~{\rm with}~a<b,~c<d.
\label{eq:U_AS}
\eeq
Finally, the Dirac operator  for the 2-index antisymmetric representation 
$D^{AS}$ is obtained by replacing 
$(U_{\mu})_{ab}$ by $(U^{AS}_{\mu})_{(ab)(cd)}$ and $Q$ by $\Psi$ in Eq.~(\ref{Eq:DiracF}).

Masses and decay constants for the mesons are extracted from the 
asymptotic behaviour of $C_{M,\,M^{\prime}}(t)$ 
at large Euclidean time.
We assume it to be dominated by a single mesonic state.
If $M=M^{\prime}$, for all  meson interpolating operators 
we can write 
\beq
C_{M,\,M}(t)\xrightarrow{t\rightarrow \infty} 
\frac{|\langle 0 | \mathcal{O}_M | M \rangle|^2}{2 m_M} 
\left[
e^{-m_M t}+e^{-m_M (T-t)}
\right],
\label{eq:corr_M}
\eeq
where $T$ is the temporal extent of the lattice. 
In our conventions, the meson states $|M\rangle$ are normalised
by writing $M=M^A T^A$, with $T^A$ the generators of the global $SU(2N_f)$ or
$SU(2n_f)$ symmetry. The  value of the pseudoscalar decay constant in
QCD in these conventions would be $f_{\rm PS}=f_\pi \simeq 93\,{\rm MeV}$. 
We also consider the correlator  defined with  $M=$ PS and  $M^{\prime}=$ AV, 
for which the  large-time behaviour is given by 
\beq
C_{{\rm PS,AV}}(t)\xrightarrow{t\rightarrow \infty} 
\frac{\langle 0 | \mathcal{O}_{\rm AV} |{\rm PS}\rangle 
\langle 0| \mathcal{O}_{\rm PS} | {\rm PS} \rangle^*}{2m_{\rm PS}}
\left[
e^{-m_{\rm PS} t}-e^{-m_{\rm PS}(T-t)}
\right]\,,
\label{eq:corr_Pi}
\eeq
having restricted attention to the components of  the AV operator with index $\mu=1,2,3$.

We parameterise the vacuum-to-meson matrix elements for fundamental fermions 
in such a way that the decay constants $f_M$ obey the following relations:
\beqs
\langle 0 | \mathcal{O}_{\rm AV} | {\rm PS} \rangle &=& \langle 0 | \overline{Q_1} \gamma_5 \gamma_\mu Q_2 | {\rm PS} \rangle  \equiv  f_{\rm PS}\, p_\mu,\nn \\
\langle 0 | \mathcal{O}_{\rm V} | {\rm V} \rangle &=& \langle 0 | \overline{Q_1} \gamma_\mu Q_2 | {\rm V} \rangle  \equiv  f_{\rm V}\, m_{\rm V}\, \epsilon_\mu,\nn \\
\langle 0 | \mathcal{O}_{\rm AV} | {\rm AV} \rangle &=& \langle 0 | \overline{Q_1} \gamma_5 \gamma_\mu Q_2 | {\rm AV} \rangle  \equiv  f_{\rm AV}\, m_{\rm AV}\, \epsilon_\mu,
\label{eq:matrix_element}
\eeqs
where the polarisation vector $\epsilon_\mu$ is transverse to the momentum $p_\mu$ 
and normalised by $\epsilon_\mu^* \epsilon^\mu=1$. 
(For  operators constituted by 
antisymmetric fermions we replace the fields $Q$ by $\Psi$.)
For spin-1 V and AV mesons we extract both masses 
and decay constants from Eqs.~(\ref{eq:corr_M}) and~(\ref{eq:matrix_element}). 
In the case of the pseudoscalar meson, we determine the masses 
and decay constants by combining Eqs.~(\ref{eq:corr_M}) with $M={\rm PS}$, Eq.~(\ref{eq:corr_Pi}) 
and \Eq{matrix_element}.

The matrix elements at finite lattice spacing $a$ have to be renormalised.
For Wilson fermions, the axial and vector currents 
receive multiplicative (finite) renormalisation. The renormalisation factors 
$Z_{\rm A}$ and $Z_{\rm V}$ are defined by the relations
\beq
f_{\rm PS}=Z_{\rm A} f^{\rm bare}_{\rm PS},~f_{\rm V}
=Z_{\rm V} f^{\rm bare}_{\rm V},~f_{\rm AV}=Z_{\rm A} f^{\rm bare}_{\rm AV}, 
\eeq
In this work we determine the renormalisation factors via  one-loop perturbative matching, 
and for Wilson fermions the relevant matching
coefficients are written as~\cite{Martinelli:1982mw}
\beq
Z_{\rm A(V)}=1+C(R) (\Delta_{\Sigma_1}+\Delta_{\Gamma})\frac{\tilde{g}^2}{16\pi^2}, 
\label{eq:matching}
\eeq
where $\Gamma=\gamma_5\gamma_\mu$ for $Z_{\rm A}$ 
and $\Gamma=\gamma_\mu$ for $Z_{\rm V}$. 
The eigenvalues of the quadratic Casimir operators
with fermions in the fundamental and antisymmetric representations 
of $Sp(4)$ are $C(F)=5/4$ and $C(AS)=2$, respectively. 
The  matching factors in \Eq{matching} are computed by one-loop integrals
 within the continuum $\overline{\rm MS}$ 
(modified minimal subtraction) regulalisation scheme. 
The resulting numerical values are 
$\Delta_{\Sigma_1}=-12.82$, $\Delta_{\gamma_\mu}=-7.75$ and 
$\Delta_{\gamma_5\gamma_\mu}=-3.0$ \cite{Martinelli:1982mw,Bennett:2017kga}. 
Following the prescription in Ref.~\cite{Lepage:1992xa}, 
in order to improve the convergence of perturbative expansion 
we replace the bare coupling $g$
by the tadpole improved coupling defined as $\tilde{g}^2= g^2/\langle 
P \rangle$. $\langle P \rangle$ is the average plaquette value,
and this procedure removes
 large tadpole-induced additive renormalisation
arising  with  Wilson fermions.

\subsection{Continuum extrapolation}
\label{Sec:continuum}

Extrapolations to the continuum limit 
are carried out following the same procedure as in Ref.~\cite{Bennett:2019jzz}.
We borrow the ideas of tree-level Wilson chiral perturbation theory (W$\chi$PT),
which we truncate at the next-to-leading order (NLO) in the double expansion in fermion mass 
and lattice spacing~\cite{Sheikholeslami:1985ij,Rupak:2002sm} 
(see also Ref.~\cite{Sharpe:1998xm}, as well as~\cite{Symanzik:1983dc, Luscher:1996sc},
though written in the
context of improvement).
Tree-level results for the full theory can be extended to (partially) quenched calculations, 
since  quenching effects only arise from  integrals in fermion loops \cite{Rupak:2002sm}. 
But we cannot {\it a priori} determine the range of validity of  tree-level W$\chi$PT at NLO. 
On the one hand, if we were too close  to the chiral limit, we would
need to include loop integrals (the well-known chiral logs). 
On the other hand, if we were in the heavy mass regime, then  we would
need to include more  higher-order terms. 
As we will discuss later, most of our data sit somewhere in between these two extrema, and 
as a consequence we can  empirically  find appropriate ranges of
fermion mass over which tree-level NLO W$\chi$PT well describes the numerical data. 

We apply the scale-setting procedure discussed in \Sec{scale}, 
and define the lattice spacing in units of the gradient-flow scale as $\hat{a}\equiv a/w_0$.
All other dimensional quantities are treated accordingly, so that 
masses are rescaled as in  $\hat{m}_M\equiv w_0 m_M$
and decay constants  as in $\hat{f}_M=w_0f_M$. 
Tree-level NLO W$\chi$PT assumes that the decay constant squared
$\hat{f}_{\rm PS}^{2 , {\rm NLO}}$ is linearly dependent on both $\hat{m}_{\rm PS}^2$ and $\hat{a}$.
We extend this assumption to all other observables as well, hence defining the ansatz
\bea
\hat{f}_M^{2,{\rm NLO}}&\equiv&\hat{f}_M^{2, \chi}
\left(1 + L_{f,M}^0 \hat{m}_{\rm PS}^2\right) + W_{f,M}^0 \hat{a}\,,
\label{eq:f2_chipt}\\
\hat{m}_M^{2,{\rm NLO}}&\equiv&\hat{m}_M^{2, \chi}
\left(1 + L_{m,M}^0 \hat{m}_{\rm PS}^2\right) + W_{m,M}^0 \hat{a}\,,
\label{eq:m2_chipt}
\eea
for decay constants squared and masses squared, respectively. 
We note that the fermion mass $m_f$ appearing in the standard W$\chi$PT 
has been replaced by the pseudoscalar mass squared by using  LO $\chi$PT results, 
according to which $\hat{m}_{\rm PS}^2=2B m_f$. 
The low-energy constant $B$ could in principle be determined 
via a dedicated study of the fermion mass, but this would go beyond our current aims.
The empirical prescription we adopt requires us to 
identify the largest possible region of lattice data showing evidence of the linear behaviour
described above and then fit the data in order to identify
the additive contribution proportional to $\hat{a}$. Extrapolation to the continuum 
is obtained by subtracting this contribution from the lattice measurements.

\subsection{Quenched spectrum: Fundamental fermions}
\label{Sec:quenched}

\begin{figure}[ht]
\begin{center}
\includegraphics[width=.45\textwidth]{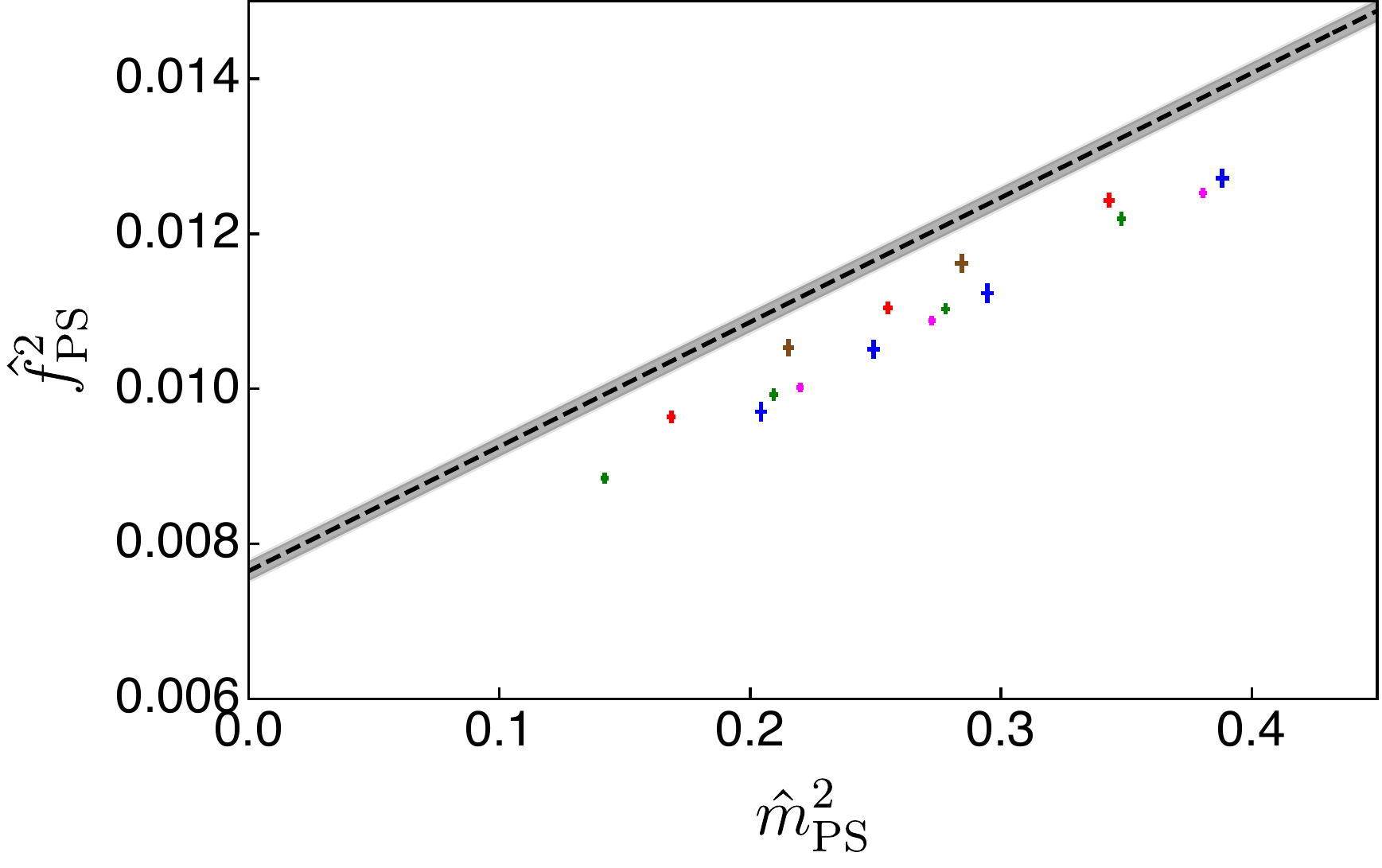}
\includegraphics[width=.45\textwidth]{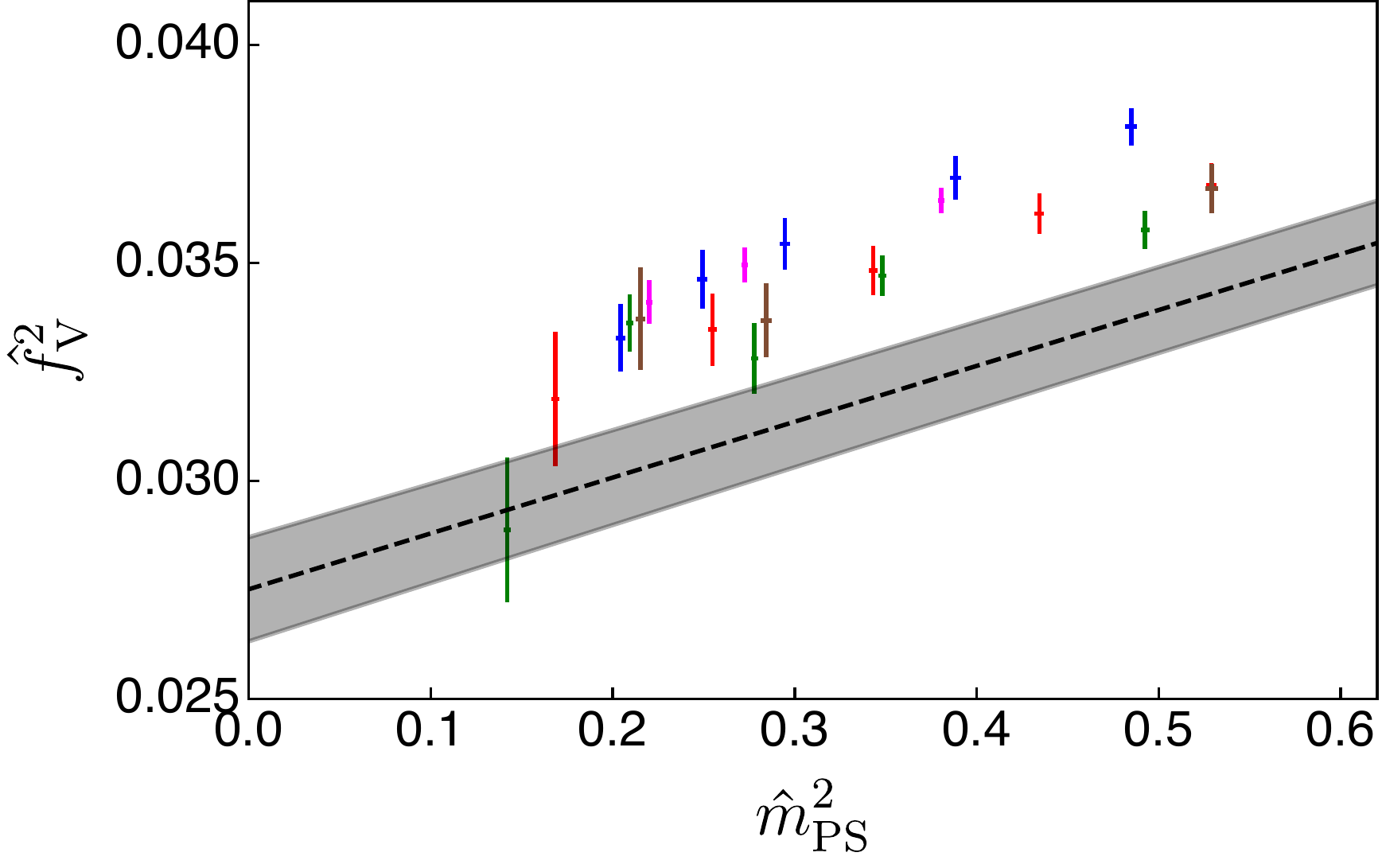}
\includegraphics[width=.45\textwidth]{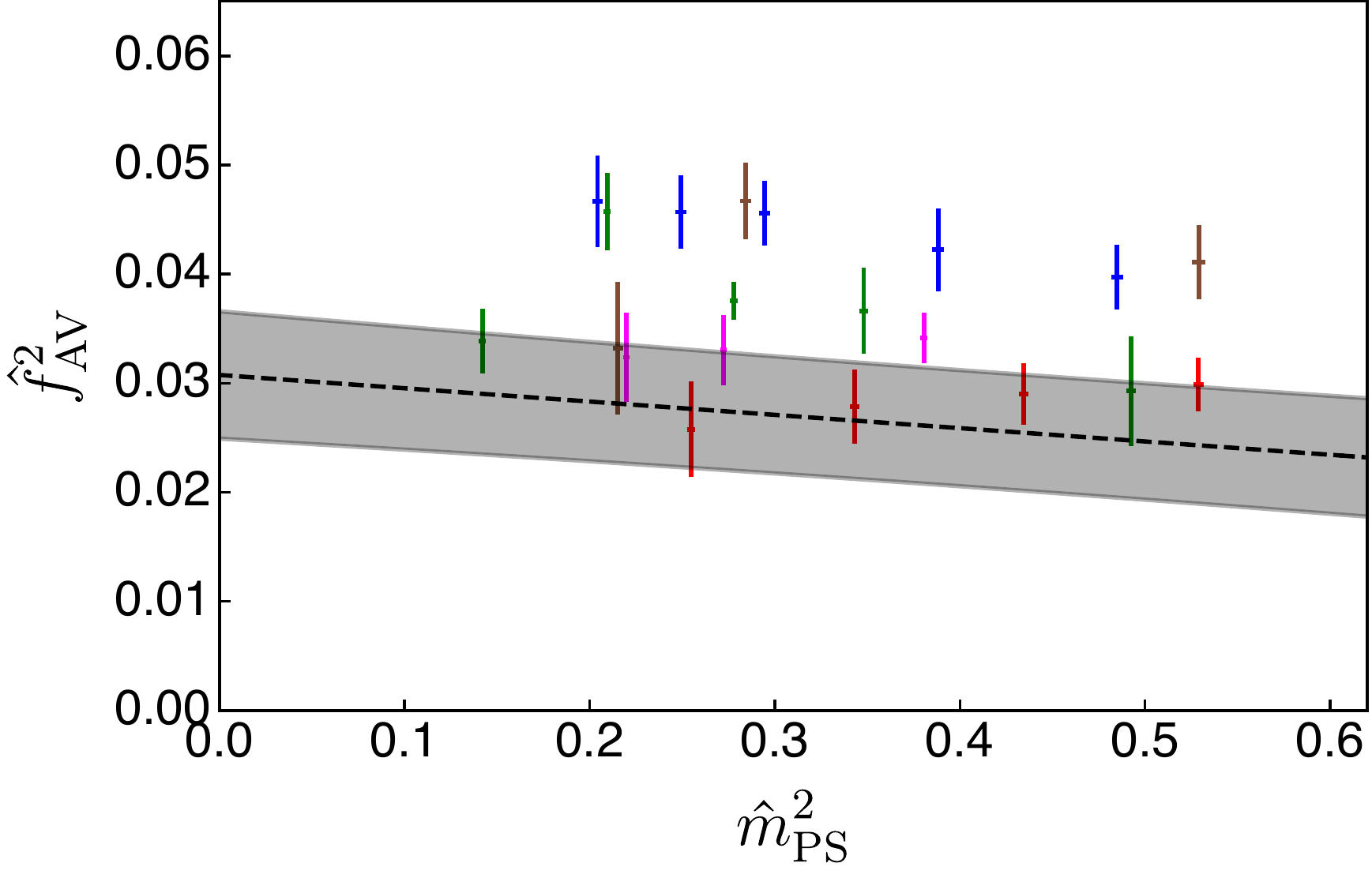}
\caption{%
\label{fig:f2_fit}%
Decay constants squared of PS, V, and AV mesons
(constituted of fermions transforming  in the fundamental representation), 
as a function of  the PS meson mass squared $\hat{m}^2_{\rm PS}$,
 for $\beta=7.62$ (blue), $7.7$ (purple), $7.85$ (green), $8.0$ (red), and $8.2$ (brown). 
 All quantities are expressed in units of the gradient-flow scale $w_0$.
The results of the continuum and massless extrapolations are represented by the grey bands. 
}
\end{center}
\end{figure}

\begin{figure}[ht]
\begin{center}
\includegraphics[width=.45\textwidth]{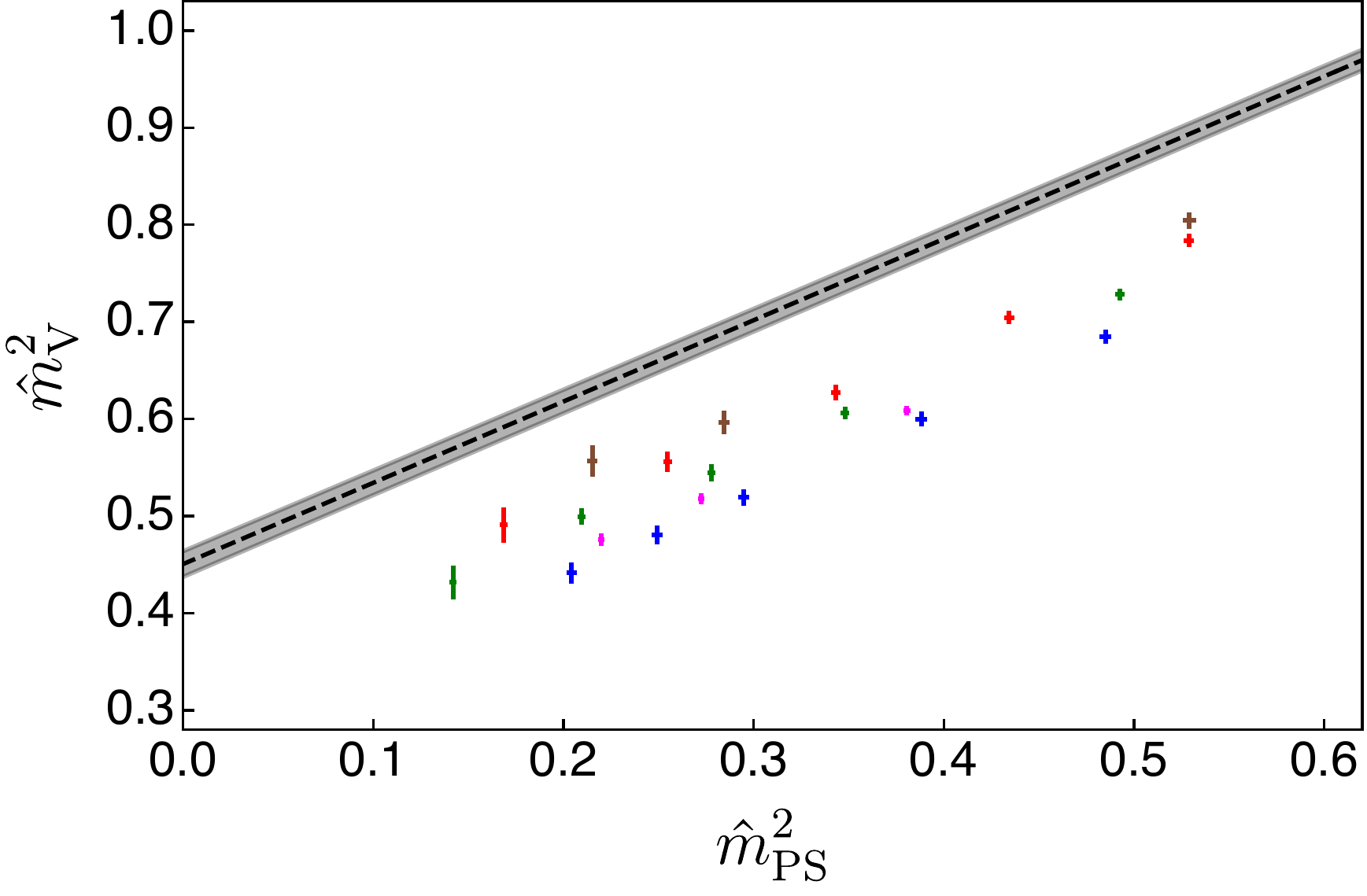}
\includegraphics[width=.45\textwidth]{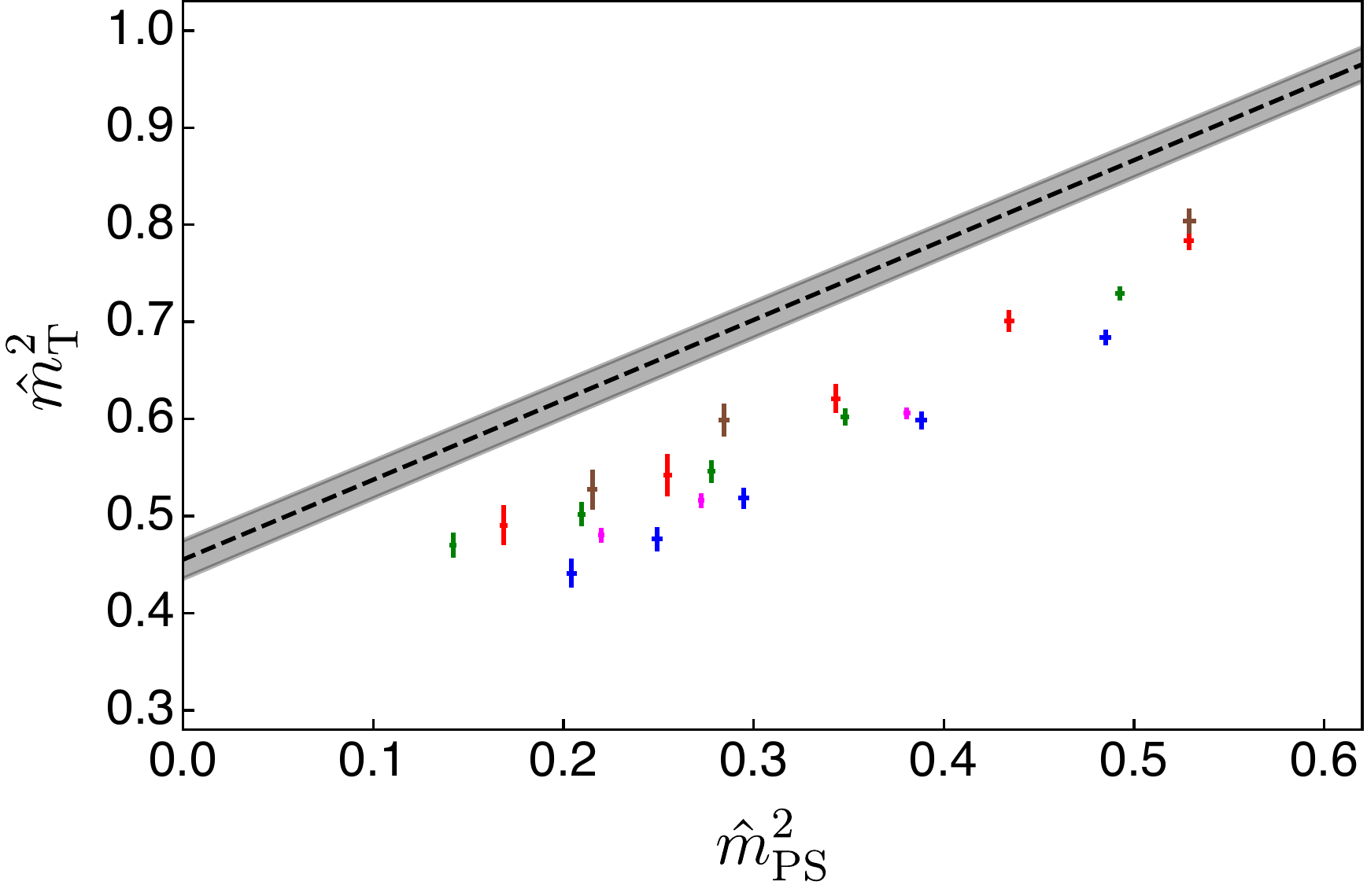}
\includegraphics[width=.45\textwidth]{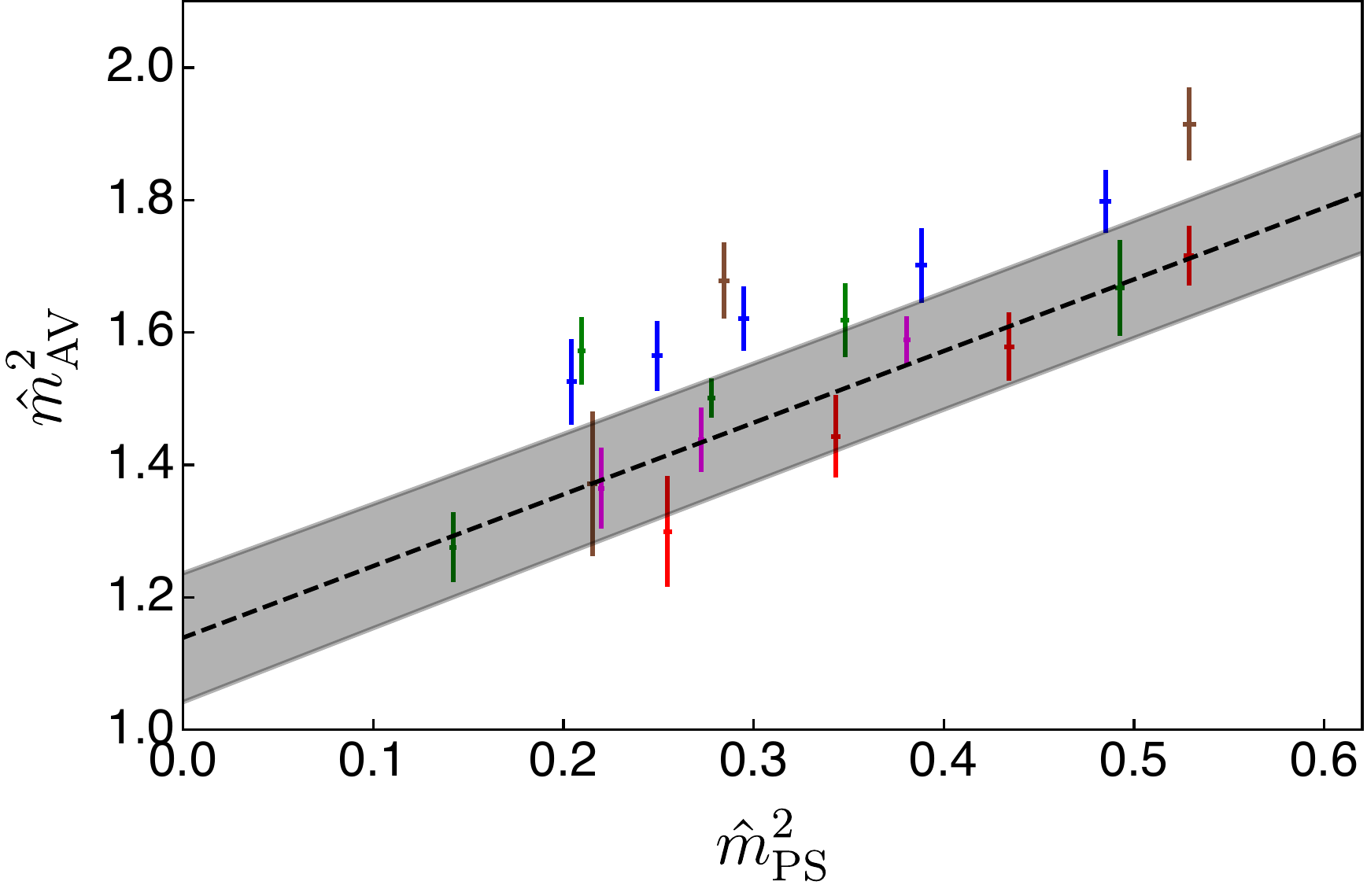}
\includegraphics[width=.45\textwidth]{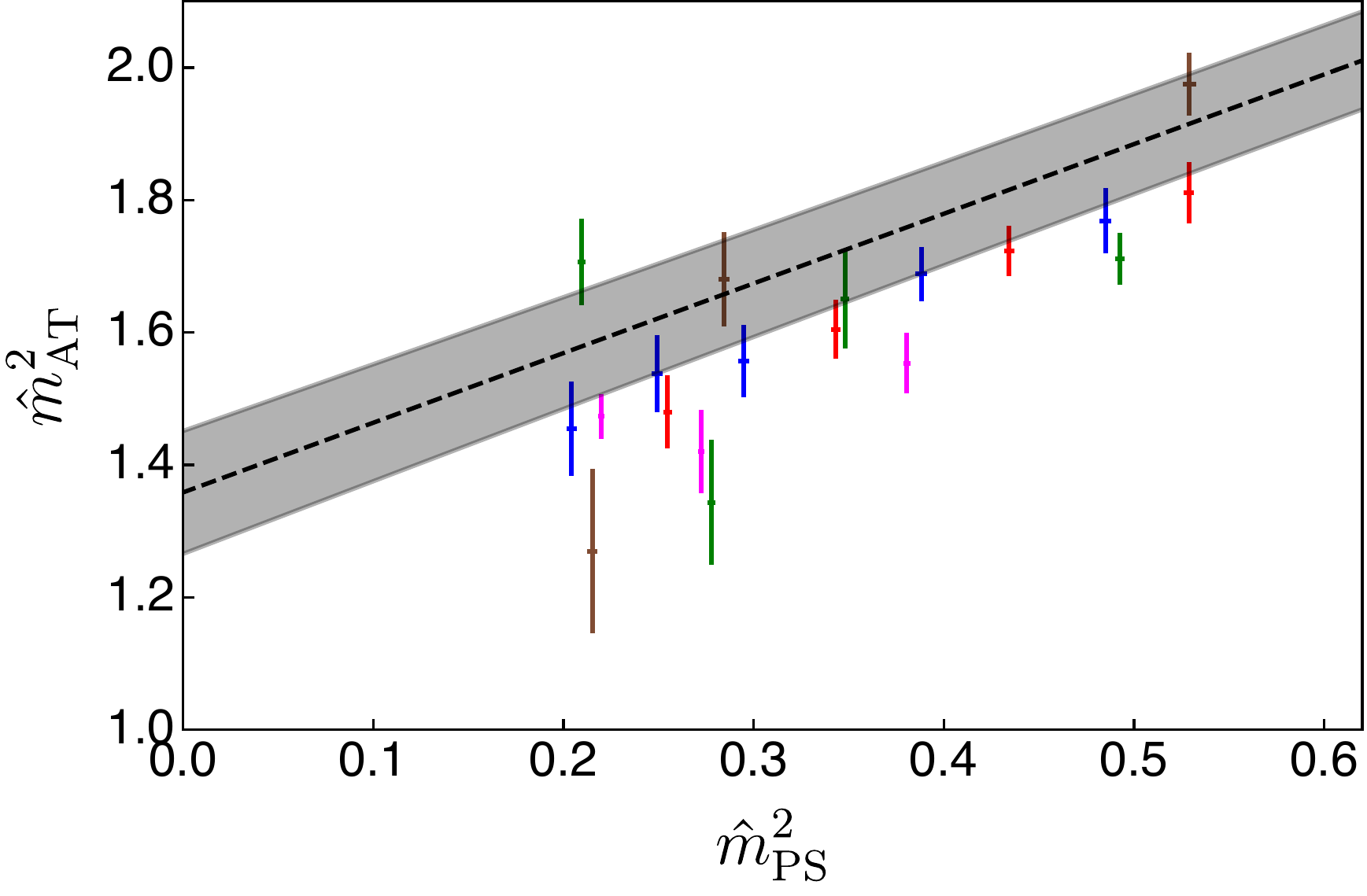}
\includegraphics[width=.45\textwidth]{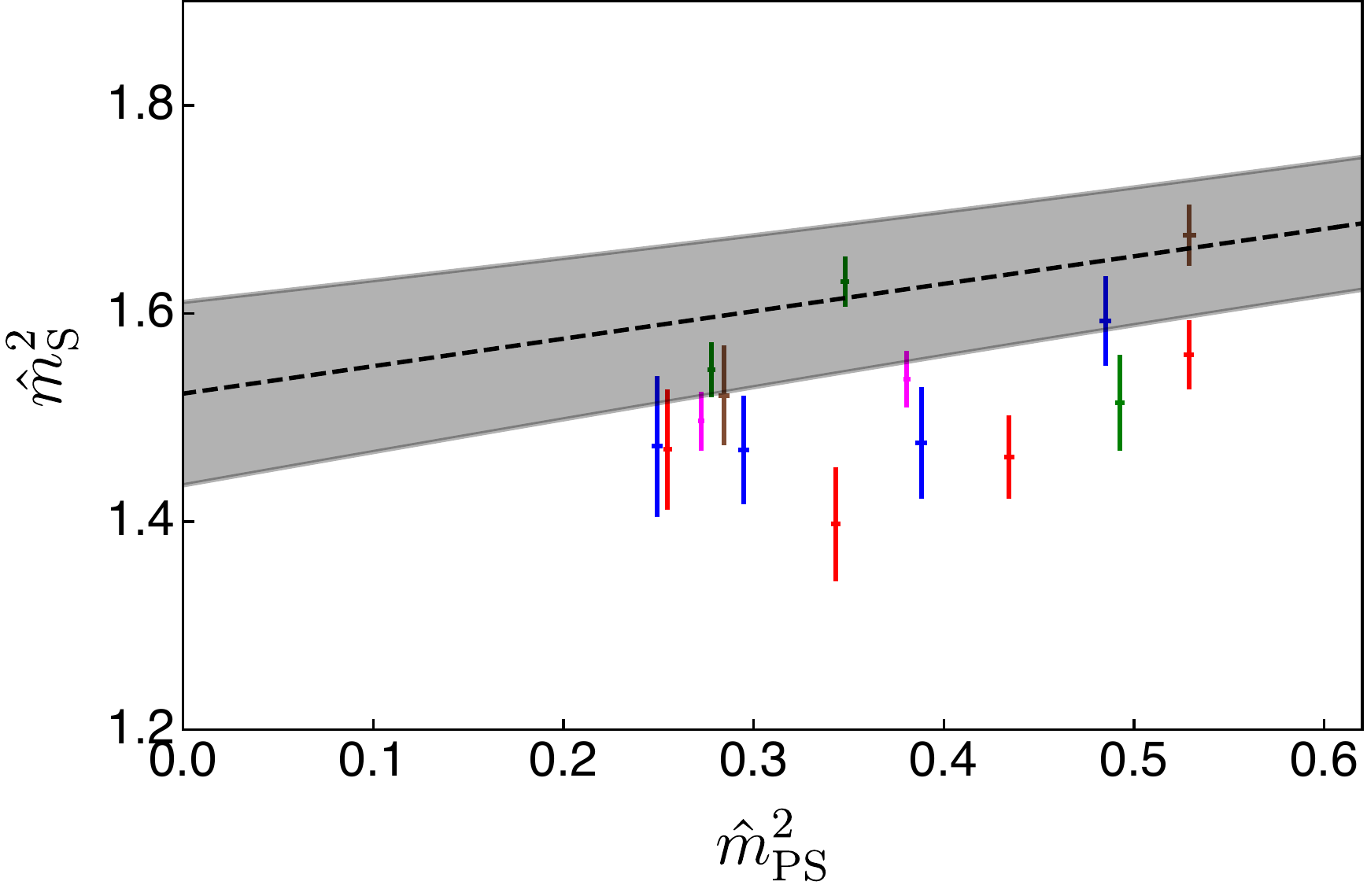}
\caption{%
\label{fig:m2_fit}%
Masses squared of V, T, AV, AT and S mesons (constituted of fermions in the fundamental representation),
as a function of the PS meson mass squared $\hat{m}^2_{\rm PS}$,
for $\beta=7.62$ (blue), $7.7$ (purple), $7.85$ (green), $8.0$ (red), and $8.2$ (brown). 
 All quantities are expressed in units of the gradient-flow scale $w_0$.
The results of the continuum and massless extrapolations are represented by the grey bands. 
}
\end{center}
\end{figure}

Reference~\cite{Bennett:2017kga} reported the quenched spectrum of 
the lightest PS, V, and AV flavoured mesons 
for two values of the lattice coupling, $\beta=7.62$ and $8.0$, 
with fermions in the fundamental representation. 
In this section, we extend the exploration of the quenched theory 
in several directions.
First, we consider three more values of the coupling, $\beta=7.7$, $7.85$, and $8.2$, 
as mentioned in \Sec{definitions}, aiming to perform continuum extrapolations,
along the lines described in Sec.~\ref{Sec:continuum}.
Second, in order to remove potential finite-volume effects, we restrict the bare 
 fermion mass $m_0$ to ensembles that satisfy the condition $m_{\rm PS} L \geq 7.5$,
in line with the results of the study with dynamical fermions~\cite{Bennett:2019jzz}.
Only part of the data in~\cite{Bennett:2017kga} meets this restriction, 
over the range of $m_0 \in [-0.7,-0.79]$ at $\beta=7.62$, 
measured on the lattice with extension $48\times 24^3$---corresponding to the ensemble 
denoted as QB1 in Table~\ref{tab:ensemble_Q}. 
For the other values of the lattice coupling we perform new calculations by using lattices with
extension $60\times 48^3$. 
The details of all the ensembles are found in \Sec{definitions} and summarised in \Tab{ensemble_Q}.

For a given ensemble QB$i$ (with $i=1\,,\,\cdots\,,\,5$) 
we introduce various choices of bare mass $a m_0$ of the fundamental fermions
(see \Tab{meson_measurement_F}), 
and calculate two-point Euclidean correlation functions of pseudoscalar PS, vector V, 
axial-vector AV, scalar S, (antisymmetric) tensor T and axial-tensor AT meson operators, 
using the interpolating operators in \Tab{mesons}. 
We follow the standard procedure described in \Sec{correlators}, 
and we extract the masses and decay constants from the correlated 
fit of the data for the correlation functions as in \Eq{corr_M}. 
In the case of the pseudoscalar meson, we simultaneously fit the data for the 
correlators $C_{\rm PS,\,PS}(t)$ and $C_{\rm PS,AV}(t)$ 
according to Eqs.~(\ref{eq:corr_M}) and~(\ref{eq:corr_Pi}). 
The fitting intervals over the asymptotic (plateau) region at large Euclidean time are chosen to
optimise the $\chi^2$ while keeping the interval as large as possible. 
Such optimised values are shown in the numerical fits presented 
in Tables~\ref{tab:meson_measurement_F} and \ref{tab:tensor_measurement_F} in \App{AppendixG}.

Notice that, in the case of AV, AT and S mesons, we are not able to find an acceptable plateau 
region for several among the lightest choices of fermion masses. 
This problem appears when approximately reaching the threshold 
for decay to three pseudoscalars. 
Similar problems have been observed before in the literature on quenched 
theories (see for example Refs.~\cite{Bernard:1995ez,Lin:2002aj, Lin:2003tn}),
and may be due to the appearance of two types of new features,
both of which are ultimately due to violations of unitarity: 
polynomial factors correct the exponential behaviour of the large-time correlation functions, 
and finite-volume effects do not decouple in the infinite-volume limit.
We pragmatically decided to ignore measurements showing evidence of these phenomena, 
and discard them from the analysis.

The resulting values of meson masses and decay constants are presented 
in Tables~\ref{tab:meson_spec_spin0_F}, \ref{tab:meson_spec_spin1_F} and \ref{tab:meson_tensor_F} in \App{AppendixG}.
In \Tab{meson_spec_spin0_F} we also present the results of $m_{\rm PS} \,L$ and $f_{\rm PS} \,L$. 
For all the listed measurements the lattice volumes are large enough 
that the finite-volume effects are expected to be negligible as $m_{\rm PS} \,L \gsim 7.5$, 
and the low-energy EFT is applicable as $f_{\rm PS} \,L \gsim 1.6$.
All fermion masses are large enough that
the decay of a V meson into two PS mesons is kinematically forbidden. 
The resulting values of the masses measured from 
the correlators involving $\mathcal{O}_{\rm V}$ and $\mathcal{O}_{\rm T}$ 
are statistically consistent with each other, 
in support of theoretical prediction: 
the V and T operators interpolate the same physical states with $J^P=1^-$ 
(identified with the  $\rho$ meson in the case of real-world QCD).

We perform simultaneous continuum and massless extrapolations 
by fitting the data for (quenched) meson masses and decay constants to 
 Eqs.~(\ref{eq:f2_chipt}) and~(\ref{eq:m2_chipt}). 
We restrict the range of masses used for the extrapolations 
to $\hat{m}_{\rm PS}^2\lesssim 0.4$  for the PS states,
and to ensembles yielding 
$\hat{m}_{\rm PS}^2 \lesssim 0.6$ for all other states, in order to retain the
largest possible range of masses within which the data 
show linear dependence on $\hat{m}_{\rm PS}^2$.
In Figs.~\ref{fig:f2_fit} and~\ref{fig:m2_fit} we show the results 
of decay constants and masses, 
with different colours being used to denote ensembles at different $\beta$ values. 
In the figures we also present the continuum-extrapolated values (denoted by grey bands 
the widths of which represent the statistical uncertainties), 
obtained after subtracting  artefacts arising from finite lattice spacing. 
We find that $\hat{f}_{\rm PS}^2$, $\hat{m}_{\rm V}^2$ and $\hat{m}_{\rm T}^2$ 
are significantly affected by the discretisation of the Euclidian space. 
And such a long continuum extrapolation can be understood from the fact that 
we have used the standard Wilson fermions. 
The size of lattice artefacts in all other quantities is comparable with 
that of statistical uncertainties.

From the numerical fits we  determine the constants appearing in
Eqs.~(\ref{eq:f2_chipt}) and~(\ref{eq:m2_chipt}), 
and we report them in \Tab{LECs_Q}.
The numbers in the first and second parentheses 
are the statistical and systematic uncertainties of the fits, respectively. 
In the table, we also present the values of $\chi^2/N_{\rm d.o.f}$. 
Some large values of $\chi^2/N_{\rm d.o.f}$ indicate that 
either the uncertainties of the individual data were underestimated or 
the fit functions are not sufficient to correctly describe the data. 
Although it would be difficult to fully account for the systematics associated with the continuum extrapolation 
with limited number of lattice spacings, 
we estimate the systematic uncertainties in the fits by 
taking the maximum and minimum values obtained from the set of data 
excluding the coarsest lattice (the ensemble with $\beta=7.62$) and 
including or excluding the heaviest measurements. 
Notice in the table that  this process yields large estimates for the
systematic uncertainty for those fits that result in 
a large value of $\chi^2/N_{\rm d.o.f}$ at the minimum.
Finally, the resulting values in the massless limit, $\hat{f}_M^{2,\,\chi}$ and $\hat{m}_M^{2,\,\chi}$, 
should be taken with a due level of caution, 
since the considered masses are still relatively heavy and only the corrections 
corresponding to the tree-level terms in the chiral expansion are used in the fits. 
We leave more dedicated studies of the massless extrapolation 
to our future work with fully dynamical and light fermions.

As seen in \Tab{meson_ratio_F}, for each given value of $\beta$ the ratio
 $\hat{m}_{\rm V}/\hat{f}_{\rm PS}$ is approximately constant 
over the mass region  $\hat{m}_{\rm V}/\hat{m}_{\rm PS}\gsim 1.3$.
From a simple linear extrapolation to the continuum of these constant 
vector masses in units of $\hat{f}_{\rm PS}$,
we find that $\hat{m}_{\rm V}/\sqrt{2} \hat{f}_{\rm PS}=5.42(5)$. 
A more rigorous, yet compatible, estimate of the massless limit is obtained by taking 
the extrapolated results in \Tab{LECs_Q} 
and yields  $\hat{m}^{\chi}_{\rm V}/\sqrt{2} \hat{f}^{\chi}_{\rm PS}=5.48(9)(4)$.

\begin{table}[ht]
\begin{center}
\begin{tabular}{|c|c|c|c|c|}
\hline\hline
 & $~~\hat{f}^{2,\,\chi}_M~~$ & $~~~L^0_{f,M}~~~$
& $~~~~~W^0_{f,M}~~~~~$ & $~~~~~\chi^2/N_{\rm d.o.f}~~~~~$ \cr \hline
PS & $0.00765(13)(11)$ & $2.101(38)(51)$ & $-0.00190(24)(15)$ & $3.1$\cr
V & $0.0275(12)(4)$ & $0.47(51)(24)$ & $0.0060(18)(4)$ & $1.5$\cr
AV & $0.031(6)(10)$ & $-0.40(18)(25)$ & $0.019(10)(20)$ & $4.2$\cr
\hline\hline
 & $~~\hat{m}^{2,\,\chi}_M~~$ & $~~~L^0_{m,M}~~~$
& $~~~~~W^0_{m,M}~~~~~$ & $~~~~~\chi^2/N_{\rm d.o.f}~~~~~$ \cr \hline
V & $0.451(13)(5)$ & $1.86(7)(4)$ & $-0.257(20)(6)$ & $0.4$\cr
T & $0.455(20)(7)$ & $1.81(8)(5)$ & $-0.256(31)(9)$ & $0.9$\cr
AV & $1.14(10)(14)$ & $0.96(14)(18)$ & $0.13(16)(29)$ & $3.8$ \cr
AT & $1.36(9)(13)$ & $0.78(10)(10)$ & $-0.19(14)(24)$ & $3.1$ \cr
S & $1.52(9)(4)$ & $0.18(6)(12)$ & $-0.14(13)(7)$ & $4.0$ \cr
\hline\hline
\end{tabular}
\end{center}
\caption{
\label{tab:LECs_Q}
Results of the fit of the coefficients in Eqs.~(\ref{eq:f2_chipt}) and~(\ref{eq:m2_chipt}), used
in the  continuum and massless extrapolations of  masses and decay constants 
of mesons in the quenched simulations involving Dirac fermions in the fundamental representation. 
The numbers in parentheses represent, respectively, statistical and systematic uncertainties due to the fit.
}
\end{table}

\subsection{Quenched spectrum: Antisymmetric fermions}
\label{Sec:antisymmetric}

We turn now our attention to the quenched spectrum of the lightest flavoured mesons 
involving the fermions transforming in the antisymmetric representation of $Sp(4)$. 
We use the same ensembles listed in \Tab{ensemble_Q}, 
but the bare masses $m_0$ of the fermions are listed in \Tab{meson_measurement_AS} of \App{AppendixG}. 
As with fundamental fermions, 
we choose the values of $a m_0$ to satisfy the condition of $m_{\rm ps} L \geq 7.5$. 
In the table, we also present the fitting intervals used for the extraction of 
the masses and the decay constants of ps, v, av, and s
 mesons as well as the resulting values of $\chi^2/N_{\rm d.o.f}$.
The results for t and at mesons are shown in \Tab{tensor_measurement_AS} of the same appendix. 
We apply to the antisymmetric case
the same numerical treatment and analysis techniques used for 
the fundamental fermions.
As in the case of the fundamental representation, we could not find an acceptable plateau region 
for some measurements at the smallest fermion masses, in the cases of v, av and s mesons. 

In \App{AppendixG} we also present the numerical results of the masses and decay constants of ps, v and av mesons, 
as well as the masses of s, t and at mesons. See Tables~\ref{tab:meson_spec_spin0_AS}, \ref{tab:meson_spec_spin1_AS} and \ref{tab:meson_tensor_AS}. 
As shown in \Tab{meson_spec_spin0_AS}, all the measurements meet the aforementioned condition $m_{\rm ps} L \geq 7.5$. 
In addition, we find that $f_{\rm ps} L \geq 2.3$, which supports 
the applicability of low-energy EFT techniques. 
Furthermore, the meson masses in units of $\hat{f}_{\rm ps}$ 
and the ratio $\hat{f}_{\rm v}/\hat{f}_{\rm ps}$ 
are presented in Tables~\ref{tab:meson_ratio_AS} and \ref{tab:meson_tensor_AS}. 
As already seen in the results for fundamental fermions, 
in all the measurements we find that the results of $\hat{m}_{\rm t}$ are 
consistent with those of $\hat{m}_{\rm v}$, given the current statistical uncertainties.

\begin{figure}[ht]
\begin{center}
\includegraphics[width=.45\textwidth]{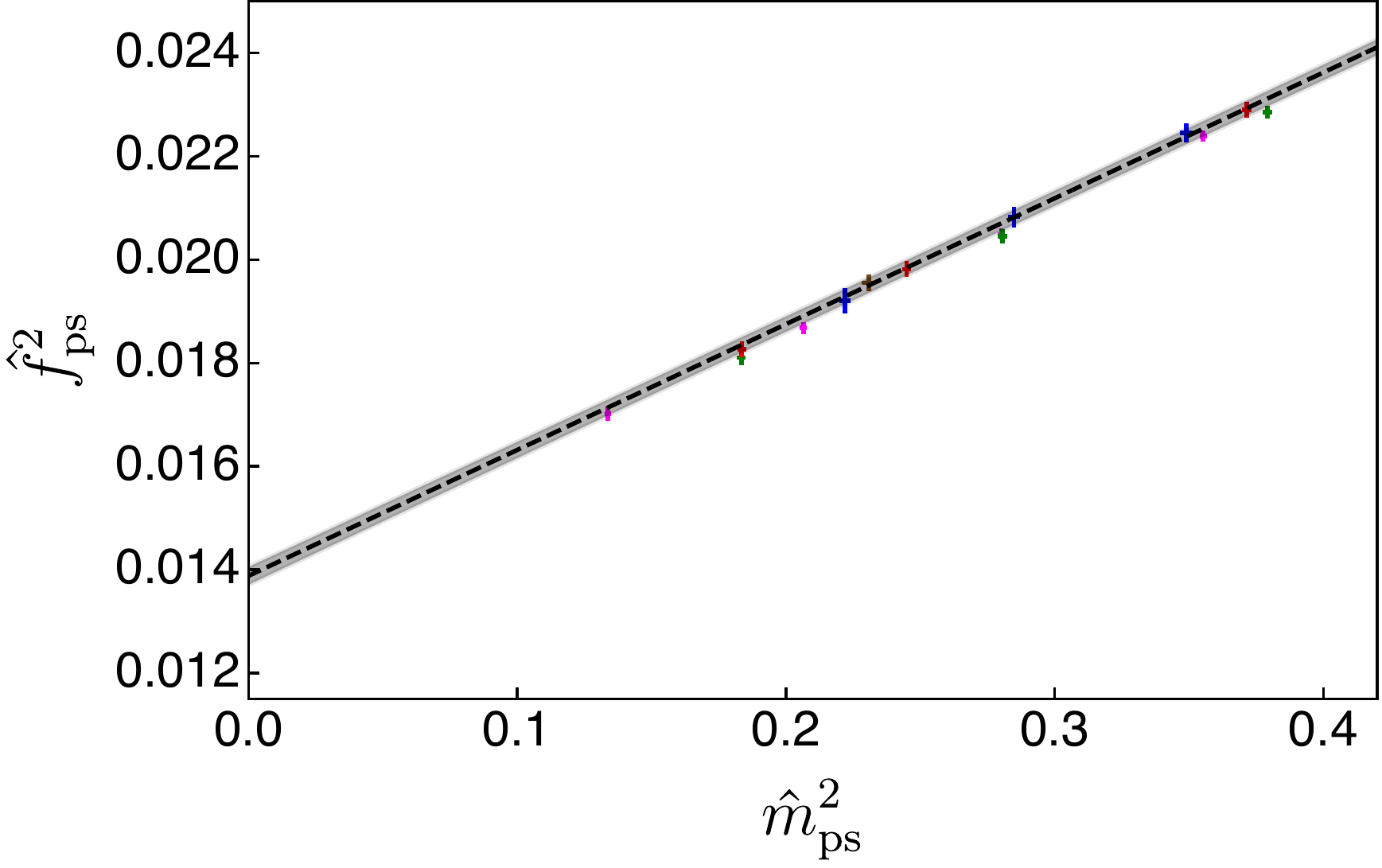}
\includegraphics[width=.45\textwidth]{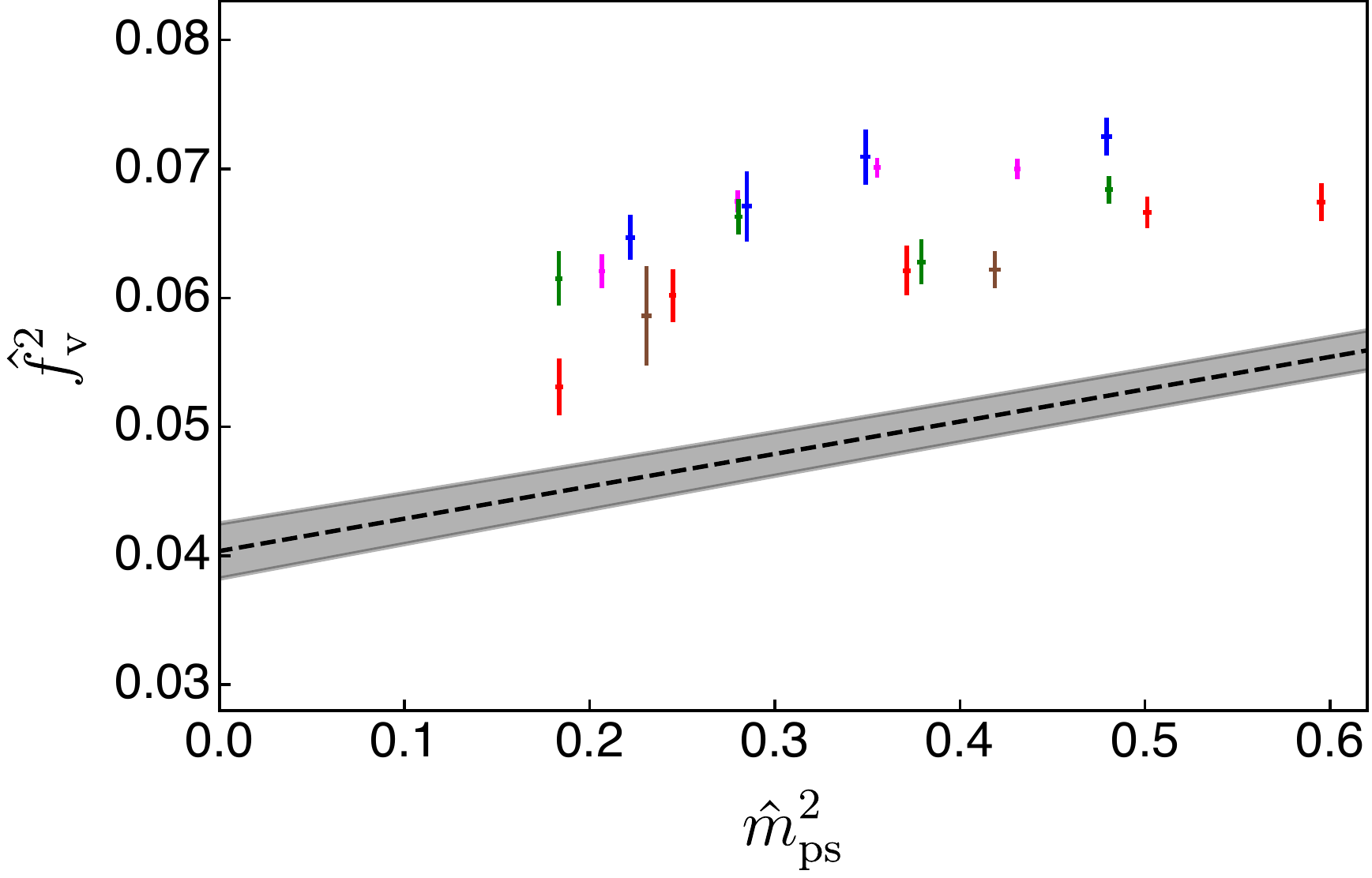}
\includegraphics[width=.45\textwidth]{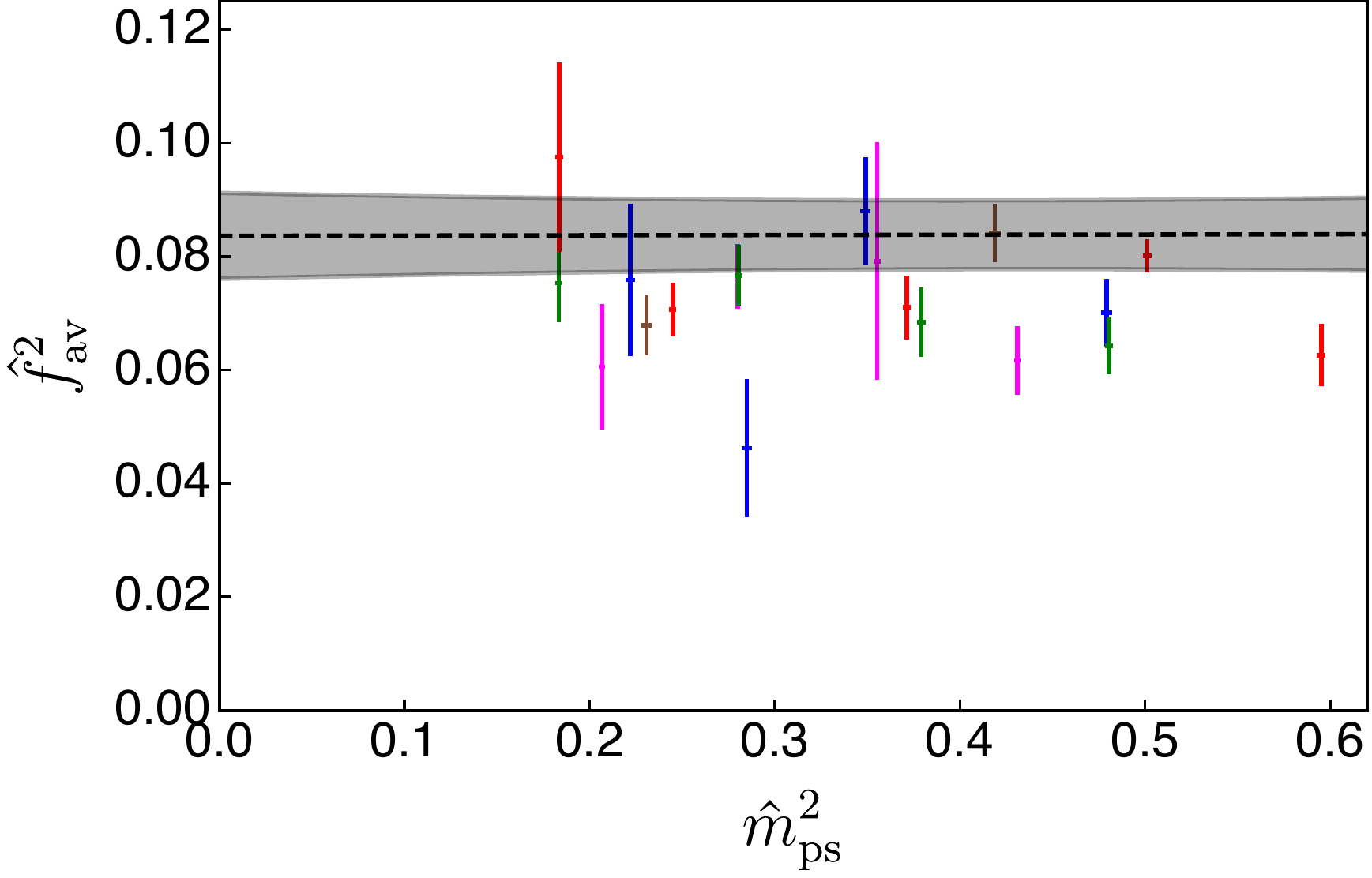}
\caption{%
\label{fig:f2_fit_AS}%
Decay constants squared of ps, v, and av mesons
(constituted of fermions in the antisymmetric representation), 
as a function of the ps meson mass squared $\hat{m}_{\rm ps}^2$, for $\beta=7.62$ (blue), 
$7.7$ (purple), $7.85$ (green), $8.0$ (red), and $8.2$ (brown). 
 All quantities are expressed in units of the gradient-flow scale $w_0$.
The results of the continuum and massless extrapolations are represented by the grey bands. 
}
\end{center}
\end{figure}

\begin{figure}[ht]
\begin{center}
\includegraphics[width=.45\textwidth]{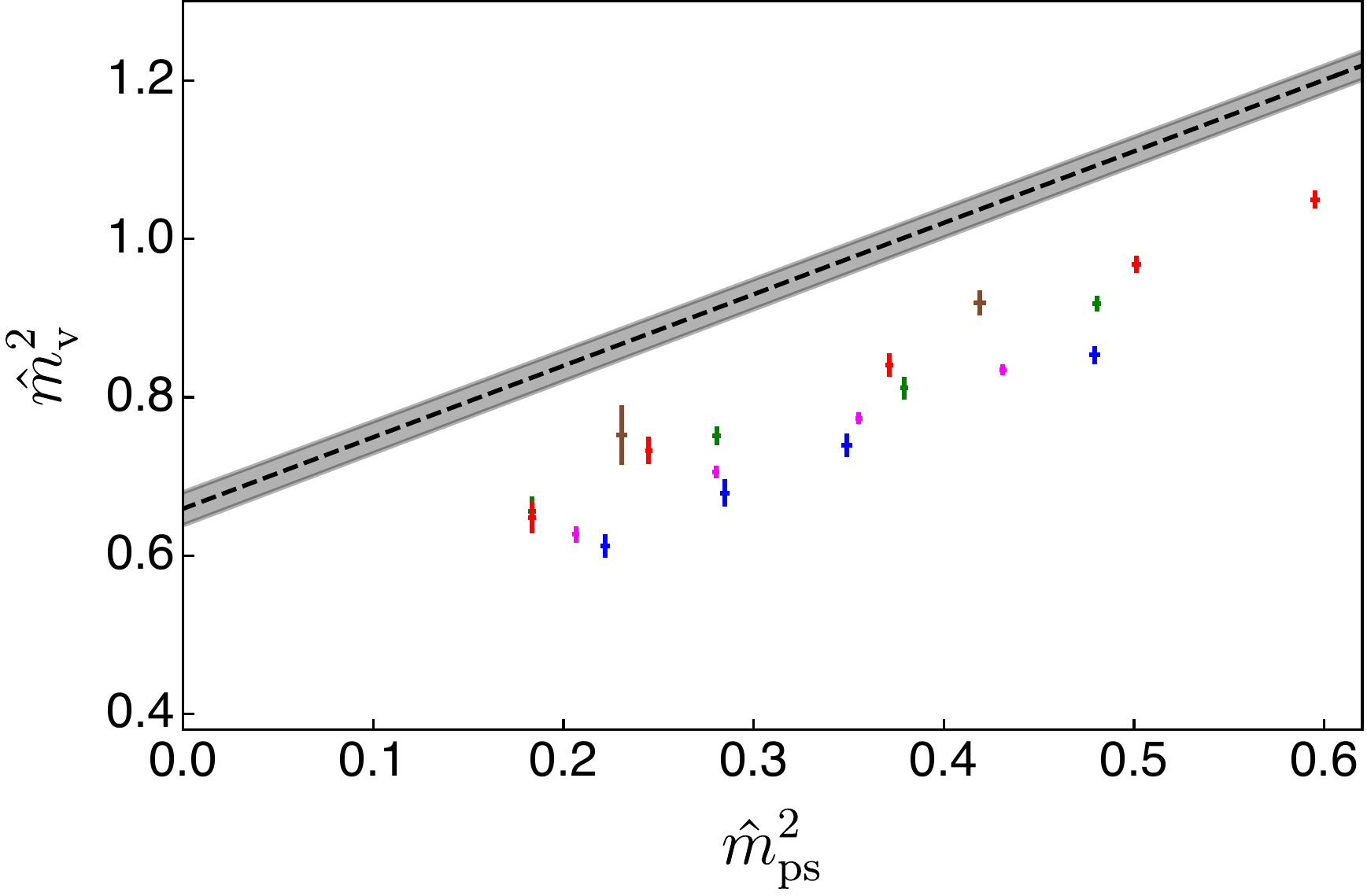}
\includegraphics[width=.45\textwidth]{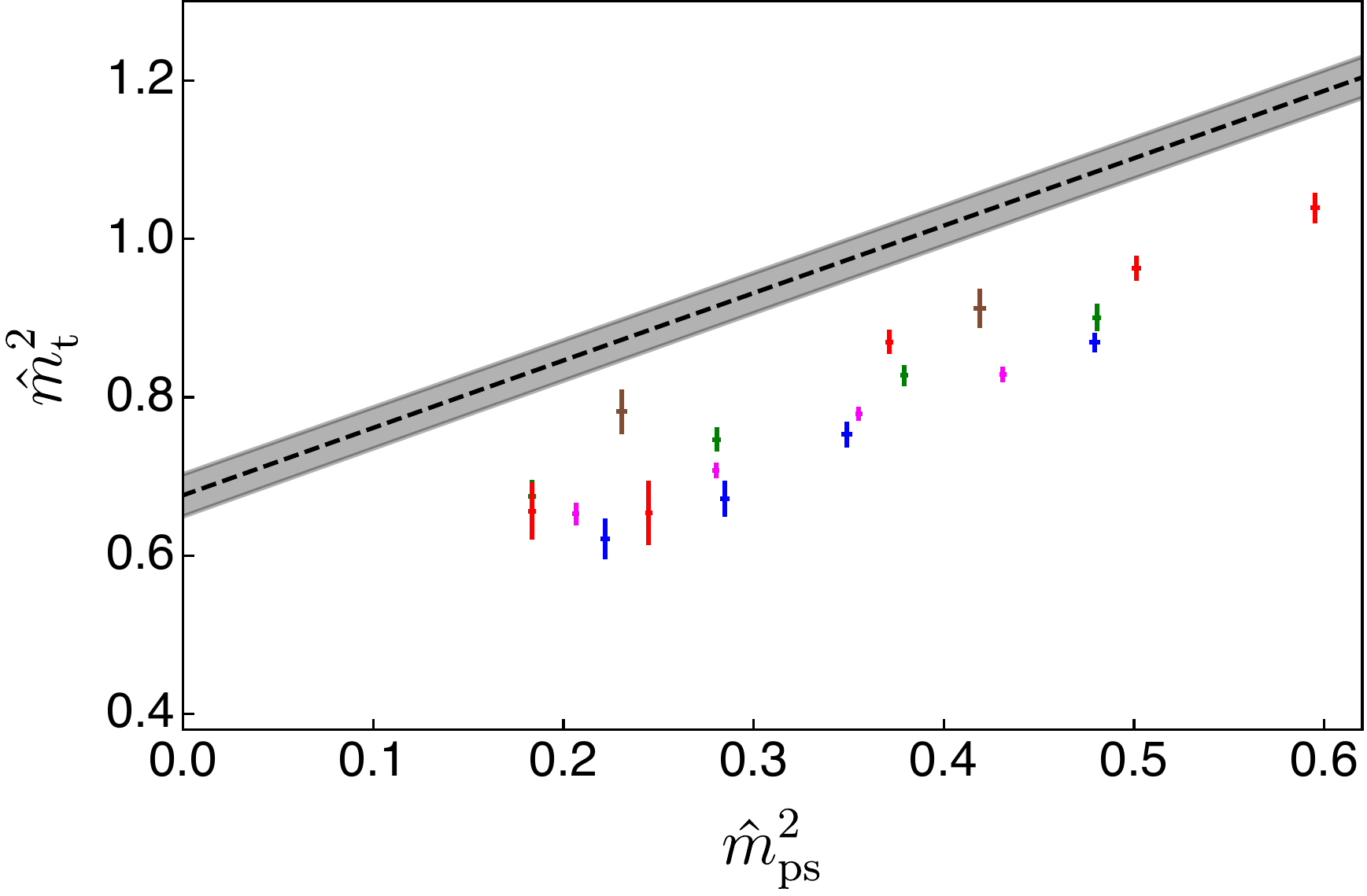}
\includegraphics[width=.45\textwidth]{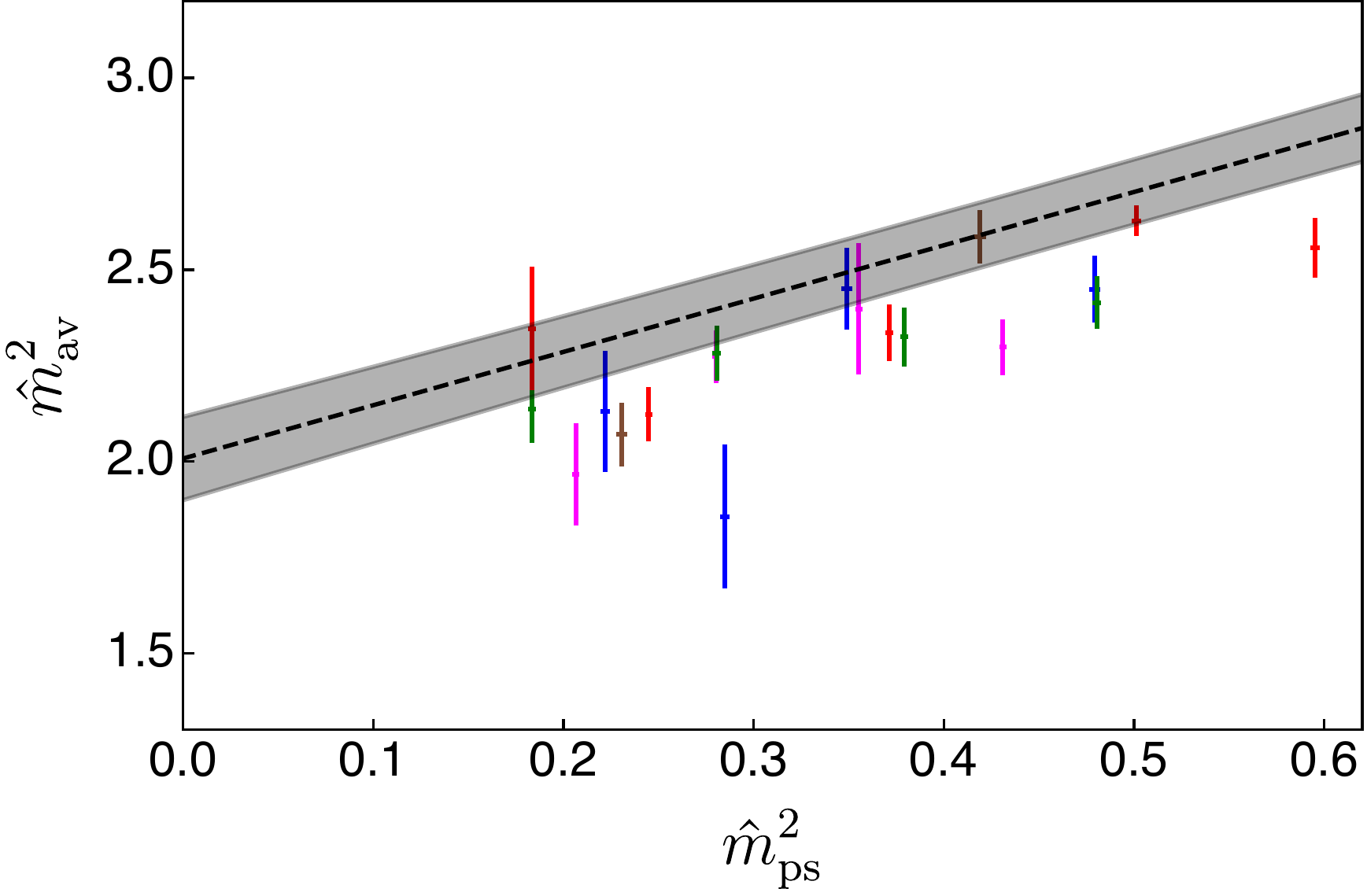}
\includegraphics[width=.45\textwidth]{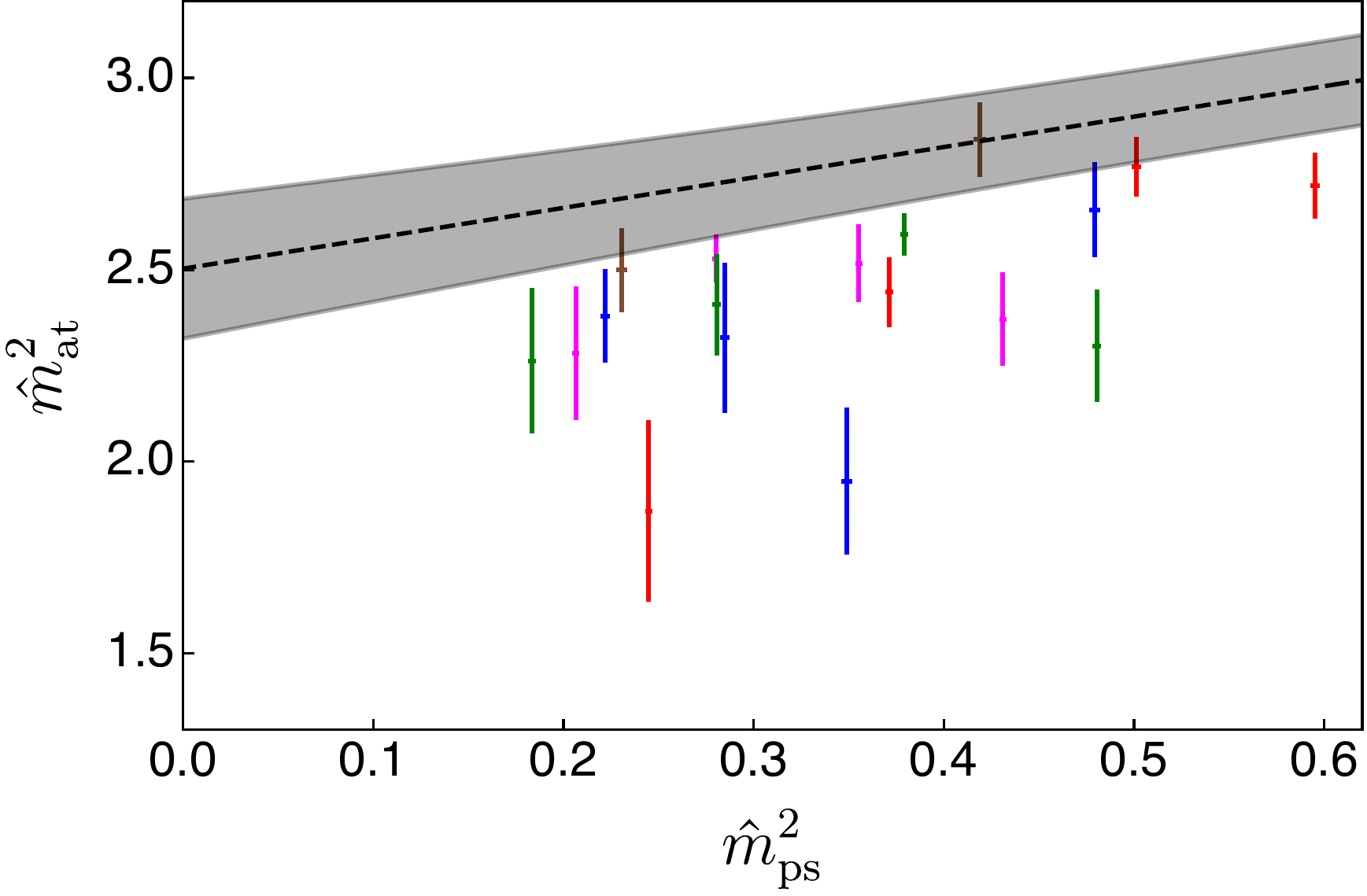}
\includegraphics[width=.45\textwidth]{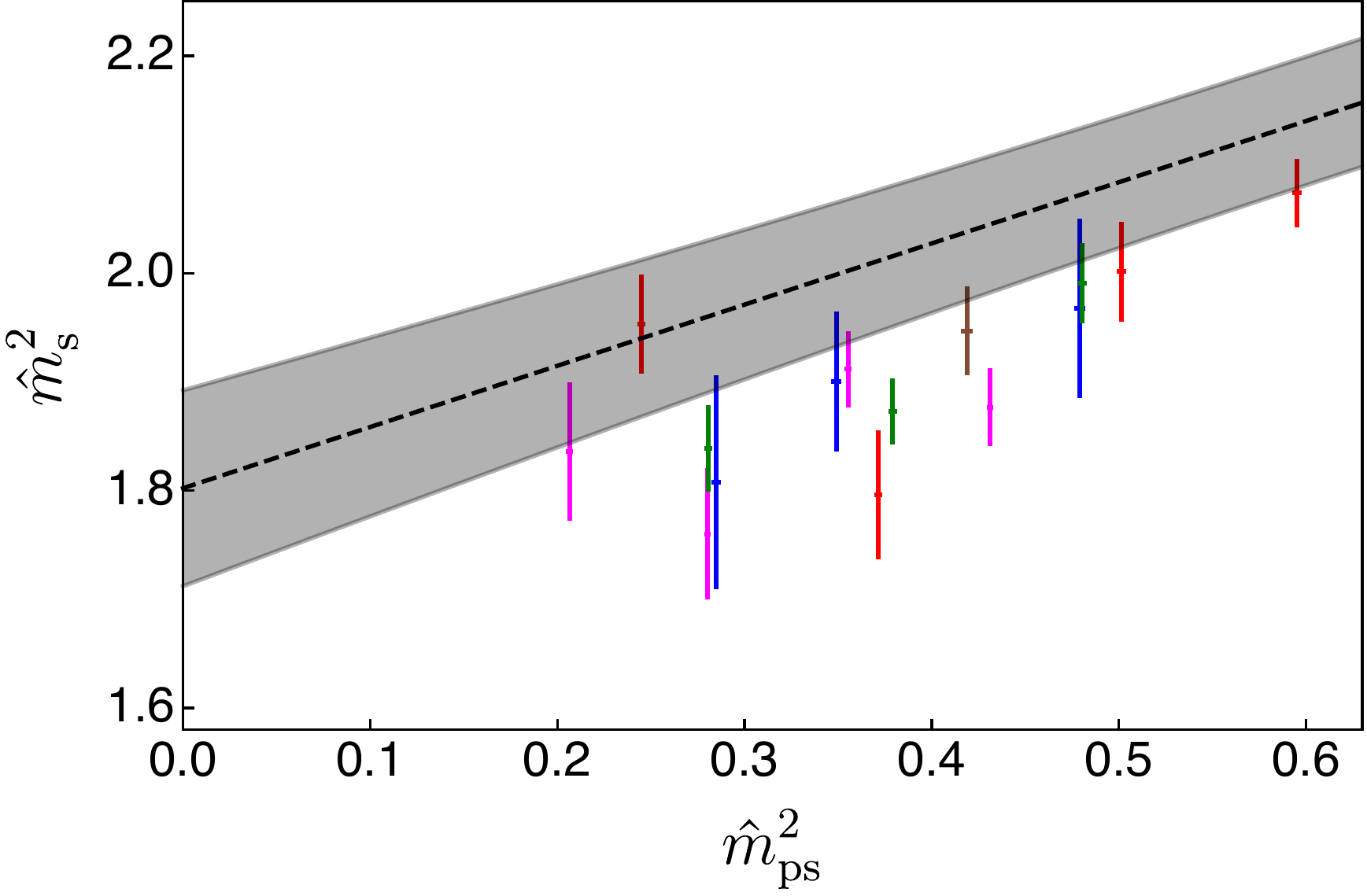}
\caption{%
\label{fig:m2_fit_AS}%
Masses squared of v, t, av, at and s mesons 
(constituted of fermions transforming in the antisymmetric representation),
as a function of the ps meson mass squared $\hat{m}_{\rm ps}^2$,
for $\beta=7.62$ (blue), $7.7$ (purple), $7.85$ (green), $8.0$ (red), and $8.2$ (brown). 
 All quantities are expressed in units of the gradient-flow scale $w_0$.
The results of the continuum and massless extrapolations are represented by the grey bands. 
}
\end{center}
\end{figure}

We perform the numerical fits of  masses and  decay constants
by using the tree-level NLO W$\chi$PT described by Eqs.~(\ref{eq:f2_chipt}) and~(\ref{eq:m2_chipt}).
In Figs.~\ref{fig:f2_fit_AS} and~\ref{fig:m2_fit_AS}, 
we present the fit results denoted by grey bands as well as 
numerical results of the masses and the decay constants 
measured at given lattice parameters. 
For the fits we consider the same ranges of $\hat{m}_{\rm ps}^2$
 taken for the case of fundamental fermions:
$\hat{m}_{\rm ps}^2 \lesssim 0.4$ and $\hat{m}_{\rm ps}^2\lesssim 0.6$, respectively,
 for the ps and all other states. 
Over these mass ranges no significant deviation from 
linearity of the data in $\hat{a}$ and $\hat{m}_{\rm ps}^2$ 
is visible in our data. 
Different colours denote different lattice couplings, 
while the widths of the bands represent the statistical uncertainties of the continuum extrapolations.

The resulting fit values are reported in \Tab{LECs_AS}.
The numbers in the first and second parentheses are the statistical 
and systematic uncertainties of the fits, respectively. 
Once more, we estimate the fitting systematics by taking the maximum and minimum values 
obtained from the set of data excluding the coarsest lattices at $\beta=7.62$ and including or excluding 
the heaviest measurements. 

As in the case of fundamental fermions $Q$, 
we find that for each $\beta$ value the vector masses in units of the pseudoscalar decay constant 
are almost constant over the range of $\hat{m}_{\rm v}/\hat{m}_{\rm ps} 
\gsim 1.3$---see \Tab{meson_ratio_AS}. 
After performing a simple linear extrapolation of these constants, 
we find that $\hat{m}_{\rm v}/\sqrt{2} \hat{f}_{\rm ps}=4.72(4)$ in the continuum limit. 
A more rigorous, yet compatible,  estimate  is obtained by making  
use of the extrapolated results in \Tab{LECs_AS}: 
we find $\hat{m}^{\chi}_{\rm v}/\sqrt{2} \hat{f}^{\chi}_{\rm ps}=4.80(12)(4)$. 
The resulting value of the ratio is smaller than that for the fundamental fermions by $13\%$.

\begin{table}[ht]
\begin{center}
\begin{tabular}{|c|c|c|c|c|}
\hline\hline
 & $~~\hat{f}^{2,\,\chi}_M~~$ & $~~~L^0_{f,M}~~~$
& $~~~~~W^0_{f,M}~~~~~$ & $~~~~~\chi^2/N_{\rm d.o.f}~~~~~$ \cr \hline
ps & $0.01388(18)(10)$ & $1.754(41)(28)$ & $-0.00028(26)(15)$ & $1.3$\cr
v & $0.0404(21)(7)$ & $0.626(91)(16)$ & $0.0310(28)(17)$ & $2.4$\cr
av & $0.084(8)(5)$ & $0.01(13)(9)$ & $-0.022(12)(9)$ & $2.3$\cr
\hline\hline
 & $~~\hat{m}^{2,\,\chi}_M~~$ & $~~~L^0_{m,M}~~~$
& $~~~~~W^0_{m,M}~~~~~$ & $~~~~~\chi^2/N_{\rm d.o.f}~~~~~$ \cr \hline
v & $0.657(21)(21)$ & $1.375(56)(5)$ & $-0.336(34)(4)$ & $0.8$\cr
t & $0.675(29)(19)$ & $1.26(7)(7)$ & $-0.326(47)(19)$ & $1.2$\cr
av & $2.01(11)(7)$ & $0.70(10)(5)$ & $-0.33(17)(11)$ & $2.1$ \cr
at & $2.50(18)(7)$ & $0.32(12)(7)$ & $-0.48(24)(8)$ & $2.4$ \cr
s & $1.80(9)(13)$ & $0.32(7)(14)$ & $-0.21(12)(15)$ & $1.6$ \cr
\hline\hline
\end{tabular}
\end{center}
\caption{
\label{tab:LECs_AS}
Results of the fit of the coefficients in Eqs.~(\ref{eq:f2_chipt}) and~(\ref{eq:m2_chipt}), used
in the  continuum and massless extrapolations of  masses and decay constants 
of mesons in the quenched simulations involving Dirac fermions transforming in the 
2-index antisymmetric representation. In parentheses we show statistical and systematic
 errors, respectively.
}
\end{table}

\subsection{Quenched spectrum: Comparison}
\label{Sec:comparison}

\begin{figure}[ht]
\begin{center}
\includegraphics[width=.45\textwidth]{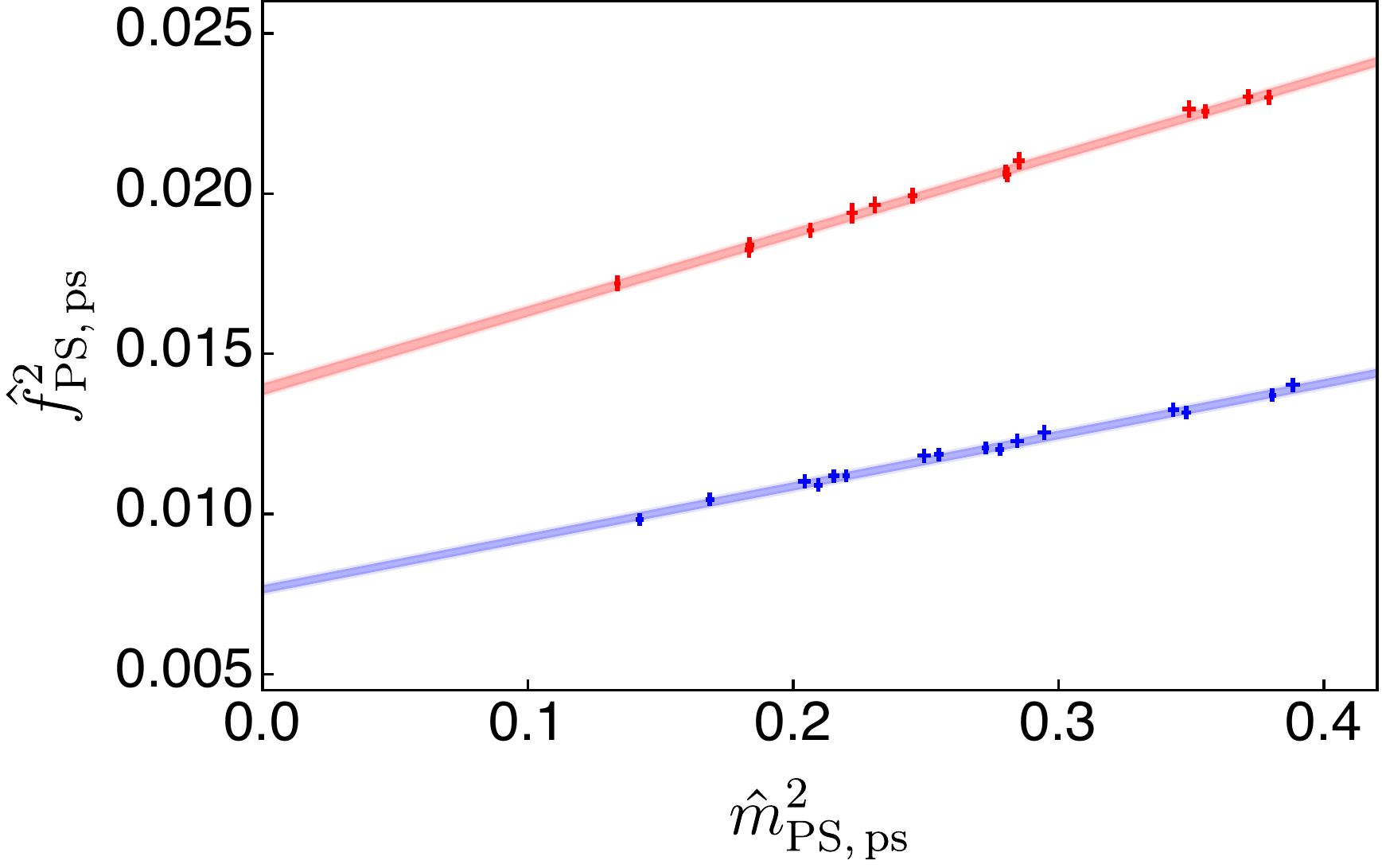}
\includegraphics[width=.45\textwidth]{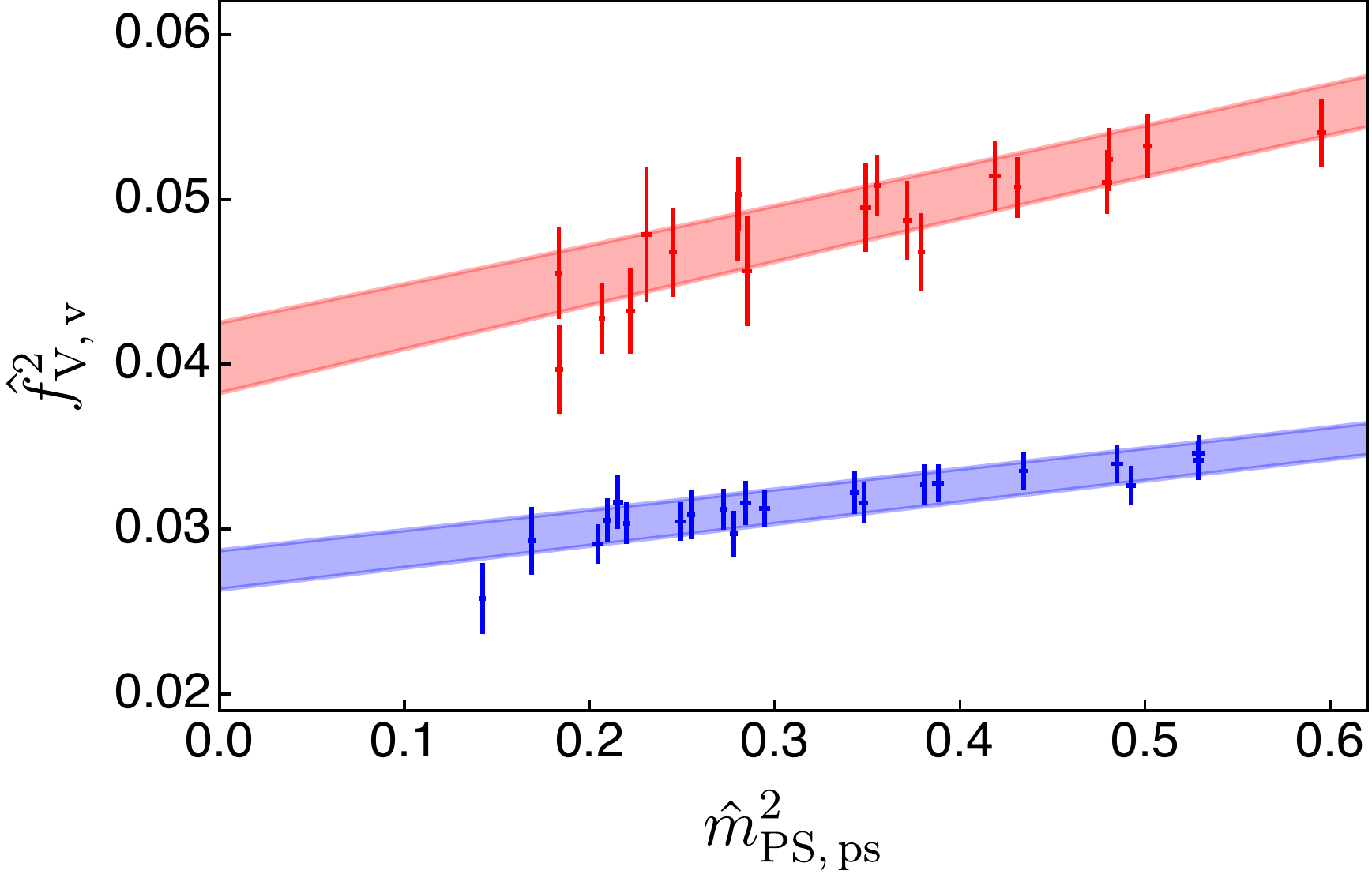}
\includegraphics[width=.45\textwidth]{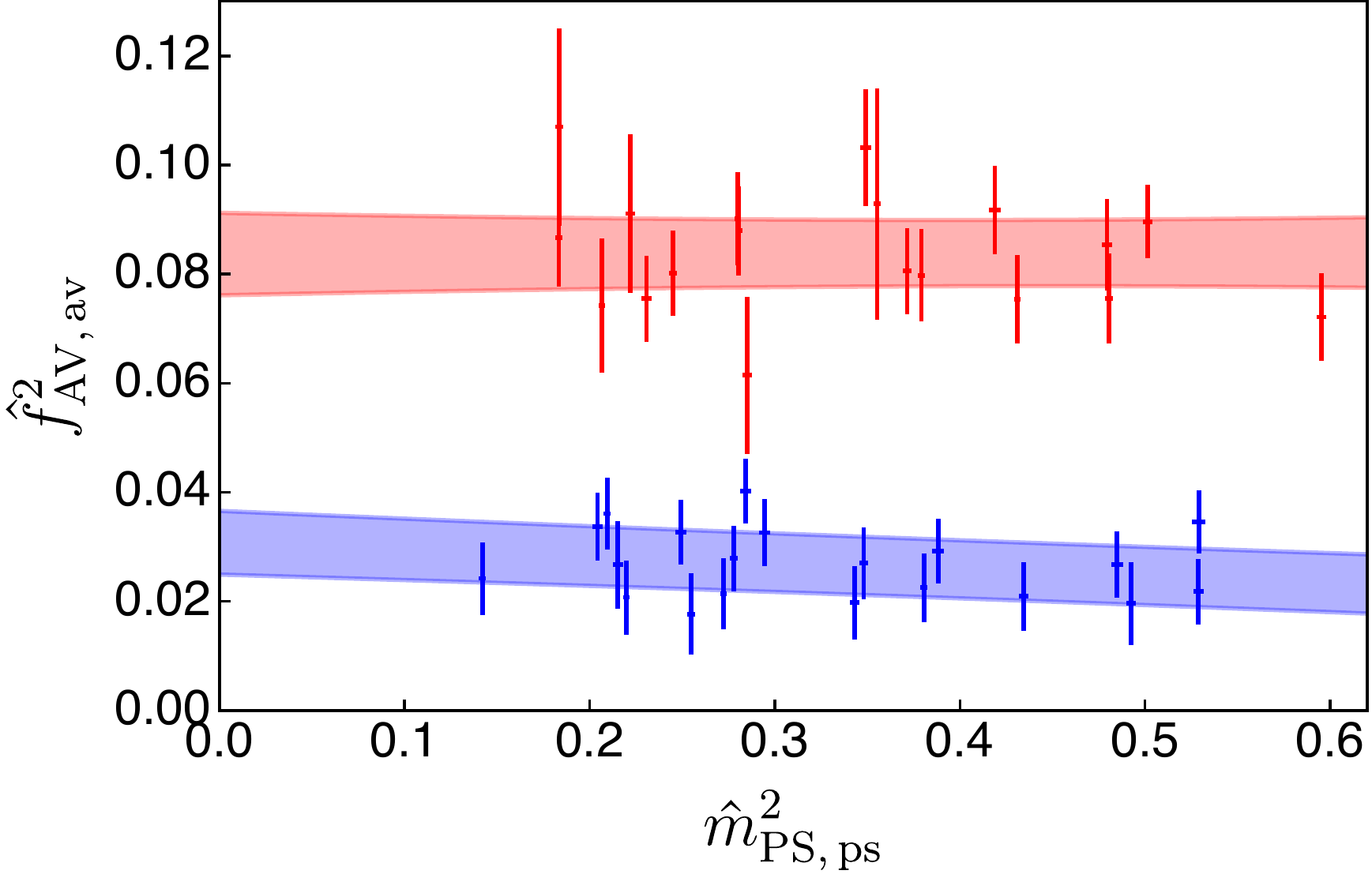}
\caption{%
\label{fig:f2_Q_F_vs_AS}%
Comparison of the decay constant squared (in the continuum limit)  of the mesons 
as a function of the pseudoscalar meson mass squared, 
in units of $w_0$,
for fermion constituents transforming in the fundamental (blue) or 2-index antisymmetric (red) 
representation. 
}
\end{center}
\end{figure}

\begin{figure}[ht]
\begin{center}
\includegraphics[width=.45\textwidth]{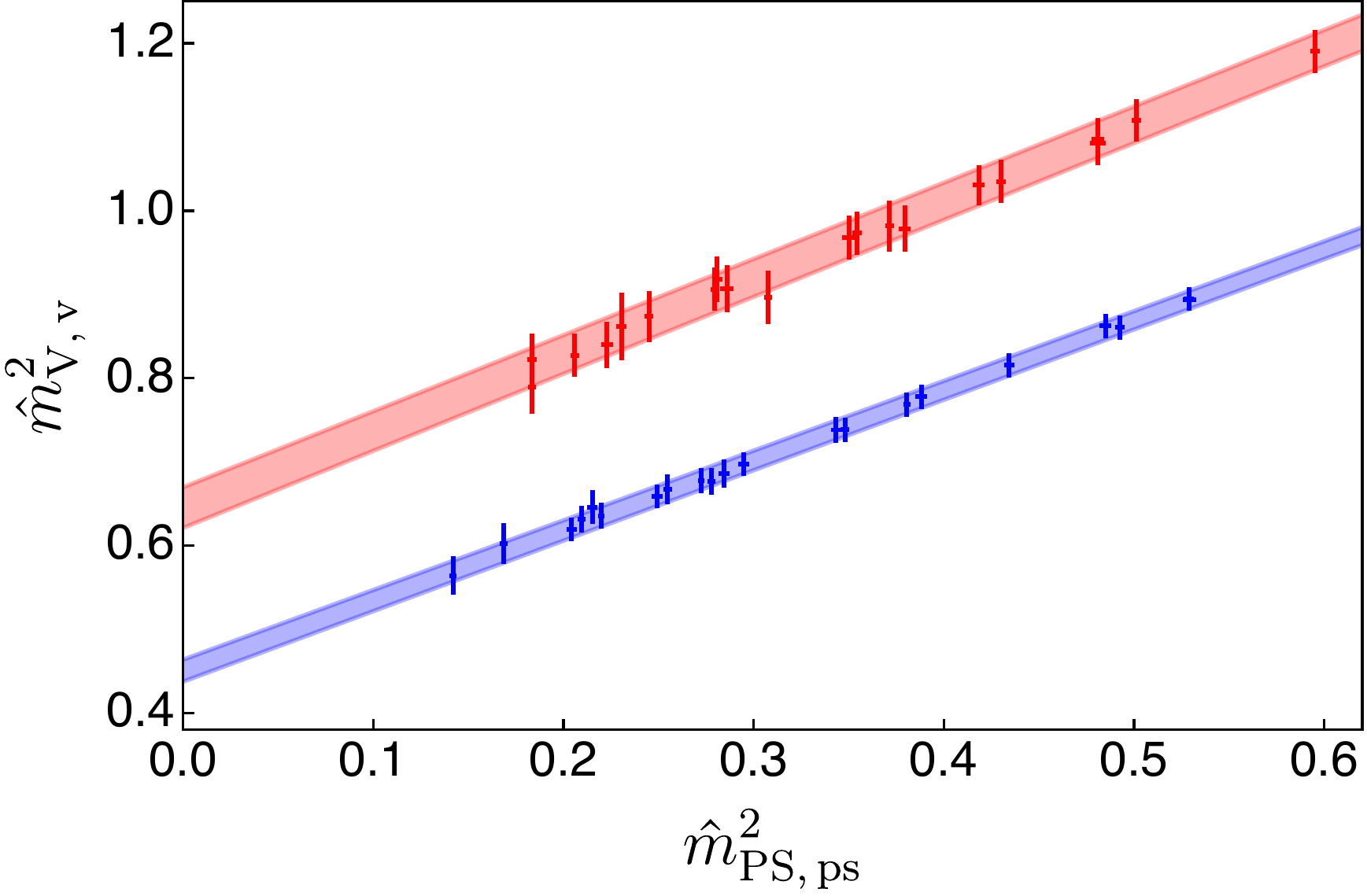}
\includegraphics[width=.45\textwidth]{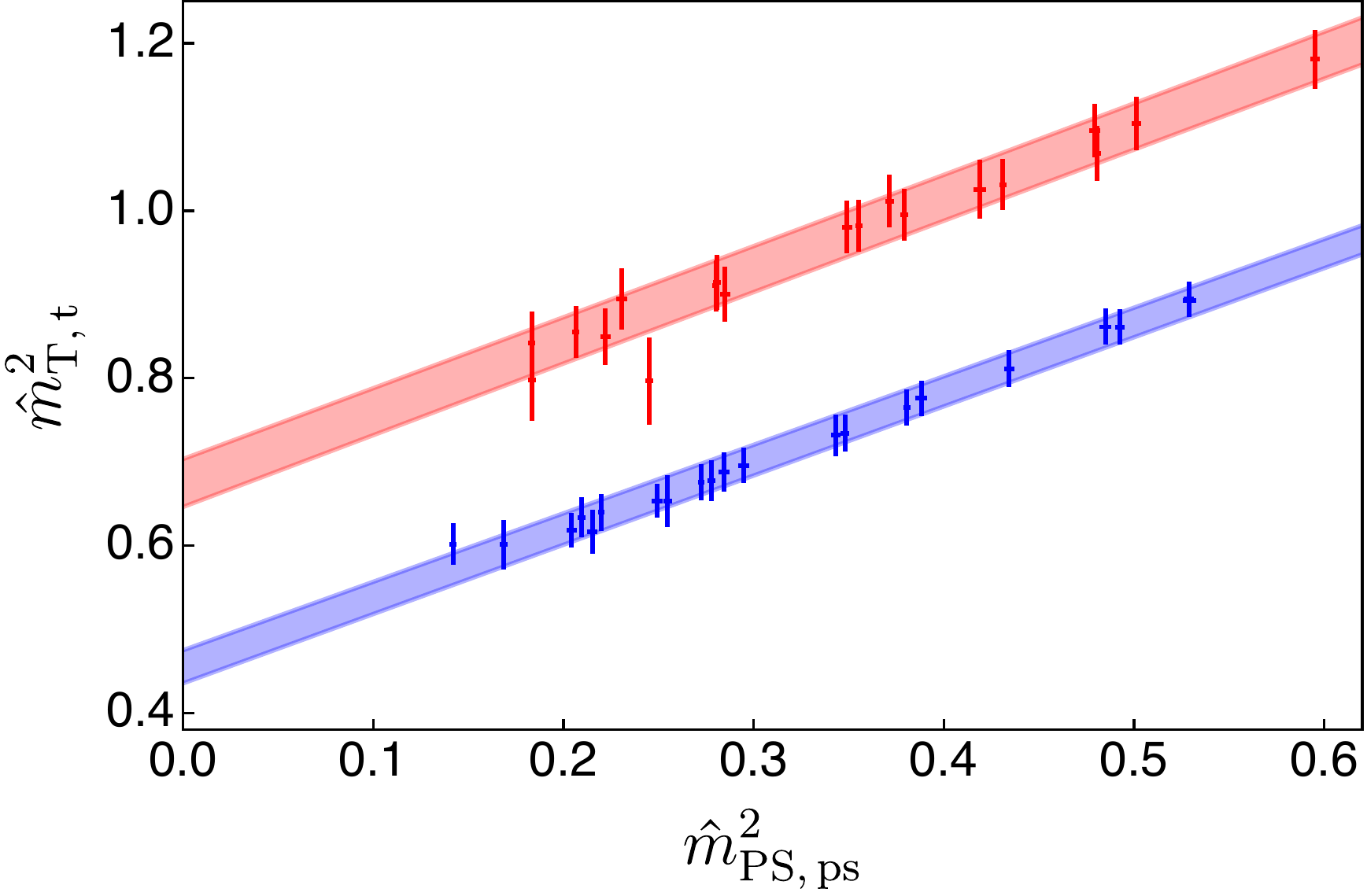}
\includegraphics[width=.45\textwidth]{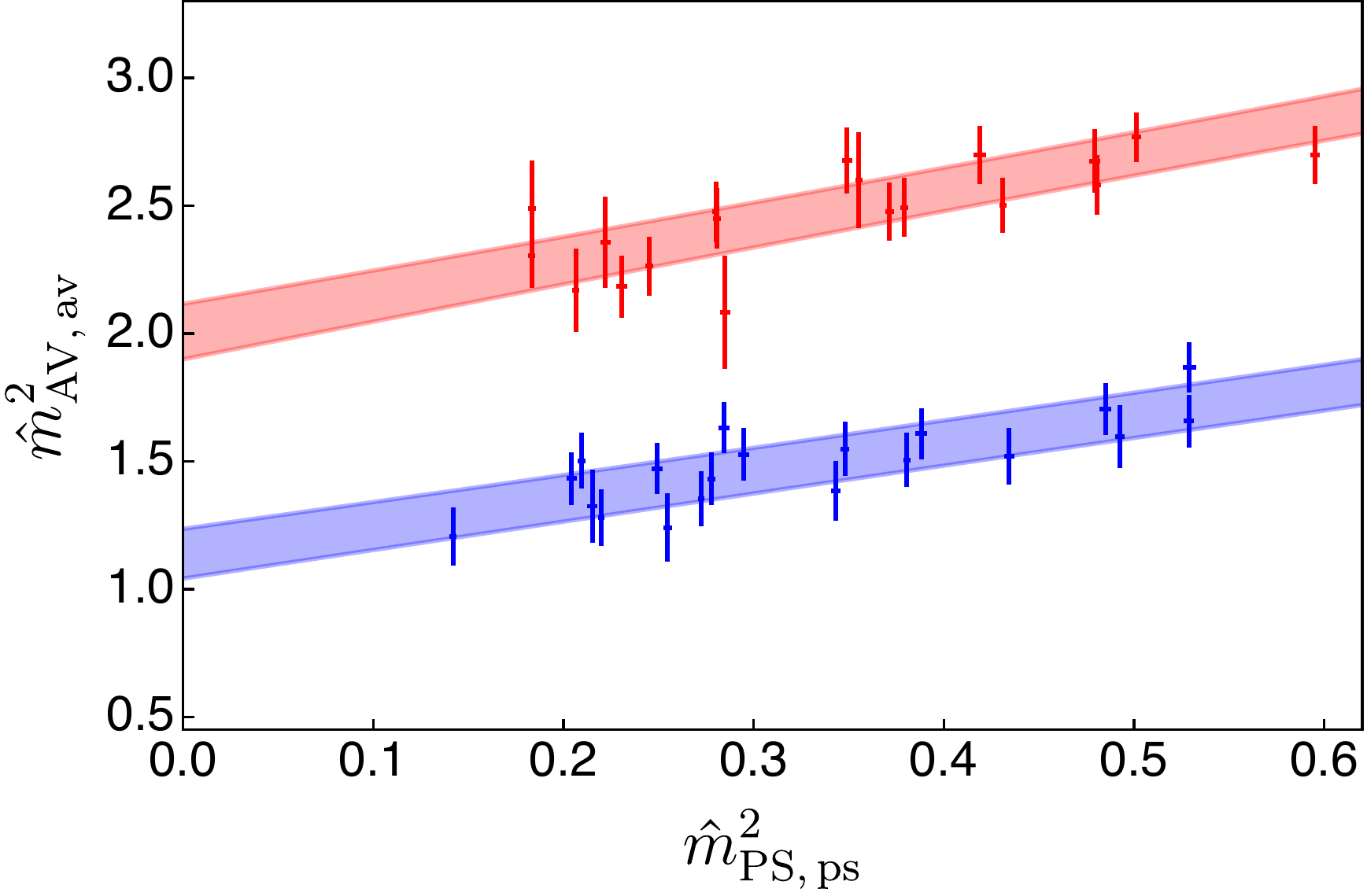}
\includegraphics[width=.45\textwidth]{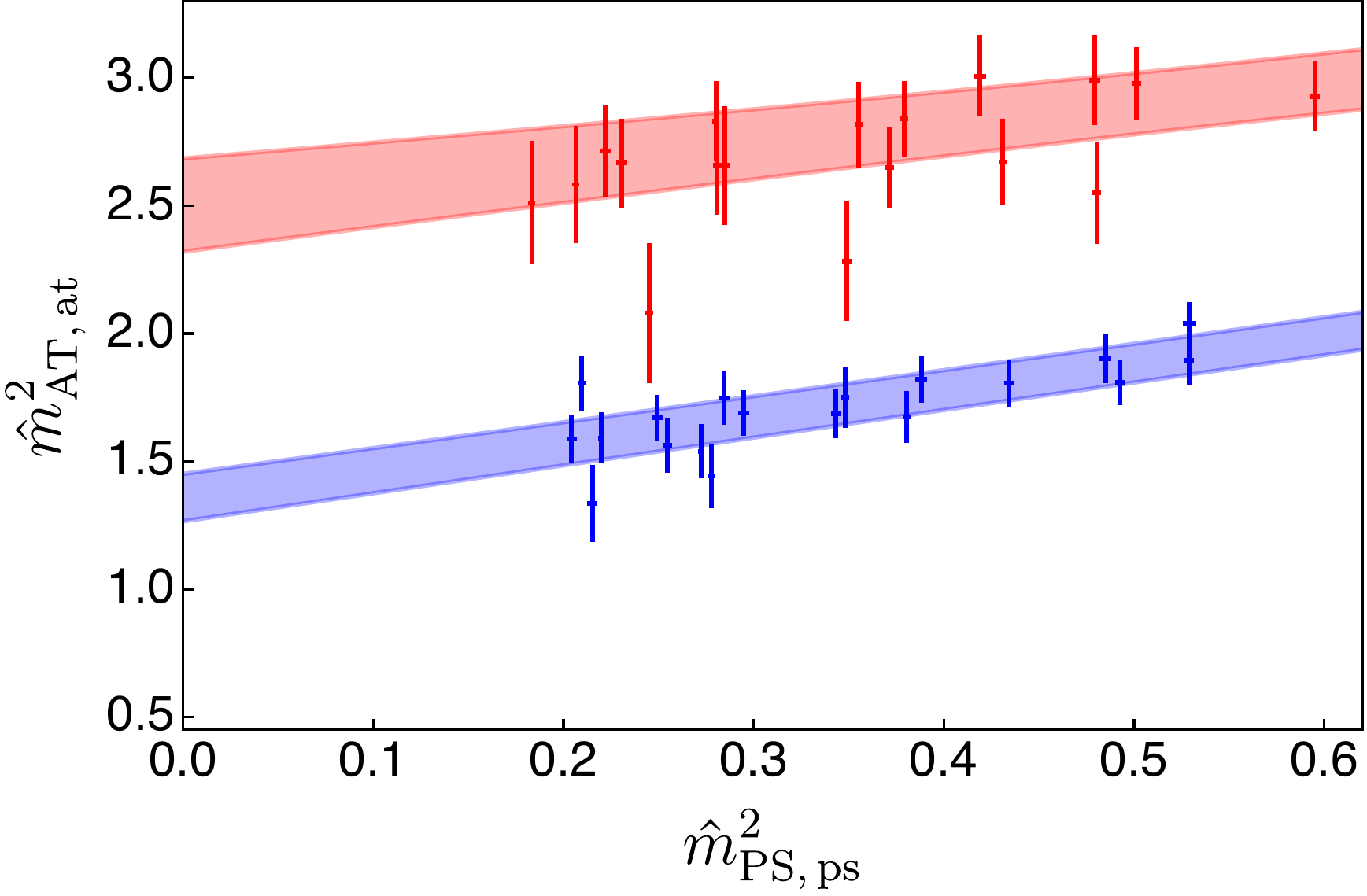}
\includegraphics[width=.45\textwidth]{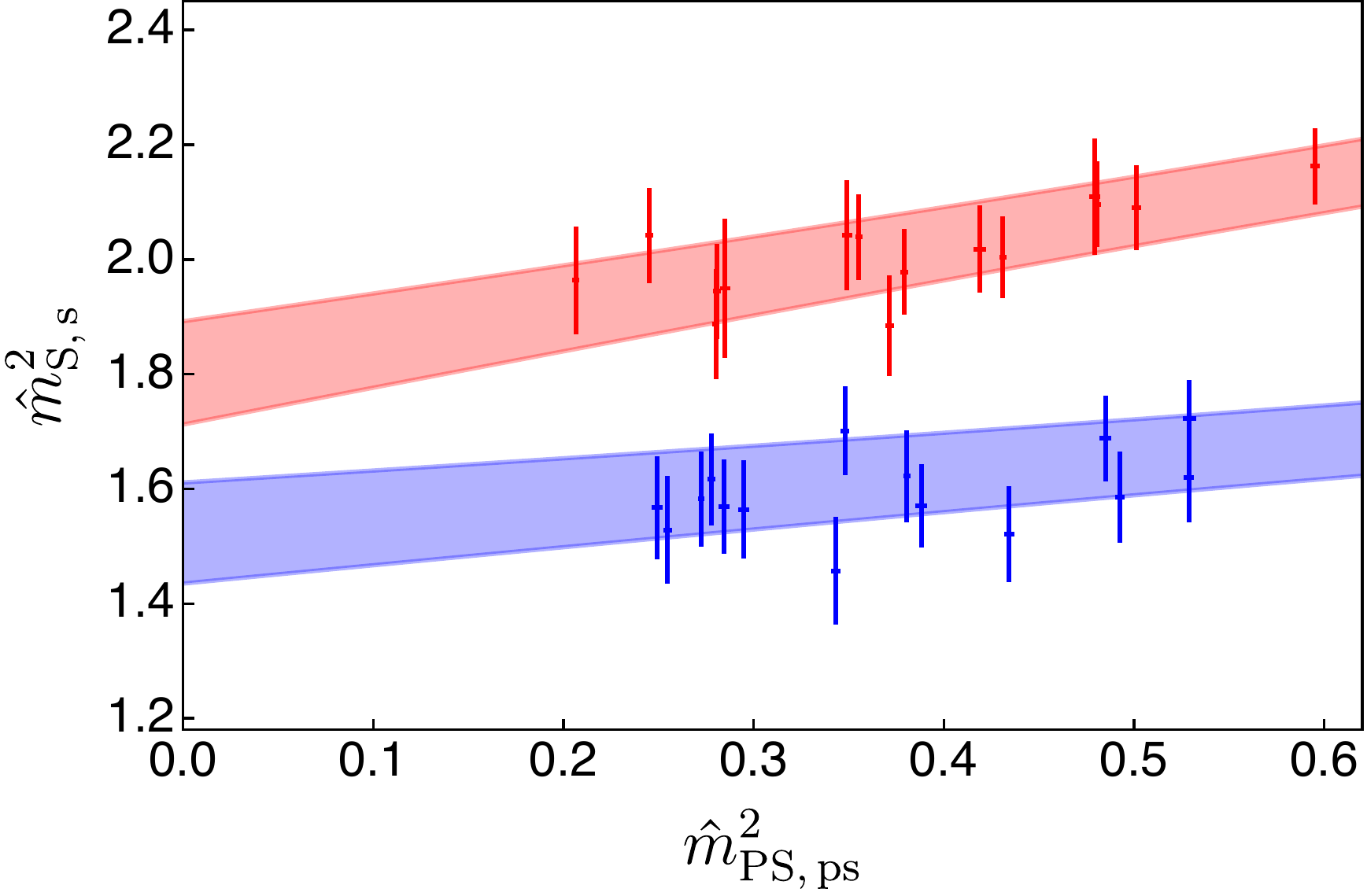}
\caption{%
\label{fig:m2_Q_F_vs_AS}%
Comparison of the mass squared (in the continuum limit) of the mesons 
as a function of the pseudoscalar meson mass squared, 
in units of $w_0$,
for fermion constituents transforming in the fundamental (blue) or 2-index antisymmetric (red) 
representation. 
}
\end{center}
\end{figure}

Fig.~\ref{fig:f2_Q_F_vs_AS} shows a visual comparison between the decay constants
of the pseudoscalar, vector, and axial-vector mesons, made of fermions transforming
in the fundamental representation of $Sp(4)$ (PS, V, AV) and in the 2-index antisymmetric 
representation (ps, v, av). In order to make the comparison, we plot the continuum-limit
results by naively identifying the masses of the
pseudoscalars $\hat{m}_{\rm PS}^2=\hat{m}_{\rm ps}^2$
as the abscissa.
The comparison at finite mass should be taken with some caution, as 
 the symmetry-breaking operators controlling the mass of PS and ps 
 states are distinct, but the massless extrapolations can be compared unambiguously.
 We repeat the exercise also for the masses of all the mesons, and 
 show the result in Fig.~\ref{fig:m2_Q_F_vs_AS}.
 
In all cases we considered, masses and decay constants of bound
 states made of fermions $\Psi$ transforming in the
2-index antisymmetric representation are considerably larger than those made of 
fermions $Q$ transforming in the fundamental representation.
Focusing on the massless limit,
we find that the ratio $\hat{f}_{\rm av}^2/\hat{f}_{\rm AV}^2 =2.7 \pm 1.1$ is the largest,
while $\hat{m}_{\rm s}^2/\hat{m}_{\rm S}^2 =1.18 \pm 0.13$ is the smallest, and
 the other results are distributed  in the range between these two values. 
 The hierarchy between the pseudoscalar 
 decay constants  is
 important  in the CHM context; we find 
 that $\hat{f}_{\rm ps}^2/\hat{f}_{\rm PS}^2 =1.81 \pm 0.04$.
 It is also to be noted that the mass of the vector states v is larger, but not substantially so,
 in respect to that of the corresponding V mesons, with 
 $\hat{m}_{\rm v}^2/\hat{m}_{\rm V}^2 =1.46 \pm 0.08$.
 
 How much of the above holds true for the dynamical calculations is not known and is an
  interesting topic for future studies.
It was shown in Ref.~\cite{Bennett:2019jzz} that, by comparing quenched and dynamical calculations
 for mesons in the fundamental representation (performed in comparable ranges of fermion mass),
 and after both the continuum and massless extrapolations were performed, 
 the discrepancies are not too large: ${\cal O}(25\%)$ for  $\hat{m}_{\rm S}^2$,  
 ${\cal O}(20\%)$ for $\hat{f}_{\rm PS}^2$, ${\cal O}(10\%)$ for  $\hat{m}_{\rm V}^2$,
 and smaller for the other measurements.
 Whether this is due to the fact that all the calculations in Ref.~\cite{Bennett:2019jzz}
are performed in a range
  of fermion masses that are comparatively large or to other reasons---the large-$N$ behaviour 
  of the theory might already be dominating the dynamics of $Sp(4)$ mesons, for example---is not 
  currently known and should be studied in future 
  dedicated investigations.
  Yet, it is suggestive that no dramatic discrepancy has emerged so far, 
  for all the observables we considered.
  
  We conclude this section by reminding the reader that the calculations performed for this paper, being
  done with the quenched approximation,
  are insensitive to the number of fundamental flavours $N_f$ and antisymmetric flavours $n_f$
  and hence apply to other models, beyond the phenomenologically relevant case with
  $N_f=2$ and $n_f=3$.
  A recent lattice study within the $SU(3)$ gauge theory~\cite{Nogradi:2019iek} of the
  ratio $m_{\rho}/f_{\pi}$ between the mass
  of the rho mesons and the decay constant of the pions  
  (corresponding to $m_{\rm V}/f_{\rm PS}$ in this paper) shows no appreciable 
  dependence on the number of flavours $N_f\lsim 6$---as long as the theory is deep inside
  the regime in which chiral symmetry breaking occurs. It would be interesting to measure
  whether this holds true
  also for other representations, in the dynamical theories.
  Meanwhile, we find that in our quenched calculation, after 
 taking both the continuum and massless limits, for the fundamental representation
 we have $\hat{m}_{\rm V}^2/\hat{f}_{\rm PS}^2=59.0\pm 2.2$, while
  for the antisymmetric representation we find 
  $\hat{m}_{\rm v}^2/\hat{f}_{\rm ps}^2=47.3\pm 2.3$.\footnote{
These data have been used in Ref.~\cite{Nogradi:2019auv} to compare these quantities with other theories.
}
  The discrepancy reaches beyond the $3\sigma$ level, 
  suggesting that this ratio---which enters 
  into the Kawarabayashi-Suzuki-Riazuddin-Fayyazuddin (KSRF) 
relation $M_{\rho}^2/(g_{\rho\pi\pi}^2f_{\pi}^2)
=2$~\cite{Kawarabayashi:1966kd,Riazuddin:1966sw}---depends on the fermion representation.
By comparison, the ratio obtained from the numerical studies with 
dynamical Dirac fermions in the fundamental representation is 
$\hat{m}_V^2/\hat{f}_{PS}^2=65.4 \pm 5.0$~\cite{Bennett:2019jzz}, 
which is slightly larger than the result  of our quenched calculation.
Once more, checking this result (as well as the KSRF relations)
 in the full dynamical theory with fermions in the antisymmetric representation
 would be of great interest.

\section{Global fits}
\label{Sec:fits}

\begin{figure}[ht]
\begin{center}
\includegraphics[width=.45\textwidth]{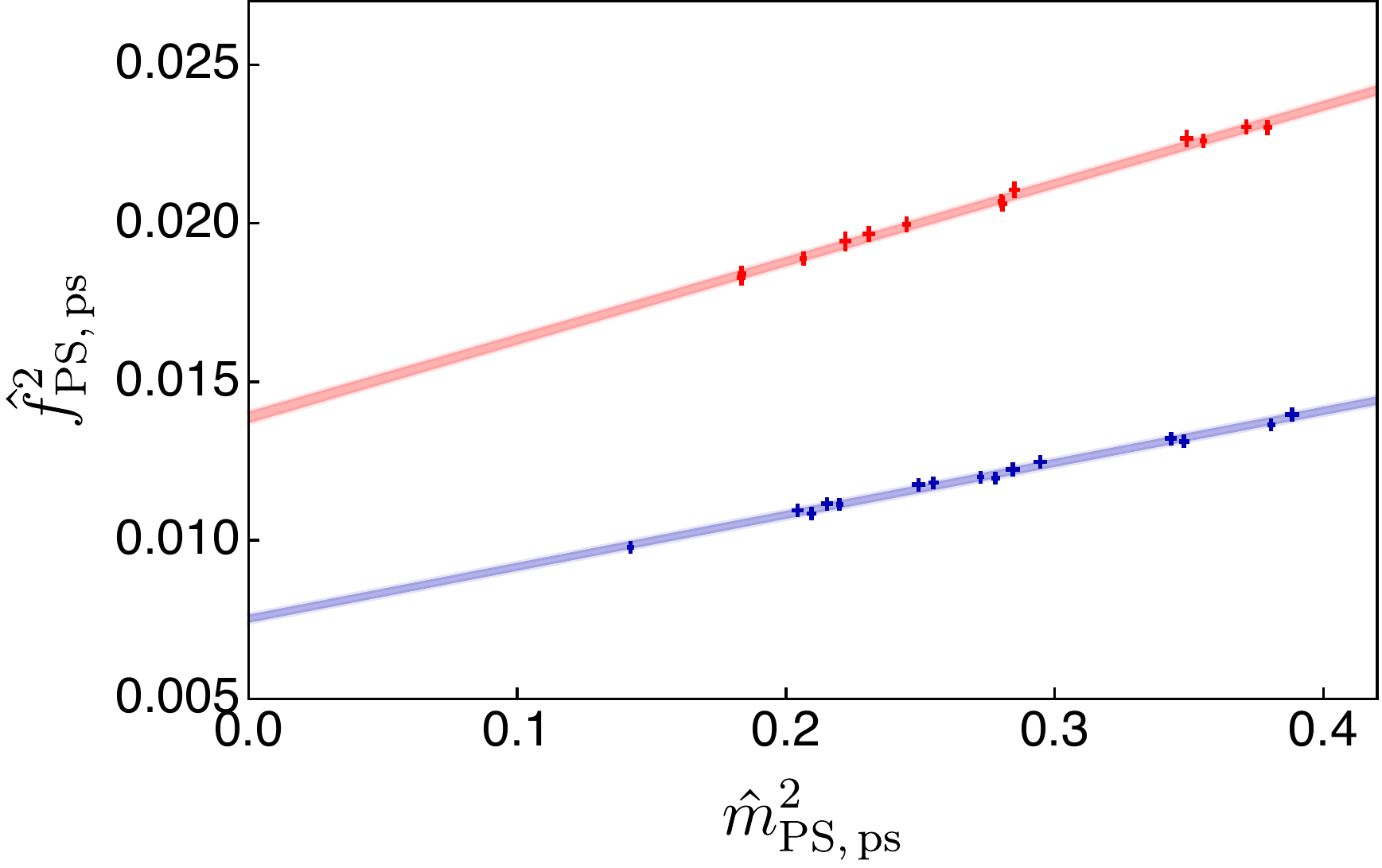}
\includegraphics[width=.45\textwidth]{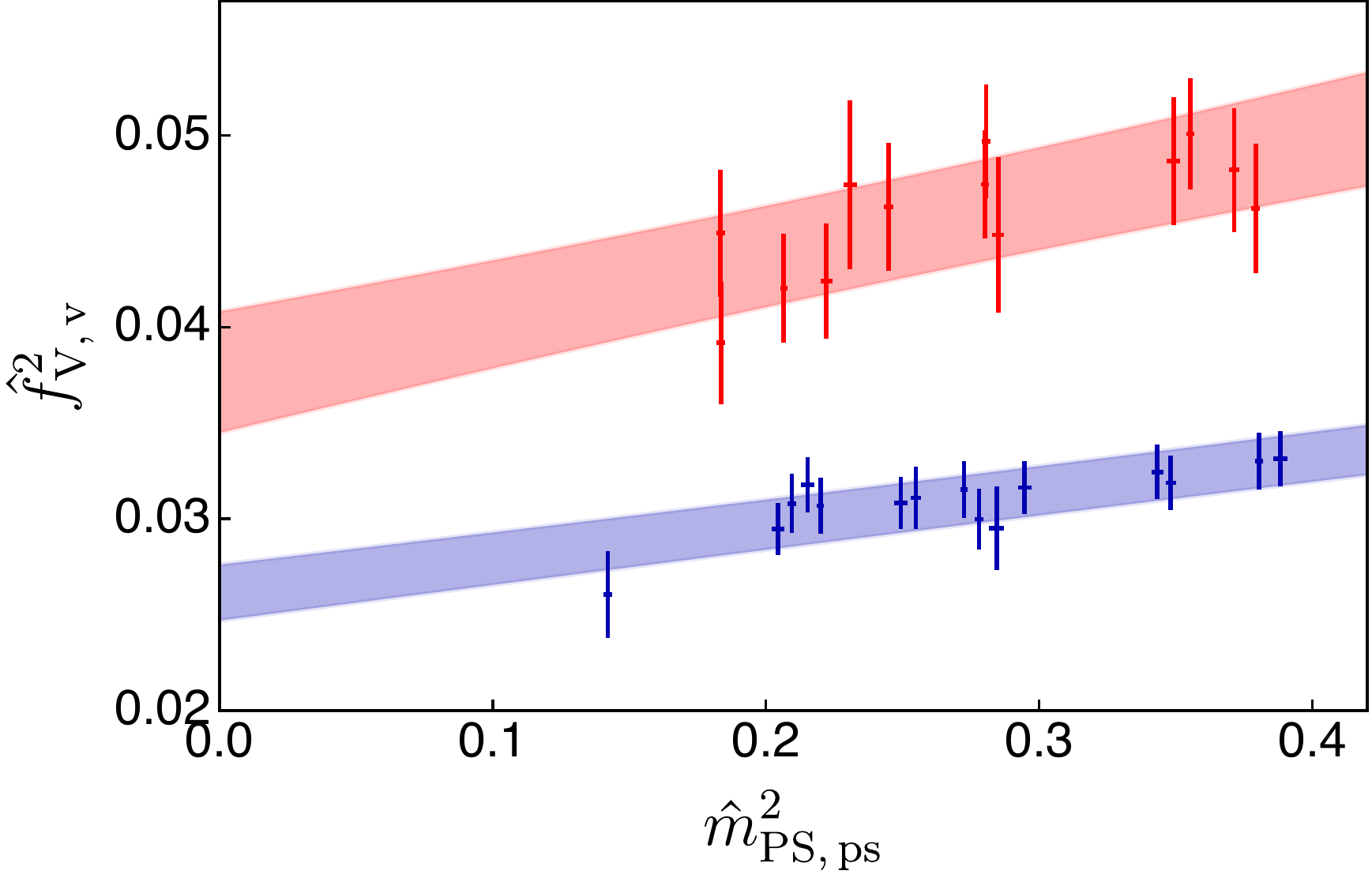}
\includegraphics[width=.45\textwidth]{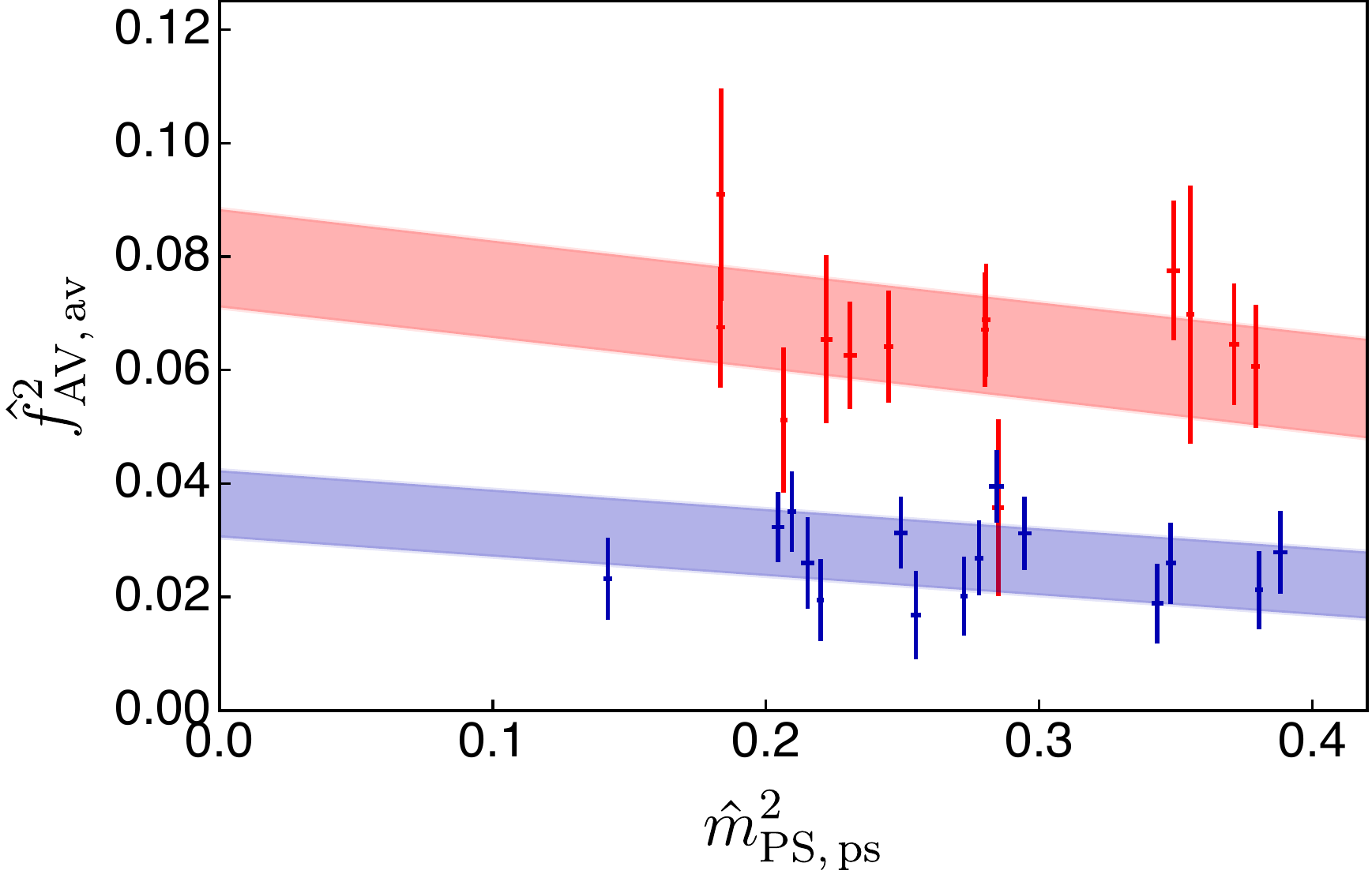}
\includegraphics[width=.45\textwidth]{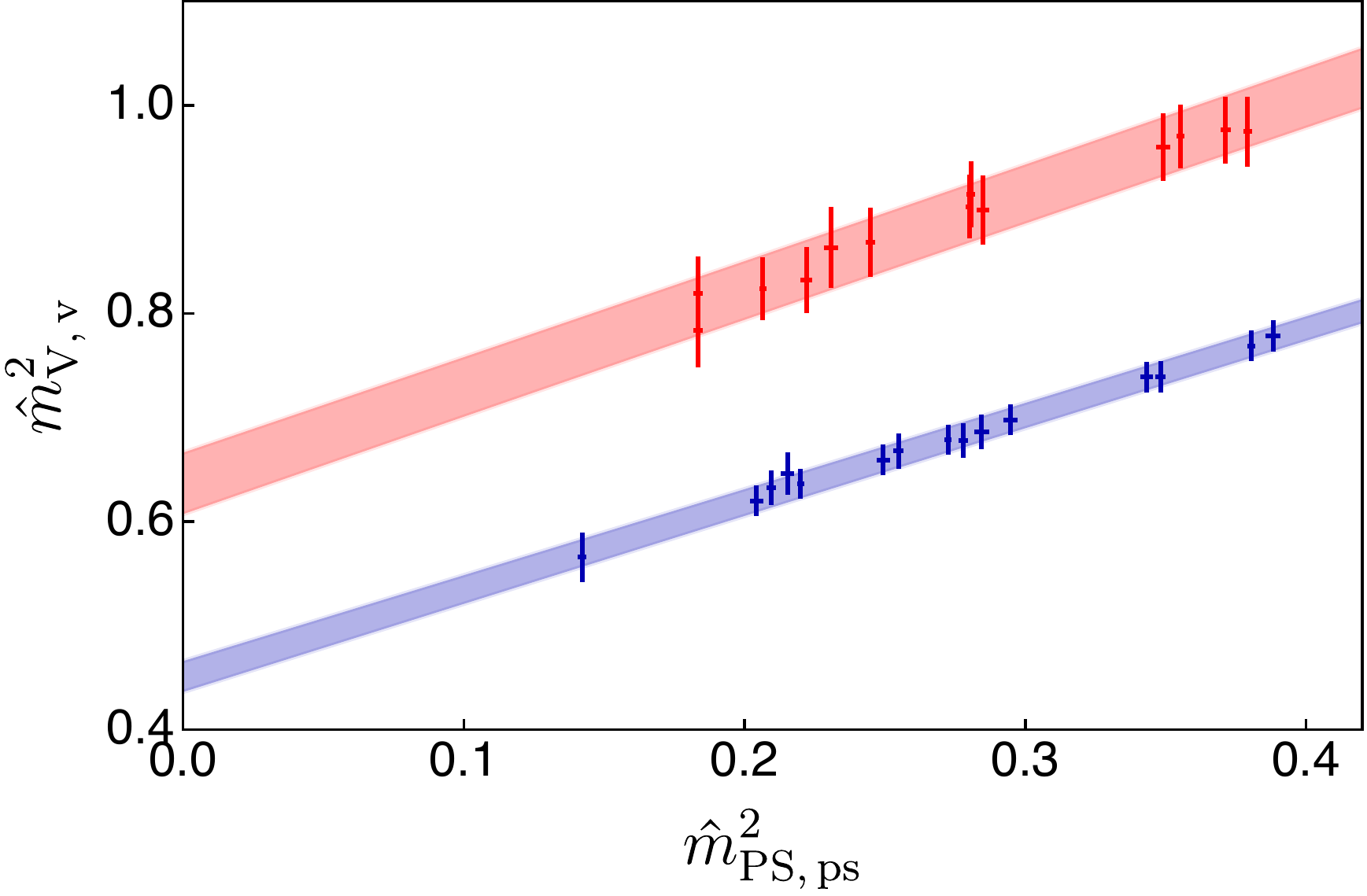}
\includegraphics[width=.45\textwidth]{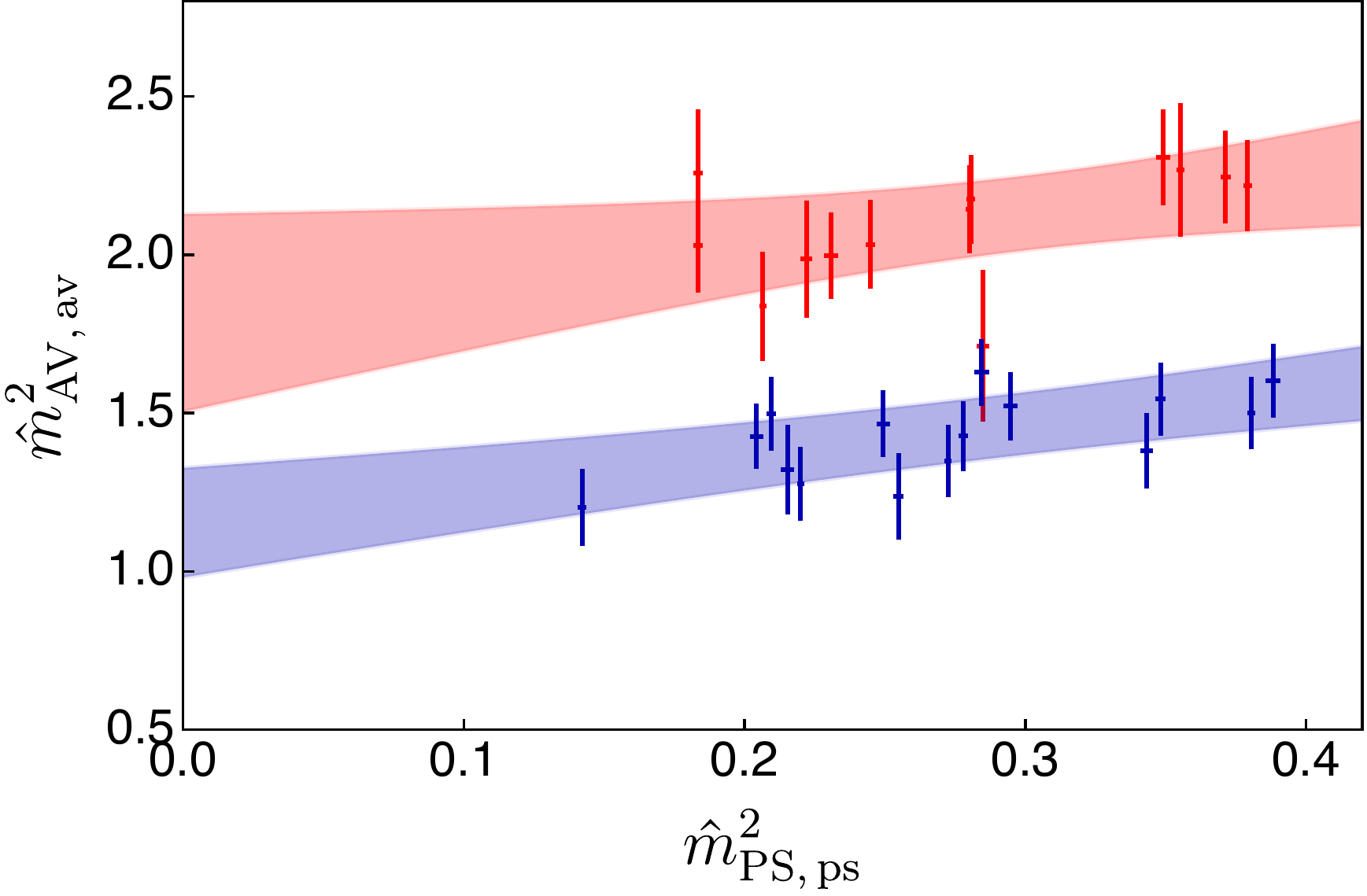}
\caption{%
\label{fig:gfit}%
Decay constants and masses in the continuum limit after subtracting
 lattice artefacts due to the finite lattice spacing. 
The global fit results are denoted by blue solid bands for the mesons constituted of fundamental 
fermions $Q$, and red bands for the ones constituted of antisymmetric 
fermions $\Psi$. The width of the bands indicates the statistical errors.
}
\end{center}
\end{figure}

\begin{figure}[ht]
\begin{center}
\includegraphics[width=.45\textwidth]{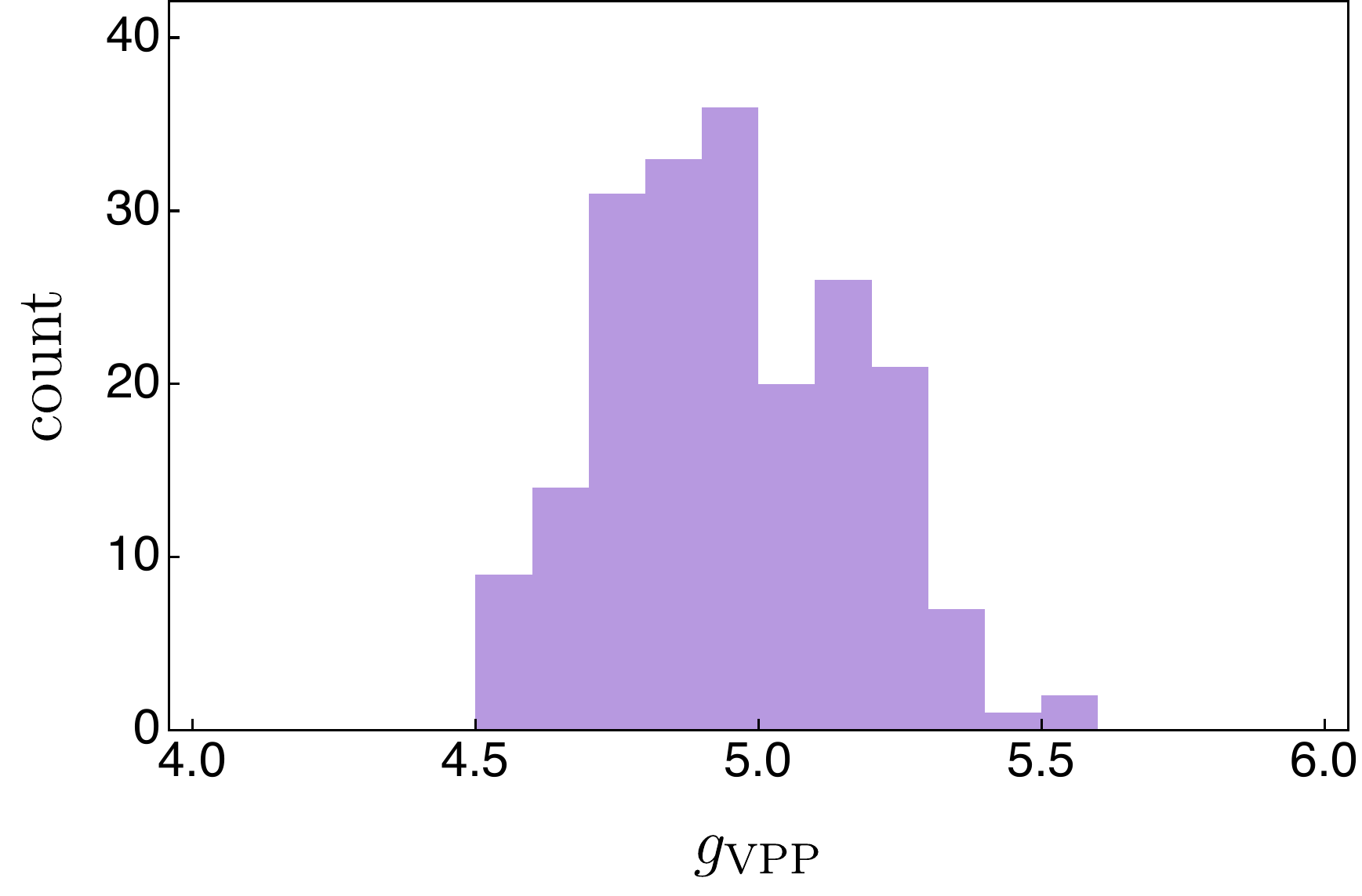}
\includegraphics[width=.45\textwidth]{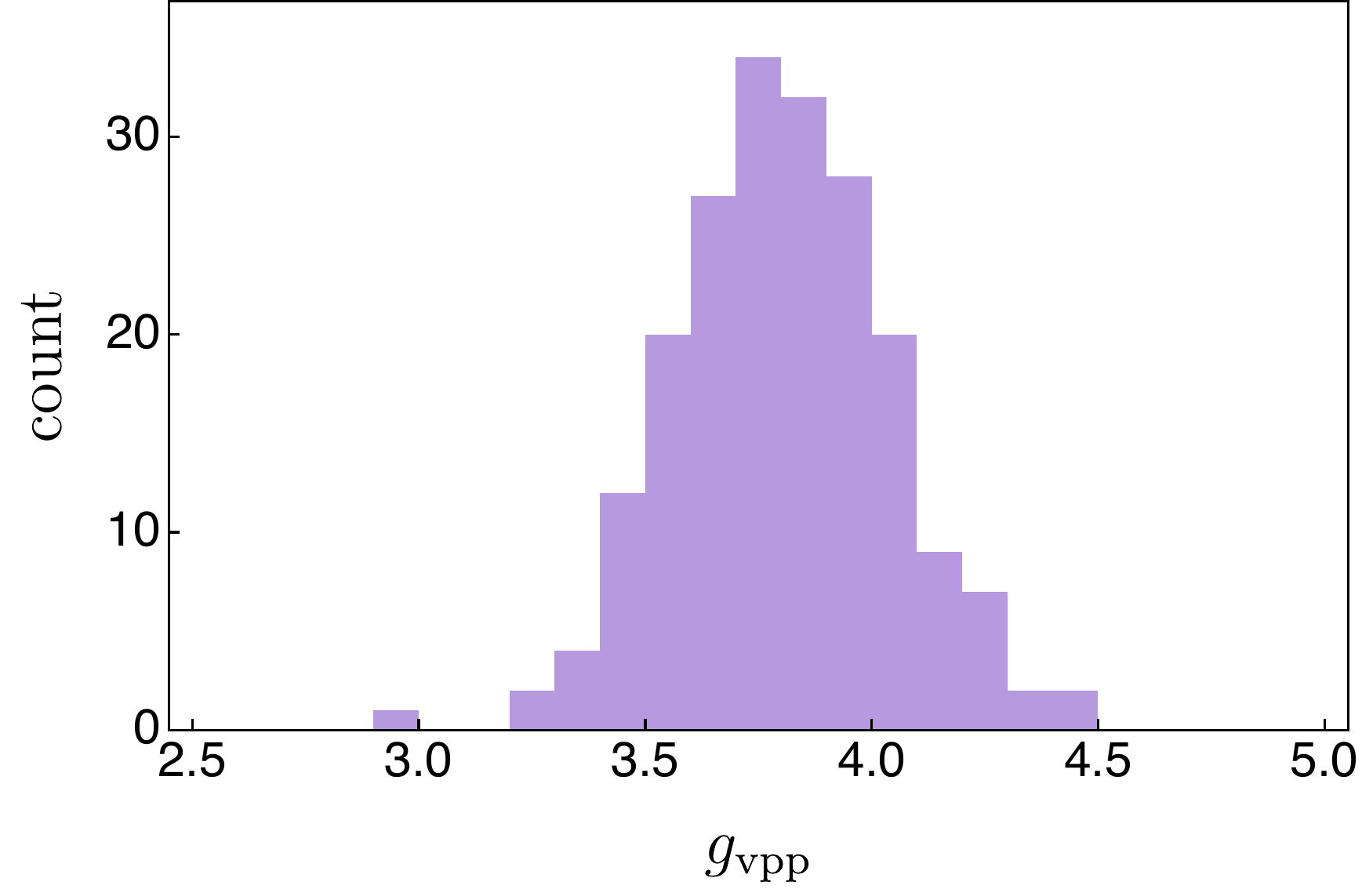}
\caption{%
\label{fig:gvpp_hist}%
Histogram distribution of the $g_{VPP}$ (left panel) and $g_{vpp}$ (right panel)
couplings,  obtained from the  quenched calculation by applying the global fit strategy discussed in the text.
In the former, $g_{VPP}$ denotes the coupling between mesons
composed of fermions in the fundamental. In the latter case the constituent fermions transform 
in the 2-index antisymmetric  representation. 
}
\end{center}
\end{figure}

In this section, we perform a global fit of the continuum-extrapolated masses and decay constants 
of PS, V, and AV mesons to the EFT described in Sec.~\ref{Sec:HLS}. 
As stated there,  the EFT equations are applicable  both to mesons constituted of fermions
 in the fundamental as well as 2-index antisymmetric representations of the $Sp(4)$ gauge group. 
We also recall from Ref.~\cite{Bennett:2017kga}
that several working assumptions have been used to arrive
at   Eqs.~(\ref{eq:gfit_mv})--(\ref{eq:gfit_fps}). 
We follow in the analysis  the  prescription introduced in Ref.~\cite{Bennett:2019jzz}.
We only repeat some of the essential features of the process,  while referring the 
reader to Ref.~\cite{Bennett:2019jzz} for details. We  focus instead on the results of the global fit.

We start by restricting the data analysed to lie 
in the mass range over which all the measured masses and decay constants 
can be extrapolated to the continuum limit using Eqs.~(\ref{eq:f2_chipt}) and~(\ref{eq:m2_chipt}).
In the case of the fundamental representation, we restrict our measurements to include
only QB1FM3$-$QB1FM6, QB2FM1$-$QB2FM3, QB3FM4$-$QB3FM7, 
QB4FM6$-$QB4FM8, and QB5FM2$-$QB5FM3.
In the case of antisymmetric representation,  we restrict to 
 QB1ASM4$-$QB1ASM6, QB2ASM3$-$QB2ASM6, QB3ASM2$-$QB3ASM4, 
 QB4ASM4$-$QB4ASM6, and QB5ASM2. 
As anticipated  in \Sec{continuum}, we use the LO $\chi$PT result for the pseudoscalar mass, 
and replace the fermion mass in Eqs.~(\ref{eq:gfit_mv})--(\ref{eq:gfit_fps})
by $\hat{m}_{\rm PS}^2=2B m_f$. 
In the mass range considered, this replacement is supported by the numerical data, 
as $\hat{m}_{\rm PS}^2$ is found to be approximately linear
with the mass of the fermion $\hat{m}_0$. 
Accordingly, we expand the EFT equations and truncate at the linear order in $\hat{m}_{\rm PS}^2$. 
The resulting fit equations have been presented as Eqs.~(6.1)--(6.5) in Ref.~\cite{Bennett:2019jzz}. 
The ten unknown low-energy constants, denoted as
 $(\hat{f},~\hat{F},~b,~c,~g_V~,\kappa,~\hat{v}_1,~\hat{v}_2,~\hat{y}_3,~\hat{y}_4)$, 
are appropriately redefined by introducing the gradient-flow scale $w_0$. 

We perform the  numerical global fits of the data to the EFTs,
 via  standard $\chi^2$ minimisation, by using $200$ bootstrapped samples 
and a simplified $\chi^2$ function that is built by just summing the individual $\chi^2$ 
functions for the five independent fit equations. 
The fit results satisfy the constraints obtained from the unitarity conditions in Eq.~(6.8) 
of Ref.~\cite{Bennett:2019jzz}. 
In practice, we guide the fits by  an initial minimisation of the full dataset.
In Fig.~\ref{fig:gfit}  we present the results of the global fit along with 
the continuum-extrapolated data 
used for the fits, by further comparing the 
results originating from fundamental and antisymmetric fermions. 
In the figure, the fit results are presented by shaded 
bands, the widths of which represent the statistical uncertainties. 
The quality of the fits is measured by the fact that $\chi^2/N_{\rm d.o.f}\sim 0.6$ 
at the minimum, although one should remember that correlations have not been taken into 
consideration in the analysis.
The results of continuum and massless extrapolations, displayed
 in Figs.~\ref{fig:f2_Q_F_vs_AS} and~\ref{fig:m2_Q_F_vs_AS}, 
 are  in good agreement, even in proximity of the massless limit, with those of this
 alternative analysis.

As pointed out in Ref.~\cite{Bennett:2019jzz}, some of the parameters in the EFTs 
are not well constrained by the global fit of measurements coming from two-point functions only.
Hence, we do not report the individual best-fit results, which are affected by flat 
directions and large correlations. 
Yet,  in the same reference it is observed that some (nontrivial) combinations of 
the parameters may be determined well. 
One of the most interesting such  quantities is 
the coupling constant associated with the decay of a vector meson V (or v) into two 
pseudoscalar mesons PS (ps). 
These couplings play the same role as the $g_{\rho \pi \pi}$ in low-energy QCD. 
The resulting values in the cases of fundamental and antisymmetric fermions are
\beq
g_{\rm VPP}^{\chi}=4.95(21)(8)~~~{\rm and}~~~
g_{\rm vpp}^{\chi}=3.80(24)(16),
\label{eq:gvpp}
\eeq
respectively, where the suffix ${\chi}$ denotes the result of simultaneous 
continuum and massless extrapolations.
As shown in~\Fig{gvpp_hist}, the distributions of this quantity exhibit a regular Gaussian shape,
from which we estimate the statistical uncertainty---the numbers in the first parentheses of \Eq{gvpp}.
The numbers in the second parentheses in \Eq{gvpp} denote the systematic errors of the fits with similar caveats 
to those discussed in \Sec{quenched}, 
that we estimated by taking the maximum and minimum values obtained
 from the set of data excluding the coarsest ensemble 
and including or excluding the heaviest measurements. 

The EFT analyses performed in this section is affected by several limitations---in particular 
by the quenched approximation and by the 
 comparatively  large fermion masses---and thus one 
should interpret the results with some caution. 
Yet, it is interesting to compare the EFT results with phenomenological models 
and with available measurements obtained with
 dynamical fermions transforming in the fundamental representation. 
We first compare the EFT results in \Eq{gvpp} with the ones predicted from the KSRF relation, 
$g_{\rm VPP}=m_{\rm V}/\sqrt{2}m_{\rm PS}$.
We  find that the left-hand side is smaller than the right-hand side of this relation
 by about $10\%$ and $23\%$, for the fundamental and antisymmetric representations, respectively. 
These discrepancies are larger than the uncertainties associated with the fits,  and might
indicate that the KSRF relation does not describe the quenched theories  accurately, 
particularly in the case of the antisymmetric representation, 
although this statement is affected by uncontrolled systematic uncertainties due  to the use of the EFT 
with such large values of $g_{\rm VPP}$ and $g_{\rm vpp}$, as well as large fermion masses.
We also find that for the fundamental representation 
the quenched value of $g_{\rm VPP}^{\chi}$ is smaller 
by $29\%$ compared to the dynamical value of 
$g_{\rm VPP}^{\chi}=6.0(4)(2)$ \cite{Bennett:2019jzz},
yielding again a discrepancy that is significantly larger than the fit uncertainties.
It would be interesting to repeat these tests with dynamical fermions in the antisymmetric representation,
and in general to explore more directly the low-mass regimes of all these theories, but these 
are tasks that we leave for future extensive studies.

\section{Conclusions and Outlook}
\label{Sec:conclusions}

Composite Higgs and (partial) top compositeness emerge naturally 
as  the low-energy EFT  description of gauge theories with fermion 
matter content in mixed representations of the gauge group.
Motivated by this framework, we considered the $Sp(4)$ gauge theory
with $N_f$  quenched Wilson-Dirac fermions $Q$ transforming in the fundamental representation of $Sp(4)$,
as well as $n_f$ quenched fermions $\Psi$ in the 2-index antisymmetric representation.
While the quenched theory is not expected to reproduce the full dynamics, 
it provides a useful comparison case for future full dynamical calculations. 
We generated lattice ensembles consisting of gauge configurations
by means of  the HB algorithm, 
modified appropriately the HiRep code~\cite{DelDebbio:2008zf},  considered 
meson operators ${\cal O}_M$ bilinear in these fermions  (see Table~\ref{tab:mesons}
for explicit definitions of the operators),
and  measured
two-point Euclidean correlation functions of such operators
on discrete lattices (and in the quenched approximation).

We hence extracted decay constants $f_M$ and masses $m_M$
 of the flavoured mesons sourced by the operators ${\cal O}_M$,
 with $M=$ PS, V, AV, S, T, and AT (and $M=$ ps, v, av, s, t, and at), defined in \Tab{mesons}.
We renormalised the decay constants, expressed all dimensional quantities in terms of the 
gradient-flow scale $w_0$, and---having restricted attention to 
ensembles for which finite-volume effects can be ignored---applied tree-level
W$\chi$PT to extrapolate toward the continuum and massless limits
 the results for mesons 
constituted  of both fermion species.
We also performed a first global fit of the continuum results that
makes use of the EFT describing the
lightest spin-1 states (besides the pseudoscalars). It is constructed by extending with the
language of hidden local symmetry the chiral-Lagrangian description of the pNGBs spanning the
$SU(2N_f)\times SU(2n_f)/Sp(2N_f)\times SO(2n_f)$ coset.

Our main results for the physical observables in the continuum limit 
are listed in the tables and plots in 
Secs.~\ref{Sec:quenched} and~\ref{Sec:antisymmetric} 
and graphically  illustrated  in Sec.~\ref{Sec:comparison} (see in particular Figs.~\ref{fig:f2_Q_F_vs_AS}
and~\ref{fig:m2_Q_F_vs_AS}).
They can be summarised as follows.
In the quenched approximation,
after extrapolation to the massless limit,
 all dimensional quantities extracted
 from two-point correlation functions involving 
operators constituted of $\Psi$ fermions are larger than the corresponding
observables involving $Q$ fermions. 
The two extremes are  $\hat{m}_{\rm s}^2/\hat{m}_{\rm S}^2 =1.18\pm 0.13$ and
$\hat{f}_{\rm av}^2/\hat{f}_{\rm AV}^2=2.7\pm 1.1$, respectively, with all other ratios
between observables in the two sectors falling between these two values. (Of particular interest
for model building are the ratios
$\hat{m}_{\rm v}^2/\hat{m}_{\rm V}^2 =1.46\pm 0.08$
and $\hat{f}_{\rm ps}^2/\hat{f}_{\rm PS}^2 =1.81\pm 0.04$.) 
The error bars comprise both
 statistical as well as systematic errors, the latter arising from the continuum 
 and massless extrapolations as discussed in details in \Sec{mesons}. 
Furthermore, we found statistically significant violations of the KSRF relations  by the 
mesons made of antisymmetric fermions, at least in the quenched approximation.
(The extraction of the $g_{\rm VPP}$ and $g_{\rm vpp}$ couplings from the global fit
 of two-point function data collected
 with large fermion mass to the EFT is affected by unknown systematic effects, and hence this should be taken as a preliminary result.)

Despite the physical limitations of the studied quenched theory, 
this paper opens the way toward addressing a number of  interesting questions 
in future related work, 
a first class of which is related to the comparison of the quenched calculations to the full dynamical ones,
in particular for the case of fermions in the antisymmetric representation.
While it was observed elsewhere~\cite{Bennett:2019jzz} that the quenched approximation captures 
remarkably well
the dynamics of fundamental fermions (at least for the range of masses hitherto explored),
there is no clear reason for this to happen also in the antisymmetric case, for which large-$N$ arguments
are less constraining.
In order to address this point, one would require to study the dynamical 
simulations with $\Psi$ fermions, in the phenomenologically relevant
low-mass ranges of the dynamical calculations,
and also to
generalise our approach to $Sp(2N)$ gauge theories.
The reader may be aware of the possibility that, with higher-dimensional representations and large
 numbers of fermion degrees of freedom, some of the $Sp(2N)$ theories we are interested in
 might be close to the edge of the conformal window
 and behave very differently. (For perturbative studies within the $Sp(2N)$ class, 
 see for instance  Refs.~\cite{Sannino:2009aw,Ryttov:2017dhd,Kim:2020yvr}, and references therein.)

The extensive line of research outlined in the previous paragraph
 complements the  development of 
our programme of studies in the context of top compositeness, 
that as outlined in Ref.~\cite{Bennett:2017kga} requires one 
to consider the dynamical theory 
in the presence of mixed representations. 
This is a novel area of exploration for lattice gauge theories,
for which the literature is somewhat limited (see for instance 
Refs.~\cite{DeGrand:2016pgq,Ayyar:2017qdf,Ayyar:2018zuk,Ayyar:2018glg,Cossu:2019hse}).
New fermion bound states, sometimes referred to as 
chimera baryons, can be sourced by operators that involve gauge-invariant combinations of
fermions in mixed representations.
 (The anomalous dimensions of chimera baryons are discussed for example
  in~\cite{DeGrand:2015yna,Cacciapaglia:2019dsq,BuarqueFranzosi:2019eee}.) 
The study of these states is necessary in the context of top compositeness, 
as they are interpreted as top partners. 
 
A third group of future research projects can be envisioned to explore the role
  of higher-dimensional operators, for which the material in the Appendixes of this paper 
  is technically useful. These operators play a role in determining the physics of vacuum (mis)alignment
  and of electroweak symmetry breaking, as their matrix elements enter 
  the calculation of the  potential in the low-energy EFT description.
These studies would  provide an additional  link to phenomenological investigations of
composite Higgs models, bringing lattice calculations in close contact with 
model-building considerations and searches for new physics at the Large Hadron Collider (LHC).

Finally, it would be interesting to investigate the finite  temperature behaviour of these theories.
As discussed in Appendix~\ref{Sec:AppendixD1}, 
it is important to characterise symmetry restoration and symmetry enhancement that appear at 
high temperature, generalising what has been studied about QCD 
to the case of real and pseudoreal representations, for which the group structure of the 
global symmetries and their breaking  is expected to be different.

\vspace{0.0cm}
\begin{acknowledgments}
\end{acknowledgments}

We acknowledge useful discussions with Axel Maas, Roman Zwicky and Daniel Nogradi. 

The work of E.B., M.M., and J.R. has been funded in part by the Supercomputing Wales project, 
which is part funded by the European Regional Development Fund (ERDF) via Welsh Government. 

The work of D.K.H. was supported by Basic Science Research Program 
through the National Research Foundation of Korea (NRF) funded by 
the Ministry of Education (NRF-2017R1D1A1B06033701).  

The work of J.-W.L. is supported in part by the National Research Foundation of Korea grant funded 
by the Korea government(MSIT) (NRF-2018R1C1B3001379) and 
in part by Korea Research Fellowship programme funded 
by the Ministry of Science, ICT and Future Planning through the National 
Research Foundation of Korea (2016H1D3A1909283).

The work of C.-J.D.L. is supported by the  Taiwanese MoST Grant No. 105-2628-M-009-003-MY4. 

The work of B.L. and M.P. has been supported in part by the STFC 
Consolidated Grants ST/L000369/1 and ST/P00055X/1. B.L. and M.P.
received funding from the European Research Council (ERC) under the
European Union's Horizon 2020 research and innovation programme under
Grant Agreement No 813942.
The work of B.L. is further supported in part by the Royal Society Wolfson Research Merit Award N0. WM170010.

J.R. acknowledges support from Academy of Finland Grant 320123

D.V. acknowledges support from the INFN HPC-HTC project.

Numerical simulations have been performed on the Swansea SUNBIRD 
system, on the local HPC
clusters in Pusan National University (PNU) and in National Chiao-Tung University (NCTU),
 and on the Cambridge Service for Data Driven Discovery (CSD3). The Swansea SUNBIRD 
system is part of the Supercomputing Wales project, which is part funded by the European Regional
Development Fund (ERDF) via Welsh Government. CSD3 is operated in part by
 the University of Cambridge Research Computing on
behalf of the STFC DiRAC HPC Facility (www.dirac.ac.uk). 
The DiRAC component of CSD3 was funded by BEIS capital funding via
STFC capital Grants No. ST/P002307/1 and No. ST/R002452/1 and 
STFC operations Grant No. ST/R00689X/1. DiRAC is part of the National e-Infrastructure.

\vspace{1.0cm}

\appendix
\section{Spinors}
\label{Sec:AppendixA}

We summarise in this Appendix  our conventions in the treatment of  spinors,
which are useful, for example, in switching between
 the two-component and the four-component notation (see also Ref.~\cite{Lee:2017uvl}).
The former is best suited to highlight the symmetries of the system, while
the latter is the formalism adopted as a starting point for the lattice numerical treatment.
We highlight some important 
symmetry aspects that offer insight  in the theories studied in this paper.

For two-component spinors, we
use  the Pauli matrices, denoted as $\tau^i$, with $i=1,2,3$, and
\beqs
\tau^1&=&\left(\begin{array}{cc}
0 & 1 \cr
1 & 0 \end{array}\right)\,,\,
\tau^2\,=\,\left(\begin{array}{cc}
0 & -i \cr
i & 0 \end{array}\right)\,,\,
\tau^3\,=\,\left(\begin{array}{cc}
1 & 0 \cr
0 & -1 \end{array}\right)\,.
\eeqs
Given a two-component spinor $u$, with no internal quantum numbers, we define the ${\cal C}$-conjugate 
$u_C\equiv i\tau^2 u^{\ast}\equiv - \tilde{C} u^{\ast}$.
Furthermore, we introduce the notation 
$\sigma^{\mu}\equiv (\mathbb{1}_2,-\tau^i)$ and $\bar{\sigma}^{\mu}\equiv (\mathbb{1}_2,\tau^i)$.

We adopt conventions in which the space-time Minkowski metric is 
\beqs
\eta_{\mu\nu}\equiv\,\left(\begin{array}{cccc}1&&&\cr
&-1&&\cr
&&-1&\cr
&&&-1\cr
\end{array}
\right)\,=\,\eta^{\mu\nu}\,.
\eeqs
 The Dirac algebra is defined by the anticommutation relation 
 \beqs
 \{\gamma^{\mu},\gamma^{\nu}\}=2\eta^{\mu\nu}\,,
 \eeqs 
with the $4\times 4$ matrix $\gamma^0$ Hermitian, while the three $\gamma^i$ are 
anti-Hermitian, so that $\gamma^0\gamma^{\mu}\gamma^0=\gamma^{\mu\,\dagger}$.
Chirality is defined by the eigenvalues of the matrix 
$\gamma_5\equiv i \gamma^0\gamma^1\gamma^2\gamma^3$, 
that satisfies the relation $\{\gamma^{\mu}\,,\,\gamma^5\}=0$.

The charge-conjugation matrix $C=i\gamma^2\gamma^0$ obeys  the defining relations
$C \gamma^{\mu} C^{-1} = - \gamma^{\mu\,\mathrm{T}}$ and $C^2=-\mathbb{1}_4=-CC^{\dagger}$. 
The chiral representation of the $\gamma^{\mu}$ matrices is
\beqs
\gamma^0&=&\left(\begin{array}{cc}
0 & \mathbb{1}_2 \cr
\mathbb{1}_2 & 0
\end{array}\right)\,,~~
\gamma^i\,=\,\left(\begin{array}{cc}
0 & -\tau^i \cr
\tau^i & 0
\end{array}\right)\,,~~
\gamma^5\,=\,\left(\begin{array}{cc}
\mathbb{1}_2 & 0 \cr
0& -\mathbb{1}_2
\end{array}\right)\,,~~
C\,=\,\left(\begin{array}{cc}
 -i\tau^2 & 0 \cr
0 & i \tau^2 
\end{array}\right)\,,
\eeqs
which implies the useful relations
\beqs
\gamma^0\gamma^{\mu}&=&
\left(\begin{array}{cc} \bar{\sigma}^{\mu} & 0 \cr 0 & \sigma^{\mu}\end{array}\right)\,,
~~~
C \gamma^0\gamma^{\mu} C^{-1}\,=\,
\left(\begin{array}{cc} {\sigma}^{\mu} & 0 \cr 0 & \bar{\sigma}^{\mu}\end{array}\right)\,.
\eeqs

We also define the matrices
\beqs
\sigma^{\mu\nu}&\equiv&\frac{i}{2}\left[\gamma^{\mu}\frac{}{},\,\gamma^{\nu}\right]\,,
\eeqs
which obey the relations $\left[\gamma_5\,,\,\sigma^{\mu\nu}\right]=0$, 
$\gamma^0\sigma^{\mu\nu}\gamma^0=(\sigma^{\mu\nu})^{\dagger}$
and $\gamma_5\sigma^{\mu\nu}=\frac{i}{2}\epsilon^{\mu\nu\rho\sigma}\sigma_{\rho\sigma}$,
where $\epsilon^{\mu\nu\rho\sigma}$ is the completely antisymmetric Levi-Civita symbol.
In the chiral representation for the $\gamma^{\mu}$ matrices, 
the six $\sigma^{\mu\nu}$ matrices are block diagonal and can be written as
\beqs
\sigma^{\mu\nu}
&=&\,\left(\begin{array}{cc}
\sigma^{\mu\nu}_{LL} & 0 \cr
0 & \sigma^{\mu\nu}_{RR}\end{array}\right)\,,
~~~~
\sigma^{\mu\nu}\gamma_5
\,=\,\,\left(\begin{array}{cc}
\sigma^{\mu\nu}_{LL} & 0 \cr
0 & -\sigma^{\mu\nu}_{RR}\end{array}\right)\,=\gamma_5\sigma^{\mu\nu}\,.
\eeqs
By isolating the spatial indices $i$, one finds that
\beqs
\sigma^{0i}&=&i\,\left(\begin{array}{cc}
\tau^i & 0 \cr
0 & -\tau^i\end{array}\right)\,,
~~~~
\sigma^{ij}\,=\,\epsilon^{ijk}\,\left(\begin{array}{cc}
\tau^k & 0 \cr
0 & \tau^k\end{array}\right)\,.
\eeqs

We introduce the notation $\bar{\lambda}\equiv \lambda^{\dagger}\gamma^0$.
A single Majorana spinor $\lambda$ obeys the relation $\lambda=\pm\lambda_C\equiv\pm C\bar{\lambda}^{\mathrm{T}} \equiv
 \pm C\gamma^0 \lambda^{\ast}=\pm i\gamma^2\lambda^{\ast}$. We conventionally
 resolve the $\pm$ ambiguity by the choice of the $+$ sign. Starting  from a two-component spinor $u$,  a four-component Majorana spinor is
\beqs
\lambda&=&\left(\begin{array}{c} u\cr i\tau^2 u^{\ast}\equiv-\tilde{C}u^{\ast}\end{array}\right)\,,
\eeqs
so that  $\lambda=\lambda_C$. 
The left-handed (LH) chiral projector is $P_L=\frac{1}{2}\left( \mathbb{1}_4+\gamma_5\right)$, so that
a four-component LH chiral spinor $\lambda_L =P_L\lambda$ satisfies $P_L\lambda_L=\lambda_L$.
Analogous definitions apply to the right-handed (RH) projector $P_R$ and spinor $\lambda_R$.
The decomposition in LH and RH four-components 
chiral Weyl spinors is given by
\beqs
\lambda_L&=&\left(\begin{array}{c} u\cr 0\end{array}\right)\,,~~~~~
\lambda_R\,=\,\left(\begin{array}{c} 0 \cr i\tau^2 u^{\ast}\equiv-\tilde{C}u^{\ast}\end{array}\right)\,,
\eeqs
and yields the relations 
 $\lambda_L=C\overline{\lambda_R}^{\,\,\mathrm{T}}$ and $\overline{\lambda_L}=\lambda_R^{\,\,\mathrm{T}} C=-\lambda_R^{\,\,\mathrm{T}} C^{-1}$.
Clearly, $u$, $\lambda$, $\lambda_L$ and $\lambda_R$ are different ways to encode the same information.

Consider two distinct, two-component spinors $u$ and $d$, with 
no additional internal degrees of freedom (aside from 
the spinor index $\alpha=1,2$).
When taken together, they naturally define the fundamental 
representation of a global $U(2)$ symmetry.
Their components are described by Grassmann variables,  satisfying the two nontrivial 
relations\footnote{The first one is the defining relation of 
the anticommuting Grassmann variable, while the second 
is required for consistency of the definition of absolute value as a real 
number $\xi^{\ast}\xi=(\xi^{\ast}\xi)^{\ast}\neq 0$.}
\beqs
u^{\alpha\,\ast}d^{\beta}\,=\,-d^{\beta}u^{\alpha\,\ast}\,,
~~~~\Big(u^{\alpha\,\ast}d^{\beta}\Big)^{\ast}\,=\,d^{\beta\,\ast}u^{\alpha}\,,
\eeqs
and analogous for all other combinations. 

A Dirac four-component spinor is obtained  by 
joining the LH projection of the Majorana spinor built starting from $u$ and the RH projection of the 
Majorana spinor corresponding to $d$, so that $Q=U_L+ D_R$ with
\beqs
U_L&\equiv&\left(\begin{array}{c} u \cr 0 \end{array}\right)\,,~~~~
D_R\,\equiv\,\left(\begin{array}{c} 0 \cr -\tilde{C}d^{\ast}
 \end{array}\right)\,,~~~~
 D_L\,\equiv\,\left(\begin{array}{c} d \cr 0 \end{array}\right)\,,~~~~
U_R\,\equiv\,\left(\begin{array}{c} 0 \cr -\tilde{C}u^{\ast}
 \end{array}\right)\,,
 \label{Eq:2to4}
\eeqs
and $Q_C\equiv C\overline{Q}^{\mathrm{T}} = C \gamma^0 Q^{\ast}=D_L\,+\,U_R$, while $\overline{Q_C}=Q^{\mathrm{T}} C$.

By inspection, one finds that the following  relations hold true:
\beqs
\overline{Q}P_L Q&=&\overline{D_R}U_L\,=\,d^{\mathrm{T}}\tilde{C}u
\,,~~~~
(\overline{Q}P_L Q)^{\ast}\,=\,\overline{Q}P_R Q\,=\,-u^{\dagger}\tilde{C}d^{\ast}\,,\nonumber
\\
\overline{Q_C}P_L Q_C&=&\overline{U_R}D_L\,=\,u^{\mathrm{T}}\tilde{C}d\,,~~~~\nonumber
(\overline{Q_C}P_L Q_C)^{\ast}\,=\,\overline{Q_C}P_R Q_C\,=\,-d^{\dagger}\tilde{C}u^{\ast}
\,,
\\ \label{Eq:scalarrelations}
\overline{Q}P_L Q_C&=&\overline{D_R}D_L\,=\,d^{\mathrm{T}}\tilde{C}d\,,~~~~
(\overline{Q}P_L Q_C)^{\ast}\,=\,\overline{Q_C}P_R Q\,=\,-d^{\dagger}\tilde{C}d^{\ast}\,,
\\
\overline{Q_C}P_L Q&=&\overline{U_R}U_L\,=\,u^{\mathrm{T}}\tilde{C}u\,,~~~~\nonumber
(\overline{Q_C}P_L Q)^{\ast}\,=\,\overline{Q}P_R Q_C\,=\,-u^{\dagger}\tilde{C}u^{\ast}\,,
\label{Eq:sc}
\eeqs
and by using the $\gamma^{\mu}$ matrices, one also finds the 
relations 
\beqs
\overline{Q}\gamma^{\mu}P_L Q &=& \overline{U_L} \gamma^{\mu} U_L\,=\, u^{\dagger} \bar{\sigma}^{\mu}u\,,~~~~
(\overline{Q}\gamma^{\mu}P_L Q)^{\mathrm{T}} \,=\,- \overline{Q_C}\gamma^{\mu}P_RQ_C\,=\,- u^{\mathrm{T}} \bar{\sigma}^{\mu\,\ast}u^{\ast}\,,~~~~\nonumber
\\
\overline{Q_C}\gamma^{\mu}P_LQ_C&=&\overline{D_L} \gamma^{\mu} D_L\,=\, d^{\dagger} \bar{\sigma}^{\mu}d\,,~~~~\nonumber
(\overline{Q_C}\gamma^{\mu}P_LQ_C)^{\mathrm{T}}\,=\,-\overline{Q}\gamma^{\mu}P_R Q \,=\,-d^{\mathrm{T}}\bar{\sigma}^{\mu\,\ast}d^{\ast}\,,
\\
\overline{Q}\gamma^{\mu}P_L Q_C &=& \overline{U_L} \gamma^{\mu} D_L\,=\, u^{\dagger} \bar{\sigma}^{\mu}d\,,~~~~
(\overline{Q}\gamma^{\mu}P_L Q_C)^{\mathrm{T}} \,=\,- \overline{Q}\gamma^{\mu}P_RQ_C\,=\,-d^{\mathrm{T}} \bar{\sigma}^{\mu\,\ast}u^{\ast}\,,~~~~~~~~
\label{Eq:vec}
\\
\overline{Q_C}\gamma^{\mu}P_L Q &=& \overline{D_L} \gamma^{\mu} U_L\,=\, d^{\dagger} \bar{\sigma}^{\mu}u\,,\nonumber~~~~
(\overline{Q_C}\gamma^{\mu}P_L Q)^{\mathrm{T}} \,=\, -\overline{Q_C}\gamma^{\mu}P_RQ\,=\,-u^{\mathrm{T}} \bar{\sigma}^{\mu\,\ast}d^{\ast}\,,
\eeqs
as well as
\beqs
(\overline{Q}\gamma^{\mu}P_L Q)^{\ast} &=& \overline{Q}\gamma^{\mu}P_LQ\,,~~~~\nonumber
\\
(\overline{Q_C}\gamma^{\mu}P_LQ_C)^{\ast} &=&\overline{Q_C}\gamma^{\mu}P_L Q_C\,,\nonumber
\\
(\overline{Q}\gamma^{\mu}P_L Q_C)^{\ast}  &=& \overline{Q_C}\gamma^{\mu}P_LQ\,,~~~~~~~~
\label{Eq:vec}
\\
(\overline{Q_C}\gamma^{\mu}P_L Q)^{\ast}  &=& \overline{Q}\gamma^{\mu}P_LQ_C\,.\nonumber
\eeqs

By definition, the transpose of a $\mathbb{C}$-number is trivial, 
and hence $\xi A \chi=(\xi A \chi)^{\mathrm{T}}\equiv -\chi^{\mathrm{T}} A^{\mathrm{T}} \xi^{\mathrm{T}}$, 
for any $\xi,\chi$ spinor written in terms of Grassmann variables
 and $A$ any matrix of $\mathbb{C}$-numbers.	
This  implies  the relation
\beqs
\overline{Q}P_L Q\,-\,\overline{Q_C}P_L Q_C&=&d^{\mathrm{T}}\tilde{C}u\,-\,u^{\mathrm{T}}\tilde{C}d\,=\,0\,,
\label{Eq:vanishing}
\eeqs 
which will be useful later.
Some algebra shows that the following identity between real numbers holds:
\beqs
\frac{1}{2}\left(\frac{}{}i\overline{Q}\gamma^{\mu}\partial_{\mu}Q\,-\,i\overline{\partial_{\mu}Q}\gamma^{\mu}Q\frac{}{}\right)&=&
\frac{1}{2}\sum_{j=1}^2\left(\frac{}{} iq^{j\,\dagger}\bar{\sigma}^{\mu}\partial_{\mu} q^{j}\,
-\,i\partial_{\mu}q^{j\,\dagger}\bar{\sigma}^{\mu}q^{j}\frac{}{}\right)\,,
\label{Eq:kinetic}
\eeqs
where $q^j=(u\,,\,d)$, 
and where the $U(2)=U(1)\times SU(2)$ global 
symmetry is now made manifest. 
This is adopted as the kinetic term of the Dirac spinor $Q$.

The Lagrangian density for the Dirac spinor $Q$ admits also a mass term.
By virtue
of the relations $\tilde{C}^{\dagger}=-\tilde{C}=\tilde{C}^{\mathrm{T}}$, 
and by the Grassmann nature of the spinors, it can be written
in terms of the symmetric matrix $\omega\equiv\tau^1$:
\beqs\nonumber
-{\cal M} \overline{Q} Q &=& -{\cal M} \left(\frac{}{}\overline{U_L} D_R \,+\,\overline{D_R} U_L\frac{}{}\right)\\
\nonumber
&=&-{\cal M}\left(\frac{}{}-u^{\dagger}\tilde{C}d^{\ast}\,+\,d^{\mathrm{T}}\tilde{C}u\frac{}{}\right)\\
\nonumber
&=&-\frac{1}{2}{\cal M}\left(\frac{}{}-u^{\dagger}\tilde{C}d^{\ast}\,-\,d^{\dagger}\tilde{C} u^{\ast}\,+\,u^{\mathrm{T}}\tilde{C}d\,+\,d^{\mathrm{T}}\tilde{C}u\frac{}{}\right)\\
&=&-\frac{1}{2}{\cal M}\sum_{jk}\omega_{jk}\left(\frac{}{}q^{j\,\mathrm{T}}\tilde{C}q^k\,-\,q^{j\,\dagger}\tilde{C}q^{k\,\ast}\frac{}{}\right)\,.
\label{Eq:mass}
\eeqs
This term breaks  the symmetry  
to the subgroup $O(2)\in U(2)$.\footnote{If the spinors have additional,
 internal degrees of freedom, their anticommuting nature, which ultimately
descends from  Fermi-Dirac statistics, might enforce to antisymmetrise over them, and can lead 
to the replacement of the symmetric $\omega$ with an antisymmetric $\Omega$. 
Such is indeed the case if $Q$
transforms in the fundamental of $Sp(2N)$, for example. Alternatively, 
if one has to antisymmetrise in two gauge indices, as in the case discussed in
Ref.~\cite{Athenodorou:2014eua} 
and also in the case relevant to the $\Psi$ spinors on the antisymmetric 
2-index representation, symmetry breaking 
is, once more, controlled by the symmetric matrix $\omega$.}

The real Lagrangian density of a single Dirac fermion is then
\beqs
{\cal L}&=&\frac{1}{2}
\left(\frac{}{}i\overline{Q}\gamma^{\mu}\partial_{\mu}Q\,-\,i\overline{\partial_{\mu}Q}\gamma^{\mu}Q\frac{}{}\right)\,-\,{\cal M}\,\overline{Q}Q
\,=\,\overline{Q}\Big(i\gamma^{\mu}\partial_{\mu} -{\cal M}\Big) Q 
\,+\,\partial_{\mu}\Big(\cdots\Big)
\\
\label{Eq:Dirac}
&=&
\frac{1}{2}\sum_j\Big(iq^{j\,\dagger}\bar{\sigma}^{\mu}\partial_{\mu} q^{j}\,
-\,i\partial_{\mu}q^{j\,\dagger}\bar{\sigma}^{\mu}q^{j}\Big)
-\frac{1}{2}{\cal M}\sum_{jk}\omega_{jk}\Big(q^{j\,\mathrm{T}}\tilde{C}q^k\,-\,q^{j\,\dagger}\tilde{C}q^{k\,\ast}\Big)\,,
\label{Eq:LDirac}
\eeqs
the first line of which (by ignoring the surface term) yields the Dirac equation:
\beqs
\left(\frac{}{}i\gamma^{\mu}\partial_{\mu} - {\cal M} \frac{}{}\right) Q&=&0\,.
\eeqs

 Equation~(\ref{Eq:LDirac}) can be generalised by adding a symmetric ${\cal M}_{jk}$ 
Majorana mass matrix via  the replacement ${\cal M}\omega_{jk}\rightarrow {\cal M}_{jk}$
in the two-component formulation:
\beqs
2{\cal L}^{\prime}_M
 &=&
\sum_j\Big(iq^{j\,\dagger}\bar{\sigma}^{\mu}\partial_{\mu} q^{j}
-i\partial_{\mu}q^{j\,\dagger}\bar{\sigma}^{\mu}q^{j}\Big)-\sum_{jk}\Big({\cal M}_{jk}q^{j\,\mathrm{T}}\tilde{C}q^k-{\cal M}^{\ast}_{jk}q^{j\,\dagger}\tilde{C}q^{k\,\ast}\Big)\,.
\label{Eq:LMajorana}
\eeqs
The Majorana mass term can then be written also in terms of four-component Dirac spinors
by applying the projector $P_L$ and  $C$ along the lines of Eq.~(\ref{Eq:scalarrelations}),  as follows
\beqs
{\cal L}^{\prime}&=&-\frac{1}{2}\left(\frac{}{}(P_LQ)^{\mathrm{T}},(P_LQ_C)^{\mathrm{T}}\right)C {\cal M} 
\left(\begin{array}{c}P_LQ\cr P_L Q_C\end{array}\right)\,+\,{\rm h.c.}\,,
\eeqs
where the matrix ${\cal M}$ is defined as
\beqs
{\cal M}&=&\left(\begin{array}{cc}
{\cal M}_{uu} & {\cal M}_{ud}=\frac{1}{2}({\cal M}_s+{\cal M}_a)\cr
{\cal M}_{du}=\frac{1}{2}({\cal M}_s-{\cal M}_a) & {\cal M}_{dd}
\end{array}\right)\,.
\eeqs
If there are no other internal degrees of freedom, ${\cal M}$ is symmetric, with ${\cal M}_{ud}={\cal M}_{du}$. 
In the language of $U(2)$, the product of two doublets  naturally 
decomposes as  $3 \oplus 1$ of $U(2)$:
\beqs
3&\sim&\left(\begin{array}{c}
\overline{Q_C}P_LQ\,=\,u^{\mathrm{T}}\tilde{C}u\cr
\frac{1}{2}\left(\overline{Q}P_LQ+\overline{Q_C}P_LQ_C\right)\,=\,\frac{1}{2}\left(d^{\mathrm{T}}\tilde{C}u+u^{\mathrm{T}}\tilde{C}d\right)\cr
\overline{Q}P_LQ_C\,=\,d^{\mathrm{T}}\tilde{C}d
\end{array}\right)\,,~~~~\\
1&\sim &
\frac{1}{2}\left(\overline{Q}P_LQ-\overline{Q_C}P_LQ_C\right)\,=\,
\frac{1}{2}\left(d^{\mathrm{T}}\tilde{C}u-u^{\mathrm{T}}\tilde{C}d\right)\,.
\eeqs
The latter  vanishes in the absence of additional degrees of freedom, due to Eq.~(\ref{Eq:vanishing}).

\section{A note about massive vectors}
\label{Sec:AppendixB}

A massive vector of mass $m$ in $D=4$ space-time dimensions 
can be described by two equivalent quantum theories, 
with different field content and  Lagrangian densities
(see for instance the detailed discussions 
in Refs.~\cite{Ecker:1989yg,Bijnens:1995ii,Bruns:2004tj,Elander:2018aub} and references therein).
\begin{itemize}
\item A vector field $A_{\mu}$ couples to a scalar field $\pi$, with Lagrangian density
\beqs
{\cal L}_0&=&-\frac{1}{4}F_{\mu\nu}F^{\mu\nu}\,-\,\frac{1}{2}\left(\partial_{\mu}\pi \frac{}{}+ m A_{\mu}\right)
\left(\partial^{\mu}\pi \frac{}{}+ m A^{\mu}\right)\,,
\eeqs
where $F_{\mu\nu}=\partial_{\mu}A_{\nu}-\partial_{\nu}A_{\mu}$.
${\cal L}_0$ is invariant under the gauge transformations
\beqs
\pi\rightarrow \pi+m \alpha\,,~~~~~~~~A_{\mu}\rightarrow A_{\mu}-\partial_{\mu}\alpha\,,
\eeqs
with $\alpha=\alpha(x)$. The gauge choice $\alpha=-\pi/m$ removes $\pi$ from the Lagrangian density,
which then depends only on  a massive vector field.
\item A 2-index antisymmetric form ${\cal B}_{\mu\nu}$ is coupled to a vector ${\cal A}_{\mu}$
(not to be confused with ${A}_{\mu}$), 
and the Lagrangian density is
\beqs
{\cal L}_1&=&-\frac{1}{12} G_{\mu\nu\rho}G^{\mu\nu\rho}-\frac{1}{4}{\cal H}_{\mu\nu}{\cal H}^{\mu\nu}\,,
\eeqs
where ${\cal F}_{\mu\nu}\equiv\partial_{\mu}{\cal A}_{\nu}-\partial_{\nu}{\cal A}_{\mu}$, 
${\cal H}_{\mu\nu}\equiv{\cal F}_{\mu\nu}+m\,{\cal B}_{\mu\nu}$ and $G_{\mu\nu\rho}\equiv
\partial_{\mu}{\cal B}_{\nu\rho}+\partial_{\rho}{\cal B}_{\mu\nu}+\partial_{\nu}{\cal B}_{\rho\mu}$. 
The Lagrangian ${\cal L}_1$ is invariant under the gauge transformation
\beqs
{\cal A}_{\mu}\rightarrow {\cal A}_{\mu}+m \alpha_{\mu}\,,
~~~~~~~~{\cal B}_{\mu\nu}\rightarrow {\cal B}_{\mu\nu} -
 \partial_{\mu} \alpha_{\nu}+\partial_{\nu} \alpha_{\mu}\,,
\eeqs
with the vector $\alpha_{\mu}=\alpha_{\mu}(x)$. The gauge choice 
$\alpha_{\mu}=-{\cal A}_{\mu}/m$ removes ${\cal A}_{\mu}$ from the Lagrangian density,
which then depends only on a massive 2-form field.
\end{itemize}
The Lagrangian ${\cal L}_1$ can also be rewritten, by defining
 ${\cal K}_{\mu\nu}\equiv \frac{1}{2m}\epsilon_{\mu\nu\rho\sigma}{\cal H}^{\rho\sigma}$,
in the form
\beqs
{\cal L}_{1}&=&\frac{1}{2}\partial^{\alpha}{\cal K}_{\mu\alpha}\partial^{\beta}{\cal K}^{\mu}_{\,\,\,\,\beta}
+\frac{m^2}{4} {\cal K}_{\mu\nu} {\cal K}^{\mu\nu}\,.
\eeqs
Gauge invariance is not manifest in this form. The Lagrangians 
${\cal L}_0$ and ${\cal L}_1$
are equivalent at the level of the path integrals
 they define~\cite{Ecker:1989yg,Bijnens:1995ii,Bruns:2004tj,Elander:2018aub}.
Hence, the use of antisymmetric massive 2-index 
tensors provides an alternative, equivalent descriptions of massive vectors.

In physical terms, there is no difference between these two (or rather, three) 
formulations. Important differences are introduced by the coupling to matter fields and sources.
For example, one can couple fermions to $A_{\mu}$ via the new term
\beqs
{\cal L}_{A}&=&i g \bar{Q}\gamma^{\mu} A_{\mu} P_L Q\,,
\eeqs
with $Q$ a Dirac fermion and $g$ the coupling. 
For the antisymmetric tensor, one may write
\beqs
{\cal L}_{\cal B}&=&g \bar{Q}\sigma^{\mu\nu} {\cal B}_{\mu\nu} P_L Q\,.
\eeqs
While ${\cal L}_A$ couples the spin-1 field to the LH component only of $Q$,
in ${\cal L}_{\cal B}$ the LH and RH projections are coupled to one another, so that while 
${\cal L}_{0}$ and ${\cal L}_{1}$ in isolation define the same theory, the addition of ${\cal L}_{\cal A}$
or ${\cal L}_{\cal B}$ leaves different global symmetries and different coupled theories.

\section{About Lie groups, algebras and SM embedding}
\label{Sec:AppendixC}

Here we summarise some group theory notions relevant for
models of composite Higgs and top 
quark compositeness based on the 
${SU(4)/Sp(4) \otimes SU(6)/SO(6)}$
coset~\cite{Ferretti:2013kya,Barnard:2013zea}. We do not repeat 
unnecessary details---in particular, our special  choice of  $SU(4)$ generators can be found 
elsewhere~\cite{Lee:2017uvl}---but we explicitly show the embedding of the SM 
gauge group (and fields, when useful).

The $SU(4)/Sp(4)$ coset governs the Higgs sector of the Standard Model.
Given the form of $\Omega$ in Eq.~(\ref{Eq:symplectic}), the 
unbroken subgroup $SO(4)\sim SU(2)_L\times SU(2)_R$ 
is the subset of the unbroken global  $Sp(4)\subset SU(4)$ that 
is generated by the following elements of the associated algebra:
\beqs
\label{Eq:SU2L}
T^{1}_L&=&\frac{1}{2}\left(\begin{array}{cccc}
0 & 0 & 1 & 0\cr
0 & 0 & 0 & 0\cr
1 & 0 & 0 & 0\cr
0 & 0 & 0 & 0\cr
\end{array}\right)\,,\,\,
T^{2}_L\,=\,\frac{1}{2}\left(\begin{array}{cccc}
0 & 0 & -i & 0\cr
0 & 0 & 0 & 0\cr
i & 0 & 0 & 0\cr
0 & 0 & 0 & 0\cr
\end{array}\right)\,,\,\,
T^{3}_L\,=\,\frac{1}{2}\left(\begin{array}{cccc}
1 & 0 & 0 & 0\cr
0 & 0  & 0 & 0\cr
0 & 0 & -1 & 0\cr
0 & 0 & 0 & 0\cr
\end{array}\right)\,,\\
T^{1}_R&=&\frac{1}{2}\left(\begin{array}{cccc}
0 & 0 & 0 & 0\cr
0 & 0 & 0 & 1\cr
0 & 0 & 0 & 0\cr
0 & 1 & 0 & 0\cr
\end{array}\right)\,,\,\,
T^{2}_R\,=\,\frac{1}{2}\left(\begin{array}{cccc}
0 & 0 & 0 & 0\cr
0 & 0 & 0 & -i\cr
0 & 0 & 0 & 0\cr
0 & i & 0 & 0\cr
\end{array}\right)\,,\,\,
T^{3}_R\,=\,\frac{1}{2}\left(\begin{array}{cccc}
0 & 0 & 0 & 0\cr
0 &1  & 0 & 0\cr
0 & 0 & 0 & 0\cr
0 & 0 & 0 & -1\cr
\end{array}\right)\,.
\label{Eq:SU2R}
\eeqs
The $T_L$ generators satisfy the $SU(2)_L$ algebra $\left[ T_L^i\,,\,T_L^j\right]=i\epsilon^{ijk}\,T_L^k$,
and similarly $\left[ T_R^i\,,\,T_R^j\right]=i\epsilon^{ijk}\,T_R^k$, while $\left[T_L^i,T_R^j\right]=0$.
In the vacuum aligned with $\Omega$ in Eq.~(\ref{Eq:symplectic}), this is the natural choice 
of embedding of the $SO(4)$ symmetries of the Higgs potential.
Following the notation in Refs.~\cite{Lee:2017uvl,Bennett:2017kga}, 
 the matrix of the five pNGB fields parametrising the $SU(4)/Sp(4)$ coset  is
\beqs
\pi(x)=\frac{1}{2\sqrt{2}}
\left(
\begin{array}{cccc}
 \pi^3(x) & \pi^1(x)-i \pi^2(x) & 0 & -i \pi^4(x)+\pi^5(x) \\
 \pi^1(x)+i \pi^2(x) & -\pi^3(x) & i \pi^4(x)-\pi^5(x) & 0 \\
 0 & -i \pi^4(x)-\pi^5(x) & \pi^3(x) & \pi^1(x)+i \pi^2(x) \\
 i \pi^4(x)+\pi^5(x) & 0 & \pi^1(x)-i \pi^2(x) & -\pi^3(x)
\end{array}
\right).
\label{Eq:pion}
\eeqs
The real fields $\pi^1$, $\pi^2$, $\pi^4$, and $\pi^5$ combine into the Higgs doublet,
while $\pi^3$ is a SM singlet.

The $SU(6)/SO(6)$ coset is relevant to top compositeness. 
The choice of  $n_f=3$ 
Dirac fermions on the 2-index  antisymmetric representation of $Sp(4)$
matches the number of colours in the  $SU(3)_c$ 
gauge group of the Standard Model.
The natural subgroup $SU(3)_L\times SU(3)_R\subset SU(6)$  is generated by
\beqs
t^B_L&=&\frac{1}{2}\left(
\begin{array}{c|c}
\lambda^B & \mathbb{0}_3 \cr
\hline
\mathbb{0}_3 &  \mathbb{0}_3
\end{array}\right)\,,
~~~~
t^B_R\,=\,\frac{1}{2}\left(
\begin{array}{c|c}
 \mathbb{0}_3 & \mathbb{0}_3 \cr
 \hline
\mathbb{0}_3 & -\lambda^{B\ast}
\end{array}\right)\,,
\eeqs
with $\lambda^B$ the eight Hermitian Gell-Mann matrices, 
normalised according to the relation $\Tr \lambda^A\lambda^B=2\delta^{AB}$ (so that
$\Tr t^A_Lt^B_L=\frac{1}{2}\delta^{AB}$).

By defining $t^B_c\equiv (t^B_L+t^B_R)$, with the choice of $\omega$ in
Eq.~(\ref{Eq:symplectic}), one can verify that $\omega t^B_c+t^{B\,\mathrm{T}}_c \omega=0$, 
that the structure constants $\left[t^A_c\,,\,t^B_c\right]=i f^{ABC} t^C_c$ are those of the $su(3)_c$ algebra, and
that $\Tr t^A_ct^B_c=\delta^{AB}$ is twice the fundamental.
The latter property is due to the fact that we are writing the $SU(3)_c$ generators
as $6\times 6$ matrices acting on two-component spinors.
We hence identify $t^B_c$ as the generators of the $SU(3)_c$ gauge symmetry of the Standard Model.
An additional, independent, unbroken generator of $SU(6)$ is given by
\beqs
X&\equiv&\left(\begin{array}{c|c}
\mathbb{1}_3 &  \mathbb{0}_3\cr
\hline
\mathbb{0}_3 & - \mathbb{1}_3
\end{array}\right)\,,
\label{Eq:X}
\eeqs
which also commutes with the generators of $SU(3)_c$. The generator $Y$ of the 
 hypercharge $U(1)_Y$ gauge symmetry of the Standard Model
 is a linear combination of $X$ and  $T^3_R$
 (see also Ref.~\cite{Cacciapaglia:2019bqz} and references therein).

\subsection{Weakly coupling  the SM gauge group}
\label{Sec:AppendixGW}

In this Appendix, we perform a technical exercise.
We compute the (divergent) contributions to the effective potential due to the gauging of the relevant SM
subgroups of the global $SU(4)\times SU(6)$ symmetry, and discuss their effects on the potential of the pNGBs.
The purpose of this exercise is to show explicitly how by gauging part of the global symmetry one 
breaks it. We also  identify the decomposition of the representations according to the
unbroken subgroup.

We adopt  the external field method 
and borrow the regulated Coleman-Weinberg potential $V_1$ from Ref.~\cite{Coleman:1973jx}, 
computed by assuming that a hard 
momentum cutoff $\Lambda$ is applied to the one-loop integrals.
With our conventions we write 
\beqs
V_1&=&\frac{\Lambda^2}{32\pi^2}{\STr}\,{\cal M}^2\,
+\,\frac{1}{64\pi^2}{\STr}\left[\frac{}{}({\cal M}^2)^2\log\frac{{\cal M}^2}{\Lambda^2}\,+\,c_i\frac{}{}\right]\,,
\eeqs
where in the trace $\STr$ fermions have negative weight 
and where $c_i$ are scheme-dependent coefficients.
The matrix ${\cal M}^2$ is obtained as follows: consider ${\cal L}_i$ 
in Eq.~(\ref{Eq:Lpi}),
 gauge the relevant subgroups, by promoting the derivatives to covariant derivatives, and
 compute the mass matrices of all the fields, as a function of the (background, external) 
scalar fields.

When applied to the $SU(4)/Sp(4)$ part of the theory (and for $M=0$), this procedure 
involves only loops of gauge bosons and yields a 
quadratically divergent 
 contribution to the mass of four of the 
 pNGBs---labeled $\pi^1$, $\pi^2$, $\pi^4$ and $\pi^5$ in Eq.~(\ref{Eq:pion}):
\beqs
\delta_4 m_{\pi}^2&=&\frac{\Lambda^2}{32\pi^2}\left(\frac{9}{2}g_L^2+\frac{3}{2}g_R^{}\right)\,,
\eeqs
where $g_L$ is the coupling associated with the $SU(2)_L$ group 
with generators in Eqs.~(\ref{Eq:SU2L}), while
$g^{}_R$ is the coupling associated with the $U(1)_R$ subgroup
 generated by  $T^3_R$ from Eqs.~(\ref{Eq:SU2R}).
The four masses are exactly degenerate, and the mass of  $\pi^3$ does not receive a correction, 
as it is associated with a generator that commutes with $SU(2)_L\times SU(2)_R$, 
and is hence left unbroken by the weak gauging of the SM gauge group---in practice,
the mass of $\pi^3$ arises for $M\neq 0$ 
due to the explicit  breaking of the global symmetry of the  Lagrangian.

When applied to the 20 pNGBs that describe the $SU(6)/SO(6)$ coset, 
the loops involve the  $SU(3)_c$ gauge bosons,
with the embedding chosen in this Appendix, and strength $g_S$, 
as well as the $U(1)_X$ gauge boson generated by Eq.~(\ref{Eq:X}), with strength $g_X$.
We find that the mass of 12  pNGBs---transforming as $6_{\mathbb{C}}$  of $SU(3)_c$---receive 
the quadratically-divergent contribution
\beqs
\delta_{12} m_{\pi}^2&=&\frac{\Lambda^2}{32\pi^2}\,\left(\frac{}{}{10}{}g_S^2+24g_X^2\right)\,,
\eeqs
and the other eight, which form the adjoint of $SU(3)_c$, receive the mass correction
\beqs
\delta_{8}m_{\pi}^2&=&\frac{\Lambda^2}{32\pi^2}\,\left(\frac{}{}9g_S^2\right)\,.
\eeqs
The complex $6_{\mathbb{C}}$ of $SU(3)_c$  has nontrivial $U(1)_X$ charge,
while the eight real components of the adjoint representation of $SU(3)_c$ have vanishing $U(1)_X$ charge.
All 20 pNGBs receive also a degenerate, explicit contribution
to their mass, which is controlled by  $m$.

 \section{Topological charge history and mesonic spectral observables}
 \label{Sec:AppendixF}

\begin{figure}
  \center
  \includegraphics[width=.80\textwidth]{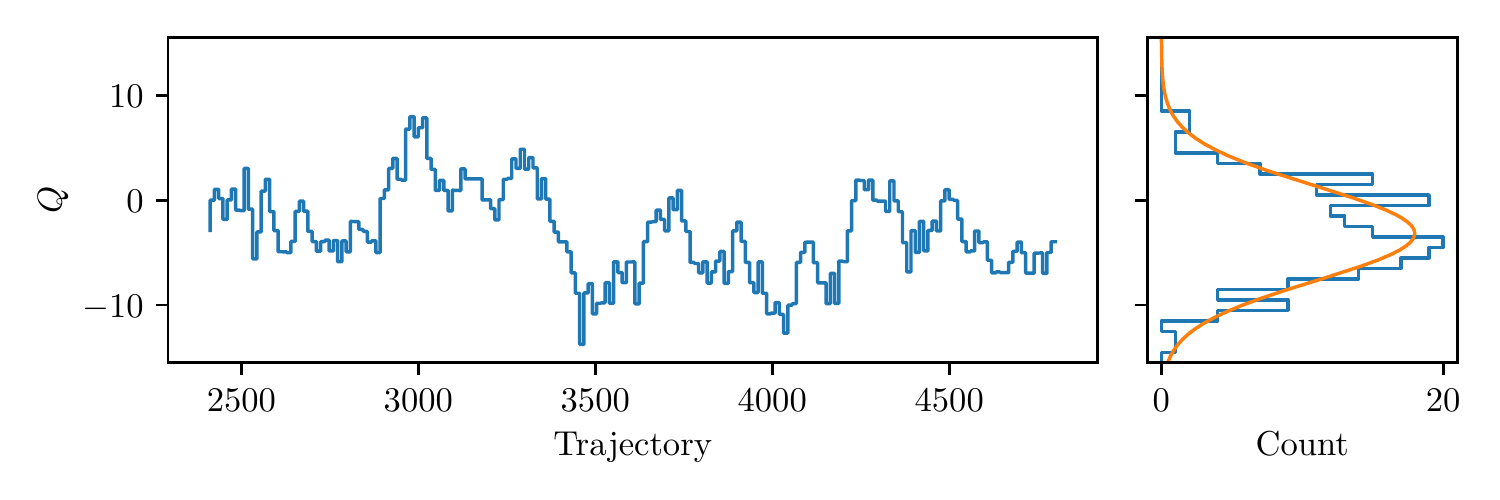}
  \caption{
Topological charge history (left), and histogram (right), for the ensemble
 QB$\overline{5}$. Fitted parameters are
$Q_0=-3.17(31)$, $\sigma=4.53(31)$, and
$\tau_{\exp} =9.2(1)$.}
  \label{fig:topchargealternative}
\end{figure}

This Appendix reports some technical details and 
supplementary numerical studies that are not used in the main body of
the paper.
We saw in Sec.~\ref{Sec:scale} that finer lattice spacings were associate with longer autocorrelation times 
of the topological charge $Q$ (see Fig.~\ref{fig:topcharge}), with the ensemble with the finest lattice spacing 
(which we denoted by QB5) showing a particularly long autocorrelation time and a marginal central value of $Q$.
To verify that this observation does not affect our main results,
we produce a second set of 2400 additional trajectories with the same lattice parameters for QB5, which we call
 QB$\bar{5}$. 
In Figure~\ref{fig:topchargealternative}
we report the topological history and statistical distribution of QB$\bar{5}$.
The behaviour of the topological charge is consistent with ergodicity.

\begin{table}
\begin{center}
\begin{tabular}{|c|c|c|}
\hline\hline
 Measurement&QB5FM2&QB$\bar{5}$FM2\\
\hline
$a \,m_{\rm PS}$ &  0.1850(4) & 0.1848(3) \\
$a \,m_{\rm V}$ &  0.2680(15) & 0.2722(17) \\
$a \,m_{\rm AV}$ &  0.449(7) & 0.437(7) \\
$a \,m_{\rm S}$ &  0.428(6) & 0.437(7) \\
$a \,m_{\rm T}$ &  0.2685(23) & 0.2676(25) \\
$a \,m_{\rm AT}$ &  0.450(9) & 0.448(9) \\
$a \,f_{\rm PS}$ &  0.03740(16) & 0.03765(13) \\
$a \,f_{\rm V}$ &  0.0637(8) & 0.0646(9) \\
$a \,f_{\rm AV}$ &  0.0749(27) & 0.0682(21) \\
\hline\hline
\end{tabular}
\caption{%
\label{tab:massesQB5QB5bar}%
Masses and (renormalised) decay constants, in lattice units, 
extracted from the measurements QB5FM2 and QB$\bar{5}$FM2.
In parentheses are reported the statistical errors.
}
\end{center}
\end{table}

We  measure, in the quenched approximation,
 the masses and (renormalised) decay constants for mesons 
built of  fermions $Q$ transforming in the fundamental representation of $Sp(4)$.
The results are shown in Table~\ref{tab:massesQB5QB5bar}.
We compare the 
measurements  in QB5FM2
(used in the main analysis in the body of the paper) 
with the ones from the ensemble QB$\bar{5}$ with  the same fermion mass
(denoted QB$\bar{5}$FM2).
The two sets of measurements are in agreement, within statistical errors, with all nine  measurements 
within $1\sigma$ and $2\sigma$ of each other. Notice that systematic uncertainties are not included.
In the body of the paper we did not include QB$\bar{5}$ in the analysis.

 \section{Global symmetries and classification  of mesons}
 \label{Sec:AppendixD}

\begin{table}
\begin{center}
\begin{tabular}{|c|c|c|c|}
\hline\hline
{\rm ~~~Particle~~~} & Mass  &  $I^GJ^{P{\cal C}}$   & Hadronic  \cr
 &  (MeV)  &   &  decay mode(s) \cr
\hline
$f_0(500)$ & $500$ & $0^+0^{++}$  & $\pi\pi$\cr
$f_0(980)$  & $980$ & $0^+0^{++}$  & $\pi\pi$\cr
$\eta$ & $548$ & $0^+0^{-+}$ &  $\pi\pi\pi$ ($\Delta {I}$)\cr 
$\eta^{\prime}$ & $960$ & $0^+0^{-+}$& $\eta\pi\pi$\cr 
$\omega$ & $783$ & $0^-1^{--}$&   $\pi\pi\pi$\cr
$\phi$ &  $1019$ & $0^-1^{--}$ & $\overline{K} K $\cr
$f_1(1285)$ & $1282$ &  $0^+1^{++}$   & $4 \pi,\eta\pi\pi$\cr
$f_1(1420)$ & $1426$ &  $0^+1^{++}$  & $\overline{K^{\ast}} K,\overline{K}K\pi$ \cr
$h_1$ &  $1170$ & $0^-1^{+-}$ &$\rho\pi$\cr
$h_1^{\prime}$ &  $1440$\cite{Abele:1997vu}  & $0^-1^{+-}$   & $\overline{K^{\ast}}K$\cr
$\omega(1420)$ & 1420 & $0^-1^{--}$ & $\rho\pi$\cr
$\phi(1680)$ & 1680 & $0^-1^{--}$ & $\overline{K^{\ast}}K$\cr
\hline
$\pi$ & $135$ & $1^-0^{-+}$   & ---\cr 
$a_0$ & $980$ &  $1^-0^{++}$   & $\eta \pi$ \cr
$\rho$ & $775$ & $1^+1^{--}$    & $\pi\pi$ \cr 
$a_1$ & $1230$ & $1^-1^{++}$   & $\rho\pi,\pi\pi\pi$\cr 
$b_1$ & $1230$ &  $1^+1^{+-}$   & $\omega \pi$ \cr
$\rho(1450)$ & $1465$ & $1^+1^{--}$   & $\pi\pi$ \cr 
\hline\hline
\end{tabular}
\end{center}
\caption{Light mesons with $S=C=B=0$
 in QCD~\cite{Tanabashi:2018oca,Abele:1997vu}, their approximate masses, and
quantum numbers. Charge conjugation  ${\cal C}$ refers to
the neutral component in multiplets including  electrically charge particles,
and $G\equiv  {\cal C}(-1)^I$ for the whole multiplet. We show also
 representative hadronic decay modes.
The three-body decay of $\eta$ 
violates isospin, and hence $G$-parity, and yields $\Gamma/M\sim 10^{-5}$.}
\label{Fig:QCD}
\end{table}

In this Appendix, we review some symmetry properties of the mesons 
in more general gauge theories of relevance as candidates for the microscopic origin of CHMs.
We discuss  the cosets that control the long-distance behaviour of the theory
at low temperatures and describe
patterns of symmetry restoration and symmetry enhancement at high temperatures. 
We keep the discussion as general as possible but occasionally exemplify our observations
with the specific case of the $Sp(4)$ gauge theory with
  $N_f=2$ and $n_f=3$.

Given a Lie group $G$ and its  subgroup $H$,
a generic element $\mathfrak{g}$ 
of the associated Lie algebra ${\cal G}$ can be decomposed
as $\mathfrak{g}=\mathfrak{h}+\mathfrak{k}$,
with $\mathfrak{h}\in {\cal H}$ an element of the algebra associated with
 $H$ and $\mathfrak{k}\in {\cal K}$
an element of the complement of ${\cal H}$ in ${\cal G}$. 
The coset space $G/H$ is said to be {\it symmetric}
if,  for all possible choices of  $\mathfrak{h}$ and $\mathfrak{k}$, the following properties are true:
\beqs
\left[\mathfrak{h}\,,\,\mathfrak{h}\right] \,\in\, {\cal H}\,,~~~~
\left[\mathfrak{h}\,,\,\mathfrak{k}\right] \,\in\, {\cal K}\,,~~~~
\left[\mathfrak{k}\,,\,\mathfrak{k}\right] \,\in\, {\cal H}\,.
\eeqs
These properties define  in an unambiguous way
 an unbroken, multiplicative $\mathbb{Z}_2$ symmetry, 
which we can call $G$-parity, which is compatible with the 
Lie algebra, and upon which $\mathfrak{k}$ is assigned $G$-parity
$-$, while $\mathfrak{h}$
is assigned $G$-parity $+$.

Three classes of cosets are  commonly considered in the CHM context
(see for instance Table I in Ref.~\cite{Cacciapaglia:2019bqz}). They all  
emerge  from gauge theories at the microscopic level.
\begin{itemize}
\item  $SU(N_f)_L\times SU(N_f)_R/SU(N_f)_V$ cosets are, for example,
 realised in $SU(N)$ gauge theories with $N_f$
fundamental Dirac fermions.
\item  $SU(2N_f)/Sp(2N_f)$ cosets are, for example, realised in $Sp(2N)$ gauge theories 
with $N_f$ fundamental Dirac fermions.
\item $SU(2N_f)/SO(2N_f)$ cosets are, for example, realised in $Sp(2N)$ gauge theories 
with $N_f$ Dirac fermions transforming in the  2-index antisymmetric representation. 
\end{itemize}
All these cosets are symmetric, and the resulting $G$-parity is a symmetry 
of the theories. It allows selection rules for scattering and decay processes to be established.
We now discuss each of these possibilities in some detail, with emphasis on the 
properties of the mesons associated with the theories they emerge from.

We begin by reviewing the case of the $SU(3)$ gauge theory with $N_f=2$ light flavours.
It describes the light mesons in QCD. The associated coset is $SU(2)_L\times SU(2)_R/SU(2)_V$.
Much of what one learns from this theory is applicable to the other symmetric cosets listed above,
 with modifications that will be discussed later.
 In Table~\ref{Fig:QCD}, we report some information about light QCD mesons
with $S=C=B=0$, taken from the Particle Data Group~\cite{Tanabashi:2018oca}.
 We found Refs.~\cite{Glozman:2007ek,Glozman:2015qva,
Denissenya:2014ywa,Rohrhofer:2017grg,Glozman:2018jkb,Rohrhofer:2019qwq,Glozman:2019fku,Engel:2011aa}, 
and Appendix B in Ref.~\cite{Gratrex:2018gmm} 
 particularly useful for the discussion that follows.

It is conventional to denote the states of QCD
by the quantum numbers $I^GJ^{P{\cal C}}$, where $I$ is the isospin
(the representation of the unbroken $SU(2)_V$)
 and $J$ the spin.
The assignment of $G$-parity for the isotriplets coincides with the traditional $G$-parity: it
 is related to charge-conjugation ${\cal C}$ of the neutral component in an isomultiplet,
and the isospin $I$ of the isomultiplet, by the relation $G\equiv {\cal C}(-1)^I$,
hence providing a link between the internal symmetry described above and a space-time symmetry.
A second subtle link between internal and space-time symmetries
involves the notion of spatial parity $P$;
the $SU(2)_{L}$ and $SU(2)_{R}$ symmetries act, respectively, on the LH and RH 
projections of the spinors, while the unbroken subgroup is the 
(vectorial) symmetric combination of the two.

We start the discussion from the isotriplets $I=1$.
The pions $\pi$  and axial vectors $a_1$ are,
respectively, the lightest spin-0 and spin-1 states 
and are associated with the broken 
generators of the global symmetry, so that they have $G=-$;
the vector mesons $\r$ are associated with the unbroken group $SU(2)_V$, and have $G=+$.
As a result, the $\rho$ decays to two $\pi$'s, 
while the $a_1$ decays to three $\pi$'s (or also one $\pi$ and one $\r$).
 $G$-parity  is a useful practical tool: while  $\r$ and $a_1$ particles
both transform in the adjoint representation of the unbroken $SU(2)_V$ (isospin)---the 
unbroken subgroup and the coset 
have the same dimension---they are distinguished unambiguously 
by the different assignments of $G$, 
and hence they decay in different ways.

The global symmetries naturally extend to $U(2)_L\times U(2)_R/U(2)_V$.
The additional unbroken vectorial $U(1)_B$ is the baryon number, and all 
mesons have vanishing $U(1)_B$ charge.
The broken, anomalous, global, axial $U(1)_A$ symmetry plays an interesting role 
in relation to parity $P$. The axial $U(1)_A$ partners of the pions $\pi$, named $a_0$,
 have the same $G$-parity but opposite $P$.
 If the $U(1)_A$ were exact, $\pi$ and $a_0$ would be degenerate.

In the $J=1$, $I=1$ sector, the role played by the $U(1)_A$ symmetry  is more subtle.
The $\rho$ and $a_1$ mesons are sourced by bilinear operators  V and  AV, 
in which spinor indices  are contracted on
the $\gamma^{\mu}$ and $\gamma^{\mu}\gamma_5$
matrices, respectively. Such operators  involve either two LH or two RH chiral spinors:
the action of $U(1)_A$ leaves them both invariant, as they are two independent 
singlets of $U(1)_A$.
But there is an important complication, as massive vectors
 in four dimensions can equivalently be described by 2-index antisymmetric tensors
 (see Appendix~\ref{Sec:AppendixB} and references therein).
Two additional sources of spin-1 states T and AT are obtained by contracting the spinor indices on
 $\sigma^{\mu\nu}$ and $\sigma^{\mu\nu}\gamma_5$, respectively. 
 The two operators T and AT couple the LH and RH chiral projections of the fermions,
  in a way that is similar to the $J=0$  isotriplet operators PS and S,
that source the $\pi$ and $a_0$ particles.
They form a doublet of the $U(1)_A\sim O(2)$ symmetry.

 Because of symmetry breaking, the operator T (built with $\sigma^{\mu\nu}$)
  has the same quantum numbers $I^GJ^{P{\cal C}}=1^+1^{--}$
 as the source V (built from $\gamma^{\mu}$).
The lightest and next-to-lightest
 states in this channel ($\r$ and $\r(1450)$ in Table~\ref{Fig:QCD}) 
can approximately  be thought of as resulting from the mixing of two states
 that have different  $SU(2)_L\times SU(2)_R$ transformation properties 
 and are sourced by different operators V and T.
The  $U(1)_A$ partner of the combination of $\r$ and $\r(1450)$
sourced by T is denoted by $b_1$ (see again Table~\ref{Fig:QCD}),
and sourced by AT.

Because we are also comparing with real-world QCD, we should notice that
the isosinglet $I=0$ sector is complicated by the fact that real-world  light mesons
are better explained by a model in which one includes $2+1$ quarks, including the 
heavier strange quark $s$.
The isosinglet mesons include an additional tower of states, due to mixing with
the $(\bar{s}s)$ singlet. 
In Table~\ref{Fig:QCD}, this
 results in the doubling of states with $I=0$ in respect to the $I=1$ case,
 as we chose to retain pairs of mesons with identical quantum numbers.
 The $G$-parity assignment of each isosinglet state is the opposite of that of the
 corresponding isotriplet with the same $J^{P{\cal C}}$.
 Notice in the table that the decay of the $\eta$ to $3\pi$ violates $G$-parity: it
 yields a suppressed rate $\Gamma$, with $\Gamma/M\sim 10^{-5}$, and originates
 from explicit breaking of isospin in real-world QCD, in which, for instance, 
 up and down quarks are not  degenerate in mass.

 The $I=0$ and $J=0$ lightest states are the $f_0(500)/f_0(980)$ and their $U(1)_A$ partners,  
 the   $\eta/\eta^{\prime}$ pair.
The $I=0$ and $J=1$ sector contains the pairs $\omega/\phi$,  $f_1(1285)/f_1(1420)$, 
 $h_1/h_1^{\prime}$, and $\omega(1420)/\phi(1680)$.
The four of them play the same roles as, respectively, the $\rho$, 
 $a_1$,   $b_1$, and $\rho(1450)$  mesons, in the isotriplet case.
One linear combination of the two $\omega/\phi$ pairs
 is sourced by the bilinear operator T, which forms a doublet
of $U(1)_A$ with the source AT of the $h_1/h_1^{\prime}$ pair.

Let us see how these considerations have to be modified for
 enlarged cosets.  (Both the $SU(2N_f)/Sp(2N_f)$ and $SU(2N_f)/SO(2N_f)$  
 contain 
the $SU(N_f)_L\times SU(N_f)_R/SU(N_f)_V$ subspace, enhanced 
because of the
(pseudo)real nature of the underlying fermion representations.)
The unbroken baryon number $U(1)_B$ is a subset of the unbroken part of these two cosets 
rather than commuting with it.
Diquark operators hence source mesons, that complete the representation of $SU(N_f)_V$
into full representations of the unbroken $Sp(2N_f)$ or $SO(2N_f)$.
The unbroken group and the coset have in these cases different dimension, 
so that  representations with different $G$-parity have different dimensionality as well 
(see Table~\ref{tab:mesons} for instance), rendering
$G$-parity redundant, at least as a way
to distinguish among them.

We conclude by discussing  explicitly the general form of the operators to be used as sources.
We start from a  two-component spinor $\chi$ transforming in
 the fundamental representation of the global $SU(2N_f)$ symmetry group. 
A spin-0 local operator $J^0$ takes the schematic form $\chi^{\mathrm{T}} \tilde{C} \chi$.
The product of two fundamental decomposes as
\beq
2N_f \otimes 2N_f =\left[ N_f(2N_f-1)\right] \oplus \left[N_f(2N_f+1)\right]\,,
\eeq
 into the 2-index antisymmetric and symmetric representations of $SU(2N_f)$, respectively.
Furthermore, depending on whether the unbroken subgroup is $Sp(2N_f)$ or $SO(2N_f)$, 
either the antisymmetric or the symmetric combination is reducible and 
further decomposes into the unbroken subgroup
 by projecting one element along  the elementary  
symplectic matrix $\Omega$ or the symmetric $\omega$, respectively.
Excitations sourced by  this operator correspond to the $f_0$ of QCD-like theories, while
those along the complement correspond to the PS flavoured states (the $\pi$ of QCD).
The representation that is irreducible, instead, 
would coincide with the adjoint of the unbroken subgroup,
except that it vanishes because of Fermi statistics---unless one considers nonlocal operators.

Along the same lines, the spin-1 local operator $J^1$ schematically 
reads ${\chi}^{\dagger} \bar{\sigma}^{\mu}\chi$,
and the decomposition in $SU(2N_f)$ takes the form 
\beqs
\overline{2N_f} \otimes 2N_f = 1 \oplus \left[4N_f^2-1\right]\,.
\eeqs
After decomposition into the  representations of the unbroken subgroup,  
the adjoint splits into its antisymmetric and symmetric parts, and hence one ultimately finds
 the same decomposition as in the spin-0 case.
This property descends from the (pseudo)real nature of 
$Sp(2N_f)$ and $SO(2N_f)$,  
that do not distinguish 
the fundamental representation $2N_f$ from its conjugate.
It is useful to notice that the four operators  V, T, AV, and AT states
 source only one state that is a singlet. This is  different from the 
$SU(N_f)_L\times SU(N_f)_R/SU(N_f)_V$, where each of the four operators
sources a singlet of the unbroken
group: the three additional $J=1$ states 
are part of the irreducible representations sourced by
V, T, and AT.
 
Summarising for the  $SU(2N_f)/Sp(2N_f)$ and $SU(2N_f)/SO(2N_f)$ cases:
 pseudoscalar PS and axial-vector AV multiplets have the same degeneracies, 
 as do the flavoured scalar S---the $U(1)_A$ partners of the PS states.    
The vector states V belong to a different representation of the unbroken group
common also to the  antisymmetric tensor T, as well as to
 its $U(1)_A$ partner AT.
 In the unbroken $Sp(2N_f)$ case, the V, T and AT mesons span a complete 
 symmetric representation of $Sp(2N_f)$, with PS, S, and  AV
 on the (antisymmetric) complement. The reverse is true in the case of $SO(2N_f)$
 (see  again Table~\ref{tab:mesons}). In both cases, the singlet sector is simpler:
 it consists of two spin-0 states forming a $U(1)_A$ doublet and of one 
 isolated spin-1 singlet state.

 \subsection{Symmetry restoration and enhancement}
 \label{Sec:AppendixD1}
 
 At high temperatures, the fermion condensates melt, leading to restoration
 of the global symmetries. Both the non-Abelian $SU(2N_f)$ 
 (or $SU(N_f)\times SU(N_f)$)
 global symmetry, as well as the Abelian $U(1)_A$ symmetry are restored~\cite{Pisarski:1983ms} 
(see also \cite{Chiu:2013wwa,Brandt:2019ksy,Suzuki:2019vzy} for progress on $N_f=2$ lattice QCD). 
 As a consequence of the former, one might find that the states sourced by V and AV operators
 become degenerate (the $\r$ and $a_1$ in the QCD-like case).
 Because of the latter, $U(1)_A$ multiplets should become degenerate, for example 
 the states sourced by PS and S operators (the $\pi$ and $a_0$ mesons in the QCD-like case).
 See for instance Ref.~\cite{Lee:2017uvl} and references therein.
 
 Recent studies have emerged suggesting that,
 because the thermal bath reduces the space-time symmetries,
 the global internal symmetry is further enhanced, with the emergence of a new 
 chiral-spin symmetry that combines with the global symmetries. 
 We refer the reader to Refs.~\cite{Glozman:2007ek,Glozman:2015qva,
Denissenya:2014ywa,Rohrhofer:2017grg,Glozman:2018jkb,Rohrhofer:2019qwq,Glozman:2019fku}
 for this research field, in which the specific case of the QCD-like,  
 $SU(3)$ theory with $N_f=2$ 
 is discussed in great detail, and numerical evidence of the emergence of a $SU(4)$ global symmetry
 is exposed, in the channels with spin $J>0$.
In the following, we limit ourselves to producing a summary of
what would be the (testable) expectations in
the three main cosets of interest to CHMs,
if the corresponding symmetry-restoration and symmetry-enhancement
  patterns were to be confirmed.

\begin{itemize}

\item The $SU(N_f)_L\times SU(N_f)_R/SU(N_f)_V$ cosets
can emerge, at $T=0$, from theories with  complex representations, 
for example the $SU(N)$ gauge 
theories with $N_f$  fundamental Dirac fermions. 
These cosets are accompanied by the anomalous
 $U(1)_A=U(1)_L\times U(1)_R/U(1)_B$ Abelian coset. At high temperatures,
the global $U(N_f)_L\times U(N_f)_R$ symmetry is restored and 
enhanced to $U(2N_f)=U(1)_A\times SU(2N_f)$.
For example, in the $N_f=2$ case that approximates QCD, in the $J=1$ sector,
four of the $I=1$ states (the two lightest $\rho$, the $a_1$, and the $b_1$)
and four of the $I=0$ states (the two $\omega/\phi$, 
the $f_1/f_1^{\prime}$, and the $h_1/h_1^{\prime}$)
have been measured to become degenerate, which would be
compatible with forming a complete $16$-dimensional adjoint 
 representation of the $U(1)\times SU(4)$ 
enhanced global symmetry group~\cite{Glozman:2007ek,Glozman:2015qva,Denissenya:2014ywa,
Rohrhofer:2017grg,Glozman:2018jkb,Rohrhofer:2019qwq,Glozman:2019fku,Engel:2011aa}.
In the $J=0$ sector, this symmetry is not manifest:
the $\pi$ and $a_0$ combine with the $f_0/f_0^{\prime}$ and the $\eta/\eta^{\prime}$
to form $2N_f^2$ degenerate states, the adjoint of the symmetry group $U(N_f)_L\times U(N_f)_R$.
(It would require an additional $2N_f^2$ components, two copies of the adjoint of $U(N_f)$,
 to complete the adjoint of $U(2N_f)$.) 

\item The $SU(2N_f)/Sp(2N_f)$ cosets 
can emerge, at $T=0$, from theories with  pseudoreal representations, 
for example the $Sp(2N)$ gauge 
theories with $N_f$ Dirac fundamental fermions. 
In addition, the anomalous $U(1)_A$ is also spontaneously broken. 
What would have been the unbroken $U(1)_B$ associated with the baryon number
in the case of complex representations is now a subgroup 
of the nonanomalous $Sp(2N_f)$. Going to a high temperature, the symmetry is expected to be 
first restored and then enhanced to $Sp(4N_f)$. 
For example, in the $N_f=2$ case, in the $J=1$ sector of the spectrum,
 the ten V and ten T mesons (which include both the 
correspondent of the $\rho$ and $\omega/\phi$ of QCD),
 the five AV mesons (corresponding to the $a_1$), the ten 
AT mesons (which include both states corresponding to the $b_1$ and $h_1/h_1^{\prime}$ of QCD)
and the singlet vector state (corresponding to the $f_1/f_1^{\prime}$) 
form a complete, $36$-dimensional adjoint
 representation of $Sp(8)$.  
 The $J=0$ sector is not expected to show high-$T$ symmetry enhancement.
  PS flavoured states  combine with the  singlet to form the antisymmetric representation 
 (in the QCD analogy, they correspond to the $\pi$ and the $f_0/f_0^{\prime}$), 
 and their $U(1)_A$ partners (the $a_0$ and the $\eta/\eta^{\prime}$ in QCD)
 combine to form the complex, antisymmetric 2-index representation
 of $SU(2N_f)$, which has dimension $2\times N_f(2N_f-1)$.
(It  would require finding an adjoint representation
of $SU(2N_f)$ of dimension $4N_f^2-1$, to yield a total of $8N_f^2-2N_f-1$, 
which is the 2-index antisymmetric of $Sp(4N_f)$. 
In the $N_f=2$ case, the antisymmetric of $Sp(8)$ decomposes as
$27=15+6_{\mathbb{C}}$ in terms of $SU(4)$, and the $15$ is missing.)
 
 \item The $SU(2n_f)/SO(2n_f)$ cosets 
can emerge, at $T=0$, from theories with  real representations, 
for example the $Sp(2N)$ gauge 
theories with $n_f$ Dirac fermions transforming as the 
2-index  antisymmetric representation.
At high temperatures, 
 the restoration of the symmetry is expected to be followed by its enhancement
to a global $SO(4n_f)$.  For example, if $n_f=3$, in the spin-1 flavoured sector
 the $15$ v, $15$ t and $15$ at mesons sourced by 
the operators in Table~\ref{tab:mesons} will be degenerate with the $20^{\prime} $ av mesons.
In addition, a flavour-singlet vector will also be degenerate, and together will yield the
$66$-dimensional antisymmetric (adjoint) representation of $SO(12)$, with spin $J=1$.
In the spin $J=0$ sector, the $20^{\prime} $ ps and $20^{\prime} $ s 
mesons, together with the $1+1$ flavor singlets
form the $21_{\mathbb{C}}$ representation of $SU(6)$. 
(It would require an additional $35$, the adjoint of $SU(6)$,
to make the 2-index  symmetric traceless $77$-dimensional representation of $SO(12)$.)
 
 \end{itemize}

 \section{Bilinear operators as sources}
 \label{Sec:AppendixE}

  We collect in this Appendix  technical clarifications about   
gauge-invariant operators ${\cal O}_{M}$, written in terms of the four-component fermions $Q$ and $\Psi$,
to be used as sources in the lattice calculations of the spectrum of composite states.
We consider the $Sp(2N)$ gauge theory, without specifying $N$.
When possible, we also write our expression in a form that applies to   $Sp(2N_f)$ and 
$SO(2n_f)$ groups with general $N_f$ fundamental and $n_f$ antisymmetric Dirac fermions.
For concreteness, in Appendix~\ref{Sec:AppendixE1} we explicitly identify the irreducible 
representations of the unbroken global $Sp(4)$ group
of relevance in the context of  CHMs,
as well as their  $SU(2)_L\times SU(2)_R$ decompositions.

 In the case of $N_f$ Dirac spinors transforming in the fundamental representation
 of the gauge group $Sp(2N)$, the fermion bilinear operators are written as in terms of 
 $N_f \times N_f$ block matrices,
 built from the $N_f$ Dirac fermions 
 \beqs
 Q^{i\,a}=\left(\begin{array}{c}q^{i\,a} \cr -\Omega^{ab}(\tilde{C} 
 q^{N_f+i\,{\ast}})_b\end{array}\right)\,,
 \eeqs 
 and, following the same lines leading to Eqs.~(\ref{Eq:2to4}), their conjugate Dirac fermion as
 \beqs
 Q^{i\,a}_{C}\equiv  -\gamma_5\Omega^{ab} C(\overline{Q^{i}}^{\mathrm{T}})_{b}= 
 -\gamma_5\Omega^{ab} C \gamma^0(Q^{i\ast})_b
 =\left(\begin{array}{c}q^{N_f+i\,a} \cr -\Omega^{ab}(\tilde{C} 
 q^{i\,{\ast}})_b\end{array}\right)\,.
 \eeqs
Notice a difference, with respect to the definition leading to Eq.~(\ref{Eq:2to4}),
in how we define the conjugate spinor: the factor of 
$-\gamma_5\Omega^{ab} $ is introduced, in order to make the decomposition in
LH and RH chiral components of $Q$ and $Q_C$ take the same form. 
 We stress that $Q_C^i$ is physically equivalent to $Q^i$, and hence one can identify the $N_f$
 Dirac fermions with $Q^i$.

 We write explicitly the form of the general $2N_f\times 2N_f$ matrices 
  built as bilinears in spinors, 
 both in two-component and four-component notation, which read
 \beqs
 J_{0}&=&\left(\begin{array}{c|c}
 \cr
\overline{Q^{i\,a}_{C}}  P_L Q^{j\,a} & 
\overline{Q^{i\,a}_{C}}  P_L Q^{j\,a}_{C}\cr
\cr
\hline
\cr
\overline{Q^{i\,a}} P_L Q^{j\,a}
&
\overline{Q^{i\,a}} P_L Q^{j\,a}_{\,C}
\cr 
 \end{array}\right)
 =\Omega_{ab} 
 \left(\begin{array}{c|c}
 \cr
q^{i\,a\,\mathrm{T}} \tilde{C} q^{j\,b}  & 
q^{i\,a\,\mathrm{T}} \tilde{C} q^{N_f+j\,b} \cr
\cr
\hline
\cr
q^{N_f+i\,a\,\mathrm{T}} \tilde{C} q^{j\,b} 
&
q^{N_f+i\,a\,\mathrm{T}} \tilde{C} q^{N_f+j\,b}
\cr 
 \end{array}\right),\\
 J_{1}^{\mu}&
 =&\left(\begin{array}{c|c}
 \cr
\overline{Q^{i\,a}}  \gamma^{\mu}P_L Q^{j\,a} & 
\overline{Q^{i\,a}}  \gamma^{\mu}P_L Q^{j\,a}_{C} \cr
\cr
\hline
\cr
\overline{Q^{i\,a}_C} \gamma^{\mu}P_L Q^{j\,a}
&
\overline{Q^{i\,a}_C}  \gamma^{\mu}P_L Q^{j\,a}_{\,C}
\cr 
 \end{array}\right)
 =
  \left(\begin{array}{c|c}
 \cr
q^{i\,a\,\dagger} \bar{\sigma}^{\mu}  q^{j\,a}  & 
q^{i\,a\,\dagger} \bar{\sigma}^{\mu}  q^{N_f+j\,a} \cr
\cr
\hline
\cr
q^{N_f+i\,a\,\dagger} \bar{\sigma}^{\mu} q^{j\,a} 
&
q^{N_f+i\,a\,\dagger} \bar{\sigma}^{\mu}  q^{N_f+j\,a}
\cr 
 \end{array}\right),
 \eeqs
 and
 \beqs
  J_{1^{\prime}}^{\mu\nu}=\left(\begin{array}{c|c}
 \cr
\overline{Q^{i\,a}_{C}}  \sigma^{\mu\nu}P_L Q^{j\,a} & 
\overline{Q^{i\,a}_{C}}  \sigma^{\mu\nu}P_L Q^{j\,a}_{C}\cr
\cr
\hline
\cr
\overline{Q^{i\,a}} \sigma^{\mu\nu}P_L Q^{j\,a}
&
\overline{Q^{i\,a}} \sigma^{\mu\nu}P_L Q^{j\,a}_{\,C}
\cr 
 \end{array}\right)
 \,=
\Omega_{ab}
 \left(\begin{array}{c|c}
 \cr
q^{i\,a\,\mathrm{T}} \tilde{C}\sigma^{\mu\nu}_{LL} q^{j\,b}  & 
q^{i\,a\,\mathrm{T}} \tilde{C}  \sigma^{\mu\nu}_{LL} q^{N_f+j\,b} \cr
\cr
\hline
\cr
q^{N_f+i\,a\,\mathrm{T}} \tilde{C} \sigma^{\mu\nu}_{LL} q^{j\,b} 
&
q^{N_f+i\,a\,\mathrm{T}} \tilde{C} \sigma^{\mu\nu}_{LL}  q^{N_f+j\,b}
\cr 
 \end{array}\right)\,.
 \eeqs
 
 The symplectic matrix $\Omega_{2N_f}$ is defined as a $2N_f\times 2N_f$ antisymmetric matrix such that 
 $(\Omega_{2N_f})^2=-\mathbb{1}_{2N_f}$.
 Because of the contraction with $\Omega_{ab}$,  $J_0$ is antisymmetric. 
 This bilinear condenses,  and hence the unbroken subgroup is $Sp(2N_f)$. 
The decomposition of $J_{0}$, $J_{1}^{\mu}$ and $J_{1^{\prime}}^{\mu\nu}$ 
in their irreducible representations
is $ 2N_f \otimes 2N_f = \left[N_f(2N_f-1) -1\right]\oplus 1 \oplus N_f(2N_f+1)$:
 \beqs
 \nonumber
 J_{0,1,1^{\prime}}&=&\left[
\frac{1}{2}\left( J_{0,1,1^{\prime}}- (J_{0,1,1^{\prime}})^{\mathrm{T}}\right)+ \frac{\Omega_{2N_f}}{2N_f}
\Tr\left\{\Omega_{2N_f}J_{0,1,1^{\prime}}\right\}\right]-\frac{\Omega_{2N_f}}{2N_f}
\Tr\left\{\Omega_{2N_f}J_{0,1,1^{\prime}}\right\}\,+\,\\
&&
 \,+\,\frac{1}{2}\left( J_{0,1,1^{\prime}} + (J_{0,1,1^{\prime}})^{\mathrm{T}}\right)
\,\equiv\,J_{0,1,1^{\prime}}^{(A)}\,+J_{0,1,1^{\prime}}^{(1)}\,+\,J_{1,1^{\prime}}^{(S)}\,.
 \eeqs
 We highlighted here the fact that the symmetric part of $J_0$ vanishes identically.
 Notice also that the singlets are antisymmetric. Both the $J_1$ and $J_{1^{\prime}}$
 decompose into symmetric and antisymmetric parts, the latter expected to be related to heavier states.
For the $SU(2N_f)/Sp(2N_f)$ coset the 
operators in Table~\ref{tab:mesons} are identified as follows: 
\beqs
{\cal O}_{\rm PS}=J_{0}^{(A)}\,,~~~
{\cal O}_{\rm V}=J_{1}^{(S)}\,,~~~
{\cal O}_{\rm AV}=J_{1}^{(A)}\,,~~~
{\cal O}_{\rm T}=J_{1^{\prime}}^{(S)}\,,
\eeqs
with the ${\cal O}_{\rm S}$ and ${\cal O}_{\rm AT}$ operators being the $U(1)_A$ conjugates
of ${\cal O}_{\rm PS}$ and ${\cal O}_{\rm T}$, respectively.

With matter content including $n_f$ Dirac fermions in the antisymmetric representation
  of $Sp(2N)$ (for $N>1$) 
we introduce the analogous $j_{0,1,1^{\prime}}$ operators built from the fermions 
\beqs
\Psi^{i\,ab}=\left(\begin{array}{c}\psi^{i\,ab} \cr 
  -\Omega^{ac}\Omega^{bd}(\tilde{C} \psi^{n_f+i\,^{\ast}})_{cd}\end{array}\right)\,,
  \eeqs
  with $i=1\,,\,\cdots\,,\,n_f$, together with their conjugates 
  \beqs
 \Psi^{i\,ab}_{C}\equiv \Omega^{ac}\Omega^{bd} 
 C(\overline{\Psi^{i}}^{\mathrm{T}})_{\,\,cd}\equiv \Omega^{ac}\Omega^{bd}C\gamma^0(\Psi^{i\ast})_{cd}
 &=&
 \left(\begin{array}{c}\psi^{n_f+i\,ab} \cr 
  -\Omega^{ac}\Omega^{bd}(\tilde{C} \psi^{i\,^{\ast}})_{cd}\end{array}\right)
 \,.
 \eeqs 
 We conventionally align the vacuum with the matrix $\omega_{2n_f}$, 
 the $2n_f\times 2n_f$ 
 symmetric matrix such that 
 $(\omega_{2n_f})^2=\mathbb{1}_{2n_f}$, generalising Eq.~(\ref{Eq:symplectic}).
We decompose $j_{0,1,1^{\prime}}$ in  irreducible representations
as $ 2n_f \otimes 2n_f = \left[n_f(2n_f+1)-1 \right]\oplus 1 \oplus n_f(2n_f-1)$:
 \beqs
 j_{0,1,1^{\prime}}&=&\left[\nonumber
\frac{1}{2}\left( j_{0,1,1^{\prime}}+ (j_{0,1,1^{\prime}})^{\mathrm{T}}\right)- 
\frac{\omega_{2n_f}}{2n_f}\Tr\left\{\omega_{2n_f}j_{0,1,1^{\prime}}\right\}\right]
+\frac{\omega_{2n_f}}{2n_j}\Tr\left\{\omega_{2n_f}j_{0,1,1^{\prime}}\right\}\,+\,\\
&&
 \,+\,\frac{1}{2}\left( j_{0,1,1^{\prime}} - (j_{0,1,1^{\prime}})^{\mathrm{T}}\right)
\,\equiv\,j_{0,1,1^{\prime}}^{(S)}\,+j_{0,1,1^{\prime}}^{(1)}\,+\,j_{1,1^{\prime}}^{(A)}\,.
 \eeqs
 We highlighted the fact that the antisymmetric part of $j_0$ vanishes.
 In this case, the singlets are symmetric matrices.

 We write these operators in $n_f\times n_f$ blocks 
as matrix representations  of the global $SU(2n_f)$ symmetry:
  \beqs
 j_{0}&=&\left(\begin{array}{c|c}
 \cr
\overline{\Psi^{i\,ab}_{C}}  P_L \Psi^{j\,ab} & 
\overline{\Psi^{i\,ab}_{C}}  P_L \Psi^{j\,ab}_{C} \cr
\cr
\hline
\cr
\overline{\Psi^{i\,ab}} P_L \Psi^{j\,ab}
&
\overline{\Psi^{i\,ab}}  P_L \Psi^{j\,ab}_{\,C}
\cr 
 \end{array}\right) \,
= \Omega_{ac}\Omega_{bd} 
  \left(\begin{array}{c|c}
 \cr
\psi^{i\,ab\,\mathrm{T}} \tilde{C}\psi^{j\,cd}  & 
\psi^{i\,ab\,\mathrm{T}} \tilde{C} \psi^{n_f+j\,cd} \cr
\cr
\hline
\cr
\psi^{n_f+i\,ab\,\mathrm{T}} \tilde{C} \psi^{j\,cd} 
&
\psi^{n_f+i\,ab\,\mathrm{T}} \tilde{C}\psi^{n_f+j\,cd}
\cr 
 \end{array}\right)\,,\\
 j_{1}^{\mu}
 &=&\left(\begin{array}{c|c}
 \cr
\overline{\Psi^{i\,ab}}   \gamma^{\mu}P_L \Psi^{j\,ab} & 
\overline{\Psi^{i\,ab}}  \gamma^{\mu}P_L \Psi^{j\,ab}_{C} \cr
\cr
\hline
\cr
\overline{\Psi^{i\,ab}_C}  \gamma^{\mu}P_L \Psi^{j\,ab}
&
\overline{\Psi^{i\,ab}_C}   \gamma^{\mu}P_L \Psi^{j\,ab}_{\,C})
\cr 
 \end{array}\right)\,
 =
  \left(\begin{array}{c|c}
 \cr
\psi^{i\,ab\,\dagger} \bar{\sigma}^{\mu}  \psi^{j\,ab}  & 
\psi^{i\,ab\,\dagger} \bar{\sigma}^{\mu}  \psi^{n_f+j\,ab} \cr
\cr
\hline
\cr
\psi^{n_f+i\,ab\,\dagger} \bar{\sigma}^{\mu} \psi^{j\,ab} 
&
\psi^{n_f+i\,ab\,\dagger} \bar{\sigma}^{\mu}  \psi^{n_f+j\,ab}
\cr 
 \end{array}\right)\,,
 \eeqs
 and 
 \beqs
  j_{1^{\prime}}^{\mu\nu}=\left(\begin{array}{c|c}
 \cr
\overline{\Psi^{i\,ab}_{C}} \sigma^{\mu\nu} P_L \Psi^{j\,ab} & 
\overline{\Psi^{i\,ab}_{C}}   \sigma^{\mu\nu} P_L \Psi^{j\,ab}_{C} \cr
\cr
\hline
\cr
\overline{\Psi^{i\,ab}}  \sigma^{\mu\nu} P_L \Psi^{j\,ab}
&
\overline{\Psi^{i\,ab}}  \sigma^{\mu\nu}  P_L \Psi^{j\,ab}_{\,C}
\cr 
 \end{array}\right)\,
 = \Omega_{ac}\Omega_{bd} \
  \left(\begin{array}{c|c}
 \cr
\psi^{i\,ab\,\mathrm{T}} \tilde{C} \sigma^{\mu\nu}_{LL}  \psi^{j\,cd}  & 
\psi^{i\,ab\,\mathrm{T}} \tilde{C} \sigma^{\mu\nu}_{LL} \psi^{n_f+j\,cd} \cr
\cr
\hline
\cr
\psi^{n_f+i\,ab\,\mathrm{T}} \tilde{C} \sigma^{\mu\nu}_{LL} \psi^{j\,cd} 
&
\psi^{n_f+i\,ab\,\mathrm{T}} \tilde{C}  \sigma^{\mu\nu}_{LL} \psi^{n_f+j\,cd}
\cr 
 \end{array}\right)\,.
 \eeqs

In the $SU(2n_f)/SO(2n_f)$ case, the 
operators in Table~\ref{tab:mesons} are identified as follows: 
\beqs
{\cal O}_{\rm ps}=j_{0}^{(S)}\,,~~~
{\cal O}_{\rm v}=j_{1}^{(A)}\,,~~~
{\cal O}_{\rm av}=j_{1}^{(S)}\,,~~~
{\cal O}_{\rm t}=j_{1^{\prime}}^{(A)}\,.
\eeqs
Again, ${\cal O}_{\rm s}$ and ${\cal O}_{\rm at}$ operators are the $U(1)_A$ conjugates
of ${\cal O}_{\rm ps}$ and ${\cal O}_{\rm t}$, respectively.

As explained also in Appendix~\ref{Sec:AppendixD}, operators $J_0$  that source spin-0 states,
and operators $J_1$ sourcing spin-1 states can be classified
 in terms of the original global $SU(2N_f)$ (enlarged) symmetry,
according to which the former transform as 
$2N_f \otimes 2N_f =\left[ N_f(2N_f-1)\right] \oplus \left[N_f(2N_f+1)\right]$,
and the latter as $\overline{2N_f} \otimes 2N_f = 1 \oplus \left[4N_f^2-1\right]$ of $SU(2N_f)$.
Only once the $SU(2N_f)$ is broken to its $Sp(2N_f)$ group, $J_0$ and $J_1$ 
decompose in the same set of irreducible  representations of the subgroup.
Similarly, operators $J_{1^{\prime}}$ couple LH to RH components of the Dirac fields, 
and hence while they
source particles with the same spin as $J_1$, they transform under $SU(2N_f)$ in the same way as $J_0$.
Once the symmetry is broken, mixing between the particles sourced 
by $J_1$ and $J_{1^{\prime}}$  will ensue,
but in general these are different sources for different $SU(2N_f)$ particles.
The same considerations apply to the $j_{0,1,1^{\prime}}$ operators, and their decompositions
in irreducible representations of the $SO(2n_f)$ subgroup.

One can now explicitly decompose the matrices $J_{0,1,1^{\prime}}$ (and $j_{0,1,1^{\prime}}$)
into a given basis of the $U(N_f)$ (and $U(n_f)$), by making a choice of generators $T^A$ (and $t^B$) 
of the group. We devote the next subsection
to showing the result of this process in the $SU(4)/Sp(4)$ case, for some of the interesting operators.

 \subsection{$SU(4)/Sp(4)$ composite operators}
\label{Sec:AppendixE1}

We focus on the  $Sp(2N)$ theory with $N_f=2$ fundamental fermions, 
and its $SU(4)/Sp(4)$ coset, which is  relevant 
 in the CHM  context (see also the discussion in~\cite{Drach:2017btk}).
 
The pions in Eq.~(\ref{Eq:pion}) are sourced by the antisymmetric 
${\cal O}_{\rm PS}=J_{0}^{(A)}$ operators.
 We can  add to the Lagrangian density in Eq.~(\ref{Eq:L6}) the following source term:
 \beqs
  {\cal L}_{\pi}&=&\frac{\sqrt{2}f}{2} \Tr\left[J_0\Sigma^{\ast}+{\rm h.c.}\right]\,,
 \eeqs
 where $\Sigma=e^{\frac{2i\pi}{f}}\Omega$ is the antisymmetric matrix defined in Eq.~(\ref{Eq:Sigmas}).
 We can expand for small $\pi/f$ and make use of the antisymmetry in flavour space of $J_0$.
We make explicit use of the generators as written in Eq.~(B.4) of Ref.~\cite{Lee:2017uvl}.
Looking at  the decomposition of the $5$ according to $SO(4)\sim SU(2)_L\times SU(2)_R \subset Sp(4)$, 
we find that the $SO(4)$-singlet $\pi^3$ is sourced by the operator 
${\cal O}_{{\rm PS}, 3}\equiv\,i\,{\delta {\cal L}_{\pi}}/{\delta \pi^3}$ that reads 
 \beqs
{\cal O}_{{\rm PS}, 3} &=&
\frac{i}{2}\left(\frac{}{}iJ_0^{13}-iJ_0^{31}-iJ_0^{24}+iJ_0^{42}\,+\,{\rm c.c.}\frac{}{}\right)
  =\left(\overline{Q^{1\,a}}  \gamma^5 Q^{1\,a}-\overline{Q^{2\,a}}\gamma^5Q^{2\,a}\right),
 \eeqs
 where we made abundant use of Eqs.~(\ref{Eq:scalarrelations}).
The $4$ of $SO(4)$ are sourced by the following operators  
${\cal O}_{{\rm PS}, A}\,\equiv\,i\,{\delta {\cal L}_{\pi}}/{\delta \pi^A}$ given by\footnote{We 
remind the reader that
 the sources for $\pi^1$, $\pi^2$ and $\pi^3$ would be the same sources 
 as for the coset ${SU(2)_L\times SU(2)_R}/{SU(2)_V}$. 
 The additional pions are due to the symmetry enhancement in replacing
 $SU(2N)$ with $Sp(2N)$ as the gauge group.}
 \beqs
 {\cal O}_{{\rm PS}, 1}
 &=&\nonumber
\frac{i}{2}\left(\frac{}{}+iJ_0^{14}-iJ_0^{41}+iJ_0^{23}-iJ_0^{32}\,+\,{\rm c.c.}\frac{}{}\right)
 =\left(\overline{Q^{1\,a}} \gamma^5 Q^{2\,a}+\overline{Q^{2\,a}} \gamma^5 Q^{1\,a}\right)\,,\\
   %
{\cal O}_{{\rm PS}, 2}&=&\nonumber
\frac{i}{2}\left(\frac{}{}-J_0^{14}+J_0^{41}+J_0^{23}-J_0^{32}\,+\,{\rm c.c.}\frac{}{}\right)
=\left(-i\overline{Q^{1\,a}} \gamma^5 Q^{2\,a}+i\,\overline{Q^{2\,a}} \gamma^5 Q^{1\,a}\right)\,,\\
 {\cal O}_{{\rm PS}, 4}&=&
\frac{i}{2}\left(\frac{}{}+J_0^{12}-J_0^{21}-J_0^{34}+J_0^{43}\,+\,{\rm c.c.}\frac{}{}\right)
  =-i\,\left(\overline{Q^{1\,a}}  Q^{2\,a}_{\,C}+\overline{Q_C^{2\,a}}  Q^{1\,a}\right)\,,\\
 {\cal O}_{{\rm PS}, 5}&=&\nonumber
\frac{i}{2}\left(\frac{}{}-iJ_0^{12}+iJ_0^{21}-iJ_0^{34}+iJ_0^{43}\,+\,{\rm c.c.}\frac{}{}\right)
   =i\,\left(-i\,\overline{Q^{1\,a}}  Q^{2\,a}_C+i\overline{Q^{2\,a}_C} Q^{1\,a}\right)\,.
 \eeqs
The additional operator given by
\beqs
{\cal O}_{{\rm PS}, 0} & \equiv & -\frac{1}{2}\left(\frac{}{}
J_0^{13}+J_0^{24}-J_0^{31}-J_0^{42}+{\rm c.c}\frac{}{}\right)\,=\,
\left(\overline{Q^{1\,a}} Q^{1\,a}+\overline{Q^{2\,a}} Q^{2\,a}\right)
\eeqs
is aligned along $\Omega$, in the internal space. It is the operator 
that develops a nontrivial VEV. 
It sources the $Sp(4)$ singlet,
which has the same role as the $f_0(500)$ of QCD (see Table~\ref{Fig:QCD}). Notice that these $6$ operators put together transform as
the antisymmetric representation  of the global $SU(4)$, before the decomposition in $Sp(4)$ irreducible representations
as $6=1+5$. 
 
 Six additional operators ${\cal O}_{{\rm S}, i}$ can be obtained by replacing 
  $\mathbb{1}_4\rightarrow i \gamma^5$ from the operators ${\cal O}_{{\rm PS}, i}$ for $i=0\,,\,\cdots\,,\, 5$.
Such operators are related to the former by the (anomalous) global $U(1)_A\sim O(2)$ symmetry. 
 (In QCD, the $U(1)_A$ partners of the $\pi$ and  $f_0(500)$ particles
 are, respectively, the $a_0$  and $\eta/\eta^{\prime}$ particles.)

  The sources of the spin-1 fields are given in terms of  the operators in $J_{1}^{\mu}$.
In Sec.~\ref{Sec:HLS} we introduced the fields $A_{\mu}=\sum_{A=1}^{15}A^{A}_{\,\,\mu}T^i$
(we drop here the subscript $6$),
 with $T^A$ the Hermitian generators of $SU(4)$ normalised as $\Tr T^A  T^B=\frac{1}{2}\delta^{AB}$.
 As in Ref.~\cite{Lee:2017uvl}, it is convenient to label the broken generators with $A=1\,,\,\cdots\,,\,5$, and the unbroken ones with $A=6\,,\,\cdots\,,\,15$.
    We hence add to the Lagrangian density the following source term:
  \beqs
 {\cal L}_{A}&=&\sqrt{2}\Tr\left[J_{1}^{\mu}A_{\mu}^{\mathrm{T}}\right]+{\rm h.c.}\,.
 \eeqs
 Starting from the AV sources, we find that
 ${\cal O}_{{\rm AV},A}^{\,\,\mu}\,\equiv\,\frac{\delta {\cal L}_A}{\delta A^{A}_{\,\,\mu}}$ are
  given by the following
 \beqs
\, {\cal O}_{{\rm AV},1}^{\,\,\mu}\,
\nonumber
&=&\frac{1}{2}\left(\frac{}{}\,J_1^{12}+\,J_1^{21}+\,J_1^{34}+\,J_1^{43}\,+\,{\rm c.c.}\right)
=
\overline{Q^{1\,a}}\gamma^{\mu}\gamma^5Q^{2\,a}
 +\,\overline{Q^{2\,a}}\gamma^{\mu}\gamma^5Q^{1\,a}\,,
 \\
 \nonumber
 \, {\cal O}_{{\rm AV},2}^{\,\,\mu}
 &=&
  \frac{1}{2}\left(\frac{}{}\,-iJ_1^{12}+i\,J_1^{21}+i\,J_1^{34}-i\,J_1^{43}\,+\,{\rm c.c.}\right)
  =
-i\,\overline{Q^{1\,a}}\gamma^{\mu}\gamma^5Q^{2\,a}
 +i\,\overline{Q^{2\,a}}\gamma^{\mu}\gamma^5Q^{1\,a}\,,\\
  %
   {\cal O}_{{\rm AV},3}^{\,\,\mu}
&=&
   \frac{1}{2}\left(\frac{}{}\,J_1^{11}-\,J_1^{22}+\,J_1^{33}-\,J_1^{44}\,+\,{\rm c.c.}\right)
    =
\overline{Q^{1\,a}}\gamma^{\mu}\gamma^5Q^{1\,a}
 -\,\overline{Q^{2\,a}}\gamma^{\mu}\gamma^5Q^{2\,a}\,,\\
    \nonumber
   {\cal O}_{{\rm AV},4}^{\,\,\mu}
&=&
    \frac{1}{2}\left(\frac{}{}-iJ_1^{14}+i\,J_1^{23}-i\,J_1^{32}+i\,J_1^{41}\,+\,{\rm c.c.}\right)
=
-i\,\overline{Q^1}\gamma^{\mu}Q_C^2+i\,\overline{Q^2_C}\gamma^{\mu}Q^1\,,
 \\
     \nonumber
     {\cal O}_{{\rm AV},5}^{\,\,\mu}
&=&
   \frac{1}{2}\left(\frac{}{}\,J_1^{14}-\,J_1^{23}-\,J_1^{32}+\,J_1^{41}\,+\,{\rm c.c.}\right)
=
\overline{Q^1}\gamma^{\mu}Q_C^2+\,\overline{Q^2_C}\gamma^{\mu}Q^1
 \,,
 \eeqs
 where we used relations such as
  $\overline{Q^{2\,a}}\gamma^{\mu}P_LQ_C^{1\,a}=(\overline{Q^{2\,a}}\gamma^{\mu}P_LQ_C^{1\,a})^{\mathrm{T}}
  =-\overline{Q^{1\,a}}\gamma^{\mu}P_RQ_C^{2\,a}$, or
    $\overline{Q^{1\,a}_C}\gamma^{\mu}P_LQ_C^{2\,a}
    =(\overline{Q^{1\,a}_C}\gamma^{\mu}P_LQ_C^{2\,a})^{\mathrm{T}}
  =-\overline{Q^{2\,a}}\gamma^{\mu}P_RQ^{1\,a}$,
  and  $(\overline{Q^{2\,a}}\gamma^{\mu}P_LQ_C^{1\,a})^{\dagger}=\overline{Q^{1\,a}_C}\gamma^{\mu}P_LQ^{2\,a}$.
  It hence turns out that the $SO(5)$ fundamental $5$
   decomposes as $4+1$ of $SO(4)$, with the singlet being sourced 
  by ${\cal O}_{{\rm AV},3}^{\,\,\mu}$, while the $4$ is sourced by 
  ${\cal O}_{{\rm AV},1}^{\,\,\mu}$, ${\cal O}_{{\rm AV},2}^{\,\,\mu}$, $
  {\cal O}_{{\rm AV},4}^{\,\,\mu}$
  and ${\cal O}_{{\rm AV},5}^{\,\,\mu}$. For completeness, 
  ${\cal O}_{{\rm AV},1}^{\,\,\mu}$, 
  ${\calO}_{{\rm AV},2}^{\,\,\mu}$ and 
  and ${\cal O}_{{\rm AV},3}^{\,\,\mu}$ would be the generators that are used in the $SU(2)_L\times SU(2)_R/SU(2)_V$ coset.

   Similar expressions hold for the ten operators sourcing the $V$ mesons. 
   We adopt the same basis as in Ref.~\cite{Lee:2017uvl},
   and within these conventions we find the following: 
   \beqs
       \nonumber
\,     {\cal O}_{{\rm V},6}^{\,\,\mu}
 &=&
     \frac{1}{2}\left(\frac{}{}-i\,J_1^{13}-i\,J_1^{24}+\,iJ_1^{31}+i\,J_1^{42}\,+\,{\rm c.c.}\right)
        =\nonumber\\
        &=&\nonumber
-\frac{i}{2}\left(\frac{}{}\overline{Q^{1\,a}}\gamma^{\mu}\gamma^5Q^{1\,a}_C+\,\overline{Q^{2\,a}}\gamma^{\mu}\gamma^5Q^{2\,a}_C
-\,\overline{Q^{1\,a}_C}\gamma^{\mu}\gamma^5Q^{1\,a}-\,\overline{Q^{2\,a}_C}\gamma^{\mu}\gamma^5Q^{2\,a}
\frac{}{}\right)
 \,,\\
        \nonumber
\,     {\cal O}_{{\rm V},7}^{\,\,\mu}
 &=&
     \frac{1}{2}\left(\frac{}{}-i\,J_1^{14}-i\,J_1^{23}+i\,J_1^{32}+i\,J_1^{41}\,+\,{\rm c.c.}\right)
               =\nonumber\\
        &=&\nonumber
-\frac{i}{2}\left(\frac{}{}\overline{Q^{1\,a}}\gamma^{\mu}\gamma^5Q^{2\,a}_C+\,\overline{Q^{2\,a}}\gamma^{\mu}\gamma^5Q^{1\,a}_C
-\,\overline{Q^{1\,a}_C}\gamma^{\mu}\gamma^5Q^{2\,a}-\,\overline{Q^{2\,a}_C}\gamma^{\mu}\gamma^5Q^{1\,a}
\frac{}{}\right)\,,\\
        \nonumber
\,     {\cal O}_{{\rm V},8}^{\,\,\mu}
 &=&
     \frac{1}{2}\left(\frac{}{}-i\,J_1^{12}-i\,J_1^{34}+i\,J_1^{21}+i\,J_1^{43}\,+\,{\rm c.c.}\right)
        =
-i\,\overline{Q^{1\,a}}\gamma^{\mu}Q^{2\,a}+\,i\,\overline{Q^{2\,a}}\gamma^{\mu}Q^{1\,a}
\,,\\
        \nonumber
\,     {\cal O}_{{\rm V},9}^{\,\,\mu}
 &=&
     \frac{1}{2}\left(\frac{}{}-i\,J_1^{13}+i\,J_1^{24}+i\,J_1^{31}-i\,J_1^{42}\,+\,{\rm c.c.}\right)
               =\nonumber\\
        &=&
-\frac{i}{2}\left(\frac{}{}\overline{Q^{1\,a}}\gamma^{\mu}\gamma^5Q^{1\,a}_C-\,\overline{Q^{2\,a}}\gamma^{\mu}\gamma^5Q^{2\,a}_C
-\,\overline{Q^{1\,a}_C}\gamma^{\mu}\gamma^5Q^{1\,a}+\,\overline{Q^{2\,a}_C}\gamma^{\mu}\gamma^5Q^{2\,a}
\frac{}{}\right)\nonumber
\,,\\
\,     {\cal O}_{{\rm V},10}^{\,\,\mu}
 &=&
     \frac{\sqrt{2}}{2}\left(\frac{}{}\,J_1^{13}+\,J_1^{31}\,+\,{\rm c.c.}\right)
        =
\frac{\sqrt{2}}{2}\left(\frac{}{}\overline{Q^{1\,a}}\gamma^{\mu}\gamma^5Q^{1\,a}_C +\,\overline{Q^{1\,a}_C}\gamma^{\mu}\gamma^5Q^{1\,a}\frac{}{}\right)
\,,\\
        \nonumber
\,     {\cal O}_{{\rm V},11}^{\,\,\mu}
 &=&
     \frac{1}{2}\left(\frac{}{}\,J_1^{14}+\,J_1^{23}+\,J_1^{32}+\,J_1^{41}\,+\,{\rm c.c.}\right)
               =\nonumber\\
        &=&\nonumber
+\frac{1}{2}\left(\frac{}{}\overline{Q^{1\,a}}\gamma^{\mu}\gamma^5Q^{2\,a}_C+\,\overline{Q^{2\,a}}\gamma^{\mu}\gamma^5Q^{1\,a}_C
+\,\overline{Q^{1\,a}_C}\gamma^{\mu}\gamma^5Q^{2\,a}+\,\overline{Q^{2\,a}_C}\gamma^{\mu}\gamma^5Q^{1\,a}
\frac{}{}\right)
\,,\\
        \nonumber
\,     {\cal O}_{{\rm V},12}^{\,\,\mu}
 &=&
     \frac{\sqrt{2}}{2}\left(\frac{}{}\,J_1^{24}+\,J_1^{42}\,+\,{\rm c.c.}\right)
        =
\frac{\sqrt{2}}{2}\left(\frac{}{}\overline{Q^{2\,a}}\gamma^{\mu}\gamma^5Q^{2\,a}_C +\,\overline{Q^{2\,a}_C}\gamma^{\mu}\gamma^5Q^{2\,a}\frac{}{}\right)
\,,\\
        \nonumber
\,     {\cal O}_{{\rm V},13}^{\,\,\mu}
 &=&
     \frac{1}{2}\left(\frac{}{}\,J_1^{12}-\,J_1^{34}+\,J_1^{21}-\,J_1^{43}\,+\,{\rm c.c.}\right)
        =
\overline{Q^{1\,a}}\gamma^{\mu}Q^{2\,a} +\,\overline{Q^{2\,a}}\gamma^{\mu}Q^{1\,a}
\,,\\
        \nonumber
\,     {\cal O}_{{\rm V},14}^{\,\,\mu}
 &=&
     \frac{1}{2}\left(\frac{}{}\,J_1^{11}-\,J_1^{22}-\,J_1^{33}+\,J_1^{44}\,+\,{\rm c.c.}\right)
        =
\overline{Q^{1\,a}}\gamma^{\mu}Q^{1\,a} -\,\overline{Q^{2\,a}}\gamma^{\mu}Q^{2\,a}
\,,\\
        \nonumber
\,     {\cal O}_{{\rm V},15}^{\,\,\mu}
 &=&
     \frac{1}{2}\left(\frac{}{}\,J_1^{11}+\,J_1^{22}-\,J_1^{33}-\,J_1^{44}\,+\,{\rm c.c.}\right)
        =
\overline{Q^{1\,a}}\gamma^{\mu}Q^{1\,a} +\,\overline{Q^{2\,a}}\gamma^{\mu}Q^{2\,a}
\,.
\eeqs
The $SO(5)$ adjoint $10$ decomposes as $4+6$ of $SO(4)$: the fundamental $4$
  is sourced by ${\cal O}_{{\rm V},7}^{\,\,\mu}$, 
  ${\cal O}_{{\rm V},8}^{\,\,\mu}$, 
  ${\cal O}_{{\rm V},11}^{\,\,\mu}$
  and ${\cal O}_{{\rm V},13}^{\,\,\mu}$, with the adjoint $6$ sourced by the other six operators.
  Again, for completeness, ${\cal O}_{{\rm V},8}^{\,\,\mu}$, ${\cal O}_{{\rm V},13}^{\,\,\mu}$
  and ${\cal O}_{{\rm V},14}^{\,\,\mu}$ would be the unbroken 
   generators  in the $SU(2)_L\times SU(2)_R/SU(2)_V$ coset, corresponding to the $\r$ mesons in QCD,
  in which case ${\cal O}_{{\rm V},15}^{\,\,\mu}$ would be 
  associated with baryon number and would source the $\omega/\phi$ in QCD.
  
  We can introduce an additional vector field $A_{\mu}^0$, 
  to complete the adjoint $SU(4)$ to the adjoint of the whole 
  $U(1)\times SU(4)$,  by adding the generator $T^0=\frac{1}{2\sqrt{2}}\mathbb{1}_4$. We hence identify the additional operator
  \beqs
    {\cal O}_{{\rm V},0}^{\,\,\mu}&\equiv&
     \frac{1}{2}\left(\frac{}{}J_1^{11}+J_1^{22}+J_1^{33}+J_1^{44}\,+\,{\rm c.c.}\right)
        =
\overline{Q^{1\,a}}\gamma^{\mu}\gamma^5Q^{1\,a} +\,\overline{Q^{2\,a}}\gamma^{\mu}\gamma^5Q^{2\,a}.
  \eeqs
One can recognise this operator to be aligned with
 the generator of the anomalous $U(1)_A$.
It sources the equivalent of the
$f_1(1285)$ of QCD (see  Table~\ref{Fig:QCD}).

These 16 operators do not have a $U(1)_A$ partner.
But a second set of sources of spin-1 states T
is built from  $J_{1^{\prime}}^{(S)}$ states, with the $\sigma^{\mu\nu}$ tensor. Their $U(1)_A$ partners
(corresponding to the more exotic states $b_1$ and $h_1/h_1^{\prime}$ mesons of QCD) 
are sourced by operators AT involving the tensor 
$\sigma^{\mu\nu}\gamma_5$  (see also~\cite{Drach:2017btk}).
We do not show this explicitly, as it requires performing again the same exercise illustrated in the 
previous pages.
Sourcing even more exotic states such as $\pi_1$ may require using 
derivatives and nonlocal operators~\cite{Engel:2011aa}, which is beyond our current purposes.

\begin{table}[t]
\begin{center}
\begin{tabular}{|c|c|cc|cc|cc|cc|}
\hline\hline
\multirow{2}{*}{Measurement} & \multirow{2}{*}{$a m_0$} 
& \multicolumn{2}{c|}{PS} & \multicolumn{2}{c|}{V} 
& \multicolumn{2}{c|}{AV} & \multicolumn{2}{c|}{S} \\
 &  & 
$I_{\rm fit}$ & $\frac{\chi^2}{N_{\rm d.o.f}}$ & $I_{\rm fit}$ & $\frac{\chi^2}{N_{\rm d.o.f}}$ 
& $I_{\rm fit}$ & $\frac{\chi^2}{N_{\rm d.o.f}}$ & $I_{\rm fit}$ & $\frac{\chi^2}{N_{\rm d.o.f}}$ \\
 &&&&&&&&&\\
\hline
QB1FM1 & -0.7 & 14-24 & 1.2 & 13-24 & 1.1 & 9-13 & 0.1 & 9-15 & 0.6\\
QB1FM2 & -0.73 & 14-24 & 1.3 & 13-24 & 1.2 & 9-16 & 0.2 & 9-16 & 0.8\\
QB1FM3 & -0.75 & 14-24 & 1.3 & 13-24 & 1.3 & 9-15 & 0.3 & 9-15 & 1.1\\
QB1FM4 & -0.77 & 14-24 & 1.0 & 12-24 & 1.8 & 8-11 & 0.2 & 8-10 & 0.3\\
QB1FM5 & -0.78 & 14-24 & 1.0 & 12-24 & 1.5 & 8-12 & 0.3 & 8-10 & 0.03\\
QB1FM6 & -0.79 & 14-24 & 0.7 & 12-24 & 1.9 & 8-11 & 0.7 &  & \\
\hline
QB2FM1 & -0.73 & 15-30 & 0.9 & 15-30 & 0.5 & 11-15 & 1.6 & 10-12 & 0.2\\
QB2FM2 & -0.75 & 15-30 & 0.9 & 15-30 & 0.6 & 11-15 & 1.4 & 9-11 & 0.4\\
QB2FM3 & -0.76 & 15-30 & 1.0 & 15-30 & 0.6 & 11-14 & 0.8 &  & \\
\hline
QB3FM1 & -0.6 & 22-30 & 0.6 & 19-30 & 0.7 & 13-26 & 0.9 & 13-28 & 0.9\\
QB3FM2 & -0.65 & 20-30 & 0.5 & 19-30 & 0.5 & 13-22 & 0.3 & 13-22 & 1.5\\
QB3FM3 & -0.68 & 22-30 & 0.9 & 21-30 & 0.7 & 15-22 & 0.8 & 14-22 & 0.8\\
QB3FM4 & -0.7 & 20-30 & 0.7 & 19-30 & 0.7 & 13-20 & 0.2 & 10-14 & 0.6\\
QB3FM5 & -0.71 & 18-30 & 1.1 & 20-30 & 0.6 & 11-15 & 0.8 & 10-13 & 0.6\\
QB3FM6 & -0.72 & 18-30 & 0.9 & 17-30 & 0.9 & 11-15 & 0.3 & & \\
QB3FM7 & -0.73 & 17-30 & 1.0 & 19-30 & 0.6 & 11-15 & 1.0 &  & \\
\hline
QB4FM1 & -0.6 & 22-30 & 2.2 & 22-30 & 1.7 & 15-23 & 0.8 & 16-25 & 0.6\\
QB4FM2 & -0.625 & 22-30 & 1.7 & 22-30 & 1.6 & 15-23 & 0.6 & 16-22 & 0.4\\
QB4FM3 & -0.64 & 22-30 & 1.5 & 22-30 & 1.1 & 15-23 & 0.7 & 15-22 & 0.6\\
QB4FM4 & -0.65 & 22-29 & 1.1 & 22-30 & 0.5 & 15-25 & 0.2 & 15-22 & 0.6\\
QB4FM5 & -0.66 & 22-29 & 1.3 & 20-30 & 0.6 & 15-24 & 0.2 & 15-22 & 0.6\\
QB4FM6 & -0.67 & 22-28 & 1.0 & 20-30 & 0.5 & 15-24 & 0.3 & 15-22 & 0.7\\
QB4FM7 & -0.68 & 19-28 & 0.8 & 20-29 & 0.6 & 15-22 & 0.3 & 13-18 & 0.2\\
QB4FM8 & -0.69 & 19-28 & 0.7 & 20-29 & 1.0 &  &  & & \\
\hline
QB5FM1 & -0.62 & 23-30 & 1.0 & 24-30 & 0.5 & 17-24 & 0.1 & 15-23 & 0.8\\
QB5FM2 & -0.64 & 21-30 & 0.6 & 21-30 & 0.6 & 15-24 & 0.3 & 14-22 & 1.4\\
QB5FM3 & -0.646 & 21-30 & 0.7 & 21-30 & 0.6 & 17-24 & 0.7 &  & \\
\hline\hline
\end{tabular}
\caption{%
\label{tab:meson_measurement_F}%
Technical details about  PS, V, AS, and S lattice correlation functions.
For each ensemble and each choice of bare mass $a m_0$, 
we show the fitting intervals of the Euclidean time $I_{\rm fit}=[t_i,t_f]$ between the minimum
 time $t_i$ and maximum time $t_f$ retained in 
the single-exponential fit to the measured correlators of mesons
made of fundamental Dirac fermions. 
We carry out a correlated fit  via  standard $\chi^2$-minimisation.
We report the values of $\chi^2$ (normalised by the number of  degrees of freedom) 
at the minima. 
In the case of  AV and S states, we leave blank some entries for which the 
numerical data do not exhibit a plateau in the effective mass plots, because of numerical noise. 
}
\end{center}
\end{table}

\begin{table}[t]
\begin{center}
\begin{tabular}{|c|cc|cc|}
\hline\hline
\multirow{2}{*}{Measurement} & \multicolumn{2}{c|}{T} & \multicolumn{2}{c|}{AT} \\
 &  $I_{\rm fit}$ & $\frac{\chi^2}{N_{\rm d.o.f}}$ 
& $I_{\rm fit}$ & $\frac{\chi^2}{N_{\rm d.o.f}}$ \\
 &&&&\\
\hline
QB1FM1 &  13-24 & 0.5 & 9-13 & 0.7 \\
QB1FM2 &  12-24 & 0.5 & 9-13 & 0.2 \\
QB1FM3 &  12-24 & 0.7 & 8-13 & 0.1 \\
QB1FM4 &  12-24 & 1.1 & 8-13 & 0.1 \\
QB1FM5 &  12-24 & 0.9 & 8-12 & 0.6 \\
QB1FM6 &  12-24 & 1.2 & 8-12 & 0.4 \\
\hline
QB2FM1 &  15-30 & 1.0 & 11-17 & 0.8 \\
QB2FM2 &  15-28 & 1.4 & 11-16 & 1.4 \\
QB2FM3 &  12-29 & 1.6 & 9-16 & 1.0 \\
\hline
QB3FM1 &  20-29 & 0.5 & 13-26 & 0.9 \\
QB3FM2 &  20-29 & 0.4 & 13-21 & 1.2 \\
QB3FM3 &  19-30 & 1.0 & 13-22 & 0.4 \\
QB3FM4 &  19-30 & 0.4 & 13-18 & 0.6 \\
QB3FM5 &  19-30 & 1.5 & 14-18 & 0.9 \\
QB3FM6 &  17-30 & 0.6 & 11-16 & 0.8 \\
QB3FM7 &  13-26 & 1.1 &  & \\
\hline
QB4FM1 &  20-30 & 1.1 & 15-26 & 0.8 \\
QB4FM2 &  20-30 & 1.1 & 15-22 & 0.3 \\
QB4FM3 &  20-28 & 1.1 & 15-20 & 0.1 \\
QB4FM4 &  22-30 & 0.9 & 14-19 & 0.3 \\
QB4FM5 &  20-30 & 0.7 & 13-19 & 0.4 \\
QB4FM6 &  22-30 & 0.5 & 13-19 & 0.6 \\
QB4FM7 &  22-30 & 0.4 & 13-18 & 0.4 \\
QB4FM8 &  17-30 & 0.6 &  & \\
\hline
QB5FM1 &  24-30 & 0.6 & 15-23 & 0.6 \\
QB5FM2 &  20-27 & 0.6 & 15-22 & 0.3 \\
QB5FM3 &  20-30 & 0.4 & 17-24 & 0.9 \\
\hline\hline
\end{tabular}
\caption{%
\label{tab:tensor_measurement_F}%
Technical details about  T and AT lattice correlation functions.
For each ensemble and each choice of bare mass $a m_0$, 
we show the fitting intervals of the Euclidean time $I_{\rm fit}=[t_i,t_f]$ between the minimum
 time $t_i$ and maximum time $t_f$ retained in 
the single-exponential fit to the measured correlators of mesons
made of fundamental Dirac fermions. 
We carry out a correlated fit  via  standard $\chi^2$-minimisation.
We report the values of $\chi^2$ (normalised by the number of  degrees of freedom) 
at the minima. In the case of the AT state, we leave blank some entries for which the 
numerical data did not exhibit a plateau in the effective mass plots, because of numerical noise.
}
\end{center}
\end{table}

\begin{table}[t]
\begin{center}
\begin{tabular}{|c|c|cc|cc|cc|cc|}
\hline\hline
\multirow{2}{*}{Measurement} & \multirow{2}{*}{$ m_0$} 
& \multicolumn{2}{c|}{ps} & \multicolumn{2}{c|}{v} & \multicolumn{2}{c|}{av} & \multicolumn{2}{c|}{s} \\
 &  & 
$I_{\rm fit}$ & $\frac{\chi^2}{N_{\rm d.o.f}}$ & $I_{\rm fit}$ & $\frac{\chi^2}{N_{\rm d.o.f}}$ 
& $I_{\rm fit}$ & $\frac{\chi^2}{N_{\rm d.o.f}}$ & $I_{\rm fit}$ & $\frac{\chi^2}{N_{\rm d.o.f}}$ \\
 &&&&&&&&&\\
\hline
QB1ASM1 & -1.05 & 14-24 & 1.2 & 13-24 & 0.5 & 7-12 & 0.1 & 10-15 & 0.5\\
QB1ASM2 & -1.08 & 12-24 & 1.9 & 12-24 & 0.5 & 7-10 & 0.1 & 7-13 & 1.8\\
QB1ASM3 & -1.1 & 12-24 & 1.1 & 12-24 & 1.0 & 7-12 & 1.1 & 8-13 & 0.7\\
QB1ASM4 & -1.12 & 12-24 & 0.8 & 12-20 & 0.5 & 7-11 & 0.5 & 7-11 & 0.2\\
QB1ASM5 & -1.13 & 11-24 & 1.6 & 12-19 & 0.3 & 8-11 & 0.6 & 7-11 & 0.4\\
QB1ASM6 & -1.14 & 13-24 & 0.7 & 10-20 & 0.5 & 7-10 & 0.3 &  & \\
\hline
QB2ASM1 & -1.05 & 16-30 & 1.2 & 14-25 & 1.1 & 9-15 & 0.7 & 11-17 & 0.6\\
QB2ASM2 & -1.08 & 15-30 & 0.7 & 13-28 & 0.8 & 9-13 & 0.1 & 9-14 & 0.8\\
QB2ASM3 & -1.09 & 15-30 & 0.9 & 12-23 & 0.5 & 10-13 & 0.5 & 8-12 & 0.2\\
QB2ASM4 & -1.1 & 15-30 & 1.0 & 12-21 & 0.4 & 8-12 & 0.2 & 9-12 & 0.2\\
QB2ASM5 & -1.11 & 15-30 & 1.3 & 12-25 & 0.7 & 9-12 & 0.2 & 8-12 & 1.0\\
QB2ASM6 & -1.12 & 16-30 & 1.3 &  &  &  &  &  & \\
\hline
QB3ASM1 & -1.03 & 16-30 & 1.5 & 16-27 & 0.4 & 10-15 & 0.2 & 10-18 & 0.6\\
QB3ASM2 & -1.04 & 14-30 & 1.1 & 17-30 & 1.5 & 10-16 & 0.1 & 9-16 & 0.5\\
QB3ASM3 & -1.05 & 16-30 & 0.8 & 14-30 & 0.5 & 9-12 & 0.2 & 9-15 & 0.5\\
QB3ASM4 & -1.06 & 18-30 & 1.1 & 14-24 & 0.2 & 9-12 & 0.3 & & \\
\hline
QB4ASM1 & -0.95 & 20-30 & 0.9 & 20-30 & 0.8 & 12-21 & 0.8 & 13-26 & 0.7\\
QB4ASM2 & -0.983 & 19-30 & 1.6 & 20-30 & 0.9 & 12-19 & 0.7 & 11-22 & 0.7\\
QB4ASM3 & -0.99 & 19-30 & 1.6 & 17-23 & 0.8 & 10-18 & 0.4 & 12-19 & 0.4\\
QB4ASM4 & -0.99 & 19-30 & 1.6 & 17-23 & 0.8 & 10-18 & 0.4 & 12-19 & 0.4\\
QB4ASM5 & -1.01 & 18-30 & 0.7 & 17-28 & 1.1 & 10-16 & 0.1 & 9-12 & 0.5\\
QB4ASM6 & -1.015 & 19-30 & 0.9 & 16-26 & 0.5 & 11-15 & 0.2 &  & \\
\hline
QB5ASM1 & -0.95 & 20-30 & 0.6 & 19-30 & 0.3 & 12-21 & 0.7 & 12-24 & 1.0\\
QB5ASM2 & -0.961 & 19-30 & 1.7 & 20-30 & 0.2 & 12-19 & 0.2 &  & \\
\hline\hline
\end{tabular}
\caption{%
\label{tab:meson_measurement_AS}%
Technical details pertaining the measurements of the correlation functions 
of operators built with 2-index antisymmetric fermions.
For each ensemble and each choice of bare mass $a m_0$, 
we show the fitting intervals of the Euclidean time $I_{\rm fit}=[t_i,t_f]$ between the minimum
 time $t_i$ and maximum time $t_f$ retained in 
the single-exponential fit to the measured correlators of mesons
made of antisymmetric Dirac fermions. 
We carry out a correlated fit  via  standard $\chi^2$-minimisation.
We report the values of $\chi^2$ (normalised by the number of the degrees of freedom) 
at the minima. 
In the case of the v, av and s states, we leave blank some entries for which the 
numerical data do not exhibit a plateau in the effective mass plots, due to numerical noise. 
}
\end{center}
\end{table}

\begin{table}[ht]
\begin{center}
\begin{tabular}{|c|cc|cc|}
\hline\hline
\multirow{2}{*}{Measurement} & \multicolumn{2}{c|}{t} & \multicolumn{2}{c|}{at} \\
 &  $I_{\rm fit}$ & $\frac{\chi^2}{N_{\rm d.o.f}}$ 
& $I_{\rm fit}$ & $\frac{\chi^2}{N_{\rm d.o.f}}$ \\
 &&&&\\
\hline
QB1ASM1 &  13-24 & 1.3 & 9-13 & 0.7 \\
QB1ASM2 &  11-24 & 0.8 & 7-11 & 1.8 \\
QB1ASM3 &  10-20 & 0.2 & 7-10 & 0.1 \\
QB1ASM4 &  10-21 & 1.0 & 8-11 & 0.1 \\
QB1ASM5 &  11-24 & 0.4 & 7-9 & 0.1 \\
QB1ASM6 &  10-15 & 1.0 & 6-10 & 0.1 \\
\hline
QB2ASM1 &  13-30 & 0.8 & 9-15 & 1.0 \\
QB2ASM2 &  14-24 & 0.7 & 9-14 & 1.3 \\
QB2ASM3 &  11-20 & 1.3 & 8-13 & 0.3 \\
QB2ASM4 &  11-26 & 1.2 & 7-12 & 0.2 \\
QB2ASM5 &  10-21 & 0.9 & 11-15 & 1.4 \\
QB2ASM6 &   &  &  & \\
\hline
QB3ASM1 &  19-30 & 0.8 & 11-15 & 0.1 \\
QB3ASM2 &  13-30 & 1.0 & 8-13 & 0.2 \\
QB3ASM3 &  12-20 & 0.9 & 9-13 & 0.2 \\
QB3ASM4 &  12-24 & 0.2 & 9-12 & 0.3 \\
\hline
QB4ASM1 &  17-30 & 1.4 & 12-16 & 1.5 \\
QB4ASM2 &  22-29 & 0.5 & 11-19 & 0.3 \\
QB4ASM3 &  17-30 & 0.7 & 10-16 & 0.7 \\
QB4ASM4 &  15-30 & 0.8 & 10-18 & 1.0 \\
QB4ASM5 &  19-27 & 0.3 & 14-19 & 0.3 \\
QB4ASM6 &  15-26 & 0.5 &  &  \\
\hline
QB5ASM1 &  17-30 & 0.6 & 11-16 & 0.3 \\
QB5ASM2 &  14-26 & 0.5 & 10-15 & 0.1 \\
\hline\hline
\end{tabular}
\caption{%
\label{tab:tensor_measurement_AS}%
Technical details pertaining the measurements of the correlation functions 
of operators built with 2-index antisymmetric fermions.
For each ensemble and each choice of bare mass $a m_0$, 
we show the fitting intervals of the Euclidean time $I_{\rm fit}=[t_i,t_f]$ between the minimum
 time $t_i$ and maximum time $t_f$ retained in 
the single-exponential fit to the measured correlators of mesons
made of antisymmetric Dirac fermions. 
We carry out a correlated fit  via  standard $\chi^2$-minimisation.
We report the values of $\chi^2$ (normalised by the number of the degrees of freedom) 
at the minima. In the case of the at state, we leave blank some entries for which the 
numerical data did not exhibit a plateau in the effective mass plots, due to numerical noise.
}
\end{center}
\end{table}

\section{Numerical results of lattice measurements}
\label{Sec:AppendixG}

\begin{table}[t]
\begin{center}
\begin{tabular}{|c|c|c|c|c|c|}
\hline\hline
Measurement & $a m_{\rm PS}$ & $a f_{\rm PS}$ & $a m_{\rm S}$ & $ m_{\rm PS}\,L$ & $ f_{\rm PS}\,L$ \\
\hline
QB1FM1 & 0.5516(4) & 0.08728(26) & 0.925(9) & 13.239(10) & 2.095(6)\\
QB1FM2 & 0.4816(5) & 0.08206(28) & 0.873(11) & 11.558(11) & 1.969(7)\\
QB1FM3 & 0.4309(5) & 0.07801(29) & 0.840(15) & 10.342(12) & 1.872(7) \\
QB1FM4 & 0.3753(6) & 0.0733(3) & 0.838(14) & 9.008(13) & 1.760(8) \\
QB1FM5 & 0.3453(6) & 0.0709(3) & 0.839(19) & 8.287(14) & 1.702(8) \\
QB1FM6 & 0.3125(7) & 0.0681(3) &  & 7.501(16) & 1.635(8) \\
\hline
QB2FM1 & 0.38340(15) & 0.06957(9) & 0.771(6) & 18.403(7) & 3.339(4) \\
QB2FM2 & 0.32442(17) & 0.06482(10) & 0.760(7) & 15.572(8) & 3.112(5) \\
QB2FM3 & 0.29148(18) & 0.06222(11) &  & 13.991(9) & 2.986(5) \\
\hline
QB3FM1 & 0.55219(15) & 0.07410(14) & 0.7980(29) & 26.505(7) & 3.557(7) \\
QB3FM2 & 0.43873(16) & 0.06705(13) & 0.711(5) & 21.059(7) & 3.218(6) \\
QB3FM3 & 0.36129(20) & 0.06099(12) & 0.633(9) & 17.342(10) & 2.928(6) \\
QB3FM4 & 0.30373(21) & 0.05644(14) & 0.657(4) & 14.579(10) & 2.728(7) \\
QB3FM5 & 0.27138(20) & 0.05406(10) & 0.640(5) & 13.026(10) & 2.595(5) \\
QB3FM6 & 0.23560(25) & 0.05128(13) &  & 11.309(12) & 2.461(6) \\
QB3FM7 & 0.19406(25) & 0.04841(12) &  & 9.315(12) & 2.324(6) \\
\hline
QB4FM1 & 0.44146(16) & 0.06150(13) & 0.649(4) & 21.190(8) & 2.952(6) \\
QB4FM2 & 0.38068(19) & 0.05752(13) & 0.560(5) & 18.272(9) & 2.761(6) \\
QB4FM3 & 0.34147(20) & 0.05471(13) & 0.577(5) & 16.390(10) & 2.626(6) \\
QB4FM4 & 0.31413(23) & 0.05293(13) & 0.540(5) & 15.078(11) & 2.541(6)\\
QB4FM5 & 0.28472(25) & 0.05064(13) & 0.522(7) & 13.666(12) & 2.431(6) \\
QB4FM6 & 0.25306(27) & 0.04816(14) & 0.511(10) & 12.147(13) & 2.312(7) \\
QB4FM7 & 0.21806(24) & 0.04539(11) & 0.523(10) & 10.467(11) & 2.179(5) \\
QB4FM8 & 0.17734(26) & 0.04240(12) &  & 8.512(12) & 2.035(6)\\
\hline
QB5FM1 & 0.2524(3) & 0.04282(15) & 0.449(4) & 12.113(14) & 2.055(7) \\
QB5FM2 & 0.1850(4) & 0.03740(16) & 0.428(6) & 8.881(17) & 1.795(7) \\
QB5FM3 & 0.1610(3) & 0.03560(14) &  & 7.727(16) & 1.709(7) \\
\hline\hline
\end{tabular}
\end{center}
\caption{%
\label{tab:meson_spec_spin0_F}%
Masses for flavoured spin-0 (PS and S) mesons, made of Dirac fermions
transforming in the fundamental representations of $Sp(4)$,
and (renormalised) decay constant of the PS states.
All results are obtained in the quenched approximation and
 presented either in units of the lattice spacing $a$ 
or volume $L=N_s a$. 
}
\end{table}

\begin{table}[t]
\begin{center}
\begin{tabular}{|c|c|c|c|c|}
\hline\hline
Measurement & $a m_{\rm V}$ & $a f_{\rm V}$ & $a m_{\rm AV}$ & $a f_{\rm AV}$ \\
\hline
QB1FM1 & 0.6259(8) & 0.1372(5) & 0.971(9) & 0.134(4) \\
QB1FM2 & 0.5721(11) & 0.1351(7) & 0.927(11) & 0.138(5) \\
QB1FM3 & 0.5356(13) & 0.1330(8) & 0.902(14) & 0.142(6) \\
QB1FM4 & 0.4981(18) & 0.1302(10) & 0.880(12) & 0.148(4) \\
QB1FM5 & 0.4793(22) & 0.1287(12) & 0.865(14) & 0.148(5) \\
QB1FM6 & 0.4592(27) & 0.1262(14) & 0.854(17) & 0.149(6) \\
\hline
QB2FM1 & 0.4848(6) & 0.1186(4) & 0.784(8) & 0.115(4) \\
QB2FM2 & 0.4474(9) & 0.1162(6) & 0.745(12) & 0.113(5) \\
QB2FM3 & 0.4288(12) & 0.1148(8) & 0.726(15) & 0.112(7) \\
\hline
QB3FM1 & 0.59445(24) & 0.10405(21) & 0.821(3) & 0.0881(15) \\
QB3FM2 & 0.4988(3) & 0.10087(29) & 0.736(5) & 0.0930(23) \\
QB3FM3 & 0.4394(7) & 0.0973(5) & 0.664(14) & 0.088(8) \\
QB3FM4 & 0.4008(8) & 0.0959(6) & 0.655(11) & 0.098(5) \\
QB3FM5 & 0.3800(16) & 0.0932(11) & 0.631(6) & 0.0997(20) \\
QB3FM6 & 0.3640(15) & 0.0944(9) & 0.645(10) & 0.110(4) \\
QB3FM7 & 0.339(4) & 0.0874(24) & 0.581(11) & 0.095(4) \\
\hline
QB4FM1 & 0.4839(5) & 0.08727(28) & 0.680(4) & 0.0789(17) \\
QB4FM2 & 0.4332(4) & 0.0851(3) & 0.633(5) & 0.0809(20) \\
QB4FM3 & 0.4023(5) & 0.0835(4) & 0.605(5) & 0.0821(24) \\
QB4FM4 & 0.3824(7) & 0.0828(5) & 0.566(7) & 0.0746(28) \\
QB4FM5 & 0.3626(7) & 0.0821(5) & 0.543(8) & 0.074(3) \\
QB4FM6 & 0.3421(10) & 0.0806(6) & 0.519(11) & 0.072(4) \\
QB4FM7 & 0.3222(16) & 0.0790(9) & 0.492(15) & 0.069(6) \\
QB4FM8 & 0.303(3) & 0.0771(18) &  &  \\
\hline
QB5FM1 & 0.3112(7) & 0.0665(5) & 0.480(6) & 0.0703(27) \\
QB5FM2 & 0.2680(15) & 0.0637(8) & 0.449(7) & 0.0749(27) \\
QB5FM3 & 0.2589(23) & 0.0637(11) & 0.406(16) & 0.063(6) \\
\hline\hline
\end{tabular}
\caption{%
\label{tab:meson_spec_spin1_F}%
Masses and decay constants, computed in the quenched approximation,
for flavoured spin-1 (V and AV) mesons, made of Dirac fermions
transforming in the fundamental representations of $Sp(4)$. 
All results are in units of the lattice spacing $a$. 
In parentheses we report the statistical uncertainties.
}
\end{center}
\end{table}

\begin{table}[t]
\begin{center}
\begin{tabular}{|c|c|c|c|c|c|c|}
\hline\hline
Measurement & $\hat{m}_{\rm V}/\hat{m}_{\rm PS}$ & $\hat{m}_{\rm PS}/\hat{f}_{\rm PS}$
 & $\hat{m}_{\rm V}/\hat{f}_{\rm PS}$ & 
$\hat{m}_{\rm AV}/\hat{f}_{\rm PS}$ & $\hat{m}_{\rm S}/\hat{f}_{\rm PS}$ & $\hat{f}_{\rm V}/\hat{f}_{\rm PS}$ \\
\hline
QB1FM1 & 1.1346(13) & 6.320(17) & 7.171(23) & 11.13(11) & 10.60(10) & 1.572(7)\\
QB1FM2 & 1.1880(20) & 5.869(17) & 6.972(27) & 11.30(14) & 10.64(13) & 1.646(9)\\
QB1FM3 & 1.2429(29) & 5.523(18)  & 6.86(3) & 11.56(18) & 10.77(18) & 1.704(11)\\
QB1FM4 & 1.327(5) & 5.120(19) & 6.79(4) & 12.01(17) & 11.43(19) & 1.776(15)\\
QB1FM5 & 1.388(6) & 4.869(20) & 6.76(4) & 12.20(20) & 11.83(26) & 1.815(17)\\
QB1FM6 & 1.469(8) & 4.587(21) & 6.74(5) & 12.54(20) &  & 1.852(21)\\
\hline
QB2FM1 & 1.2646(14) & 5.511(7) & 6.969(12) & 11.26(12) & 11.08(10) & 1.705(6)\\
QB2FM2 & 1.3791(27) & 5.005(7) & 6.902(17) & 11.50(18) & 11.73(11) & 1.793(9)\\
QB2FM3 & 1.471(4) & 4.685(7) & 6.891(22) & 11.67(25) &  & 1.845(12)\\
\hline
QB3FM1 & 1.0765(4) & 7.452(14) & 8.022(16) & 11.08(5) & 10.76(5) & 1.404(3)\\
QB3FM2 & 1.1370(7) & 6.544(12) & 7.440(15) & 10.97(8) & 10.61(8) & 1.505(4)\\
QB3FM3 & 1.2161(18) & 5.924(10) & 7.204(18) & 11.70(11) & 10.39(15) & 1.596(9)\\
QB3FM4 & 1.3196(27) & 5.343(12) & 7.051(22) & 11.52(19) & 11.56(8) & 1.687(10)\\
QB3FM5 & 1.400(6) & 5.020(8) & 7.03(3) & 11.66(11) & 11.84(10) & 1.725(20)\\
QB3FM6 & 1.545(7) & 4.595(11) & 7.10(3) & 12.41(15) &  & 1.841(21)\\
QB3FM7 & 1.745(21) & 4.009(9) & 7.00(8) & 12.01(24) &  & 1.81(5)\\
\hline
QB4FM1 & 1.0961(6) & 7.179(14) & 7.868(17) & 11.06(7) & 10.55(7) & 1.419(4)\\
QB4FM2 & 1.1381(10) & 6.618(14)& 7.532(18) & 11.01(8) & 10.42(10) & 1.480(6)\\
QB4FM3 & 1.1782(14) & 6.242(14) & 7.354(20) & 11.06(10) & 10.54(11) & 1.527(7)\\
QB4FM4 & 1.2175(20) & 5.945(12) & 7.226(21) & 10.69(13) & 10.19(10) & 1.565(10)\\
QB4FM5 & 1.2735(25) & 5.634(12) & 7.160(24) & 10.71(16) & 10.31(13) & 1.621((10)\\
QB4FM6 & 1.352(4) & 5.255(12) & 7.104(29) & 10.77(22) & 10.60(20) & 1.674(13)\\
QB4FM7 & 1.478(7) & 4.804(11) & 7.10(4) & 10.8(3) & 11.53(21) & 1.741(20)\\
QB4FM8 & 1.708(19) & 4.183(11) & 7.14(9) &  &  & 1.82(4)\\
\hline
QB5FM1 & 1.2332(27) & 5.894(18) & 7.268(27) & 11.21(15) & 10.49(9) & 1.552(10)\\
QB5FM2 & 1.449(8) & 4.948(19) & 7.17(4) & 12.02(20) & 11.44(18) & 1.703(19)\\
QB5FM3 & 1.608(14) & 4.522(18) & 7.27(7) & 11.4(4) &  & 1.789(29)\\
\hline\hline
\end{tabular}
\caption{%
\label{tab:meson_ratio_F}%
Some useful ratios of (quenched) masses and decay constants of mesons made of Dirac fermions
transforming in the fundamental representation.
In parentheses we report the statistical uncertainties.
}
\end{center}
\end{table}

\begin{table}[t]
\begin{center}
\begin{tabular}{|c|c|c|c|c|}
\hline\hline
Measurement & $a\,m_{\rm T}$ & $\hat{m}_{\rm T}/\hat{f}_{\rm PS}$ & $a\,m_{\rm AT}$ 
& $\hat{m}_{\rm AT}/\hat{f}_{\rm PS}$  \\
\hline
QB1FM1 & 0.6257(10) & 7.169(23) & 0.963(10) & 11.03(11) \cr
QB1FM2 & 0.5719(14) & 6.969(28) & 0.920(12) & 11.21(15) \cr
QB1FM3 & 0.5350(18) & 6.86(3) & 0.899(10) & 11.52(13) \cr
QB1FM4 & 0.4978(25) & 6.79(4) & 0.863(14) & 11.77(20) \cr
QB1FM5 & 0.477(3) & 6.73(5) & 0.858(15) & 12.09(23) \cr
QB1FM6 & 0.459(4) & 6.74(7) & 0.834(19) & 12.2(3) \cr
\hline
QB2FM1 & 0.4838(9) & 6.954(15) & 0.775(10) & 11.13(15) \cr
QB2FM2 & 0.4465(15) & 6.887(24) & 0.741(15) & 11.42(24) \cr
QB2FM3 & 0.4307(15) & 6.922(27) & 0.754(8) & 12.12(13) \cr
\hline
QB3FM1 & 0.5944(3) & 8.021(16) & 0.8201(4) & 11.07(6) \cr
QB3FM2 & 0.4986(5) & 7.437(17) & 0.735(6) & 10.96(9) \cr
QB3FM3 & 0.4397(9) & 7.209(12) & 0.673(7) & 11.04(12) \cr
QB3FM4 & 0.3997(14) & 7.03(3) & 0.661(14) & 11.63(26) \cr
QB3FM5 & 0.3806(23) & 7.04(4) & 0.596(20) & 11.0(4) \cr
QB3FM6 & 0.3649(26) & 7.12(5) & 0.672(12) & 13.11(24) \cr
QB3FM7 & 0.3532(29) & 7.30(6) &  &  \cr
\hline
QB4FM1 &  0.4844(4) & 7.877(17) & 0.675(4) & 10.97(8) \cr
QB4FM2 &  0.4336(5) & 7.538(20) & 0.626(6) & 10.88(10) \cr
QB4FM3 &  0.4023(7) & 7.354(23) & 0.595(7) & 10.87(13) \cr
QB4FM4 &  0.3824(11) & 7.226(26) & 0.581(7) & 10.98(13) \cr
QB4FM5 &  0.3617(15) & 7.14(3) & 0.657(6) & 11.19(11) \cr
QB4FM6 &  0.3405(23) & 7.07(5) & 0.547(7) & 11.36(15) \cr
QB4FM7 &  0.318(4) & 7.01(9) & 0.525(9) & 11.59(20) \cr
QB4FM8 &  0.303(4) & 7.14(9) &  &  \cr
\hline
QB5FM1 & 0.3111(14) & 7.27(4) & 0.488(5) & 11.39(13) \cr
QB5FM2 & 0.2685(23) & 7.18(6) & 0.450(9) & 12.03(25) \cr
QB5FM3 & 0.252(3) & 7.08(9) & 0.391(19) & 11.0(5) \cr
\hline\hline
\end{tabular}
\caption{%
\label{tab:meson_tensor_F}%
Masses of $T$ and $AT$ states, in units of $a$ and $f_{\rm PS}$, for
each of the ensembles and bare masses $m_0$. 
The Dirac fermions are in the fundamental representation. 
In parentheses we report the statistical uncertainties.
}
\end{center}
\end{table}

\begin{table}[t]
\begin{center}
\begin{tabular}{|c|c|c|c|c|c|}
\hline\hline
Measurement & $a m_{\rm ps}$ & $a f_{\rm ps}$ & $a m_{\rm s}$ & $ m_{\rm ps}\,L$ 
& $ f_{\rm ps}\,L$ \\
\hline
QB1ASM1 & 0.6254(4) & 0.1249(4) & 1.045(28) & 15.009(9) & 2.997(9)\\
QB1ASM2 & 0.5413(4) & 0.1166(3) & 1.036(10) & 12.991(10) & 2.798(8)\\
QB1ASM3 & 0.4789(4) & 0.1107(4) & 0.970(20) & 11.495(9) & 2.657(9)\\
QB1ASM4 & 0.4087(5) & 0.1036(3) & 0.953(16) & 9.809(11) & 2.487(8) \\
QB1ASM5 & 0.3693(5) & 0.0998(4) & 0.930(25) & 8.863(12) & 2.396(8) \\
QB1ASM6 & 0.3260(5) & 0.0958(5) &  & 7.823(13) & 2.300(12) \\
\hline
QB2ASM1 & 0.50776(12) & 0.10646(13) & 0.894(11) & 24.372(6) & 5.110(6) \\
QB2ASM2 & 0.40809(14) & 0.09668(18) & 0.851(8)  & 19.588(7) & 4.641(6) \\
QB2ASM3 & 0.37047(16) & 0.09300(14) & 0.860(7)  & 17.782(8) & 4.464(7) \\
QB2ASM4 & 0.32896(16) & 0.08898(14) & 0.824(14) & 15.790(8) & 4.271(7) \\
QB2ASM5 & 0.28241(19) & 0.08494(16) & 0.842(14) & 13.556(9) & 4.077(8) \\
QB2ASM6 & 0.22727(22) & 0.08108(22) &  & 10.909(10) & 3.892(10) \\
\hline
QB3ASM1 & 0.35682(16) & 0.08149(13) & 0.726(6) & 17.127(8) & 3.912(6) \\
QB3ASM2 & 0.31698(16) & 0.07781(13) & 0.704(5) & 15.215(8) & 3.735(6) \\
QB3ASM3 & 0.27265(21) & 0.07361(18) & 0.698(7) & 13.087(10) & 3.533(8) \\
QB3ASM4 & 0.22041(27) & 0.06926(19) &  & 10.580(13) & 3.325(10) \\
\hline
QB4ASM1 & 0.44487(15) & 0.08239(15) & 0.692(4) & 21.354(7) & 3.945(7) \\
QB4ASM2 & 0.33323(16) & 0.08239(15) & 0.623(4) & 15.995(8) & 3.444(6) \\
QB4ASM3 & 0.30578(19) & 0.06921(15) & 0.611(7) & 14.678(9) & 3.322(7)\\
QB4ASM4 & 0.26323(18) & 0.06536(15) & 0.579(9) & 12.635(9) & 3.137(7) \\
QB4ASM5 & 0.21375(20) & 0.06080(17) & 0.604(7) & 10.260(10) & 2.918(8) \\
QB4ASM6 & 0.18506(25) & 0.05838(17) &  & 8.883(12) & 2.802(8) \\
\hline
QB4ASM1 & 0.22454(27) & 0.05392(13) & 0.484(5) & 10.778(13) & 2.588(6) \\
QB4ASM2 & 0.1666(3) & 0.04851(15) &  & 7.999(15) & 2.329(7) \\
\hline\hline
\end{tabular}
\end{center}
\caption{%
\label{tab:meson_spec_spin0_AS}%
Masses for flavoured spin-0 (ps and s) mesons, made of Dirac fermions
transforming in the 2-index  antisymmetric representations of $Sp(4)$,
and decay constant of the PS states.
All results are obtained in the quenched approximation and presented 
either in units of the lattice spacing $a$ 
or volume $L$.  In parentheses we report the statistical uncertainties.
}
\end{table}

\begin{table}[t]
\begin{center}
\begin{tabular}{|c|c|c|c|c|}
\hline\hline
Measurement & $a m_{\rm v}$ & $a f_{\rm v}$ & $a m_{\rm av}$ & $a f_{\rm av}$ \\
\hline
QB1ASM1 & 0.7457(11) & 0.1970(11) & 1.216(10) & 0.196(5) \\
QB1ASM2 & 0.6836(12) & 0.1924(11) & 1.146(13) & 0.192(6) \\
QB1ASM3 & 0.6393(19) & 0.1862(17) & 1.082(18) & 0.183(7) \\
QB1ASM4 & 0.595(3) & 0.1842(25) & 1.083(22) & 0.205(11) \\
QB1ASM5 & 0.571(4) & 0.179(3) & 0.94(5) & 0.148(18) \\
QB1ASM6 & 0.542(4) & 0.1759(21) & 1.01(4) & 0.190(16) \\
\hline
QB2ASM1 & 0.6378(7) & 0.1726(7) & 1.020(9) & 0.159(5) \\
QB2ASM2 & 0.5679(8) & 0.1645(7) & 0.942(14) & 0.154(7) \\
QB2ASM3 & 0.5466(10) & 0.1646(7) & 0.96(3) & 0.174(22) \\
QB2ASM4 & 0.5222(12) & 0.1615(8) & 0.937(13) & 0.172(6) \\
QB2ASM5 & 0.4921(22) & 0.1548(14) & 0.871(28) & 0.152(13) \\
QB2ASM6 &  &  &  &  \\
\hline
QB3ASM1 & 0.4933(11) & 0.1346(9) & 0.800(10) & 0.130(5) \\
QB3ASM2 & 0.4637(19) & 0.1290(16) & 0.785(12) & 0.135(6) \\
QB3ASM3 & 0.4461(18) & 0.1325(12) & 0.778(11) & 0.142(5) \\
QB3ASM4 & 0.417(3) & 0.1276(20) & 0.752(14) & 0.141(6) \\
\hline
QB4ASM1 & 0.5287(6) & 0.1221(6) & 0.770(6) & 0.1060(29) \\
QB4ASM2 & 0.4425(12) & 0.1122(11) & 0.691(10) & 0.108(4) \\
QB4ASM3 & 0.4249(12) & 0.1115(9) & 0.700(5) & 0.1222(19) \\
QB4ASM4 & 0.3960(20) & 0.1076(15) & 0.660(10) & 0.115(4) \\
QB4ASM5 & 0.3697(25) & 0.1060(16) & 0.629(10) & 0.115(4) \\
QB4ASM6 & 0.348(4) & 0.0995(19) & 0.662(22) & 0.134(11) \\
\hline
QB5ASM1 & 0.3327(14) & 0.0865(9) & 0.558(7) & 0.1006(28) \\
QB5ASM2 & 0.301(5) & 0.0840(26) & 0.499(9) & 0.090(3) \\
\hline\hline
\end{tabular}
\caption{%
\label{tab:meson_spec_spin1_AS}%
Masses and decay constants, computed in the quenched approximation,
for flavoured spin-1 (v and av) mesons, made of Dirac fermions
transforming in the antisymmetric representations of $Sp(4)$. 
All results are in units of the lattice spacing $a$. 
In parentheses we report the statistical uncertainties.
}
\end{center}
\end{table}

\begin{table}[t]
\begin{center}
\begin{tabular}{|c|c|c|c|c|c|c|}
\hline\hline
Measurement & $\hat{m}_{\rm v}/\hat{m}_{\rm ps}$ & $\hat{m}_{\rm ps}/\hat{f}_{\rm ps}$
 & $\hat{m}_{\rm v}/\hat{f}_{\rm ps}$ & 
$\hat{m}_{\rm av}/\hat{f}_{\rm ps}$ & $\hat{m}_{\rm s}/\hat{f}_{\rm ps}$ 
& $\hat{f}_{\rm v}/\hat{f}_{\rm pv}$ \\
\hline
QB1ASM1 & 1.1924(17) & 5.009(14) & 5.972(21) & 9.74(8) & 8.37(22) & 1.578(10)\\
QB1ASM2 & 1.2629(23) & 4.643(12) & 5.864(19) & 9.83(12) & 8.89(9) & 1.651(10)\\
QB1ASM3 & 1.335(4) & 4.326(14) & 5.774(25) & 9.77(17) & 8.76(18) & 1.682(14)\\
QB1ASM4 & 1.457(8) & 3.944(11)  & 5.75(3) & 10.45(21) & 9.20(15) & 1.778(24)\\
QB1ASM5 & 1.545(11) & 3.699(12)  & 5.72(5) & 9.4(5) & 9.31(25) & 1.79(3)\\
QB1ASM6 & 1.654(15) & 3.401(16) & 5.65(5) & 10.5(4) &  & 1.835(23)\\
\hline
QB2ASM1 & 1.2561(13) & 4.770(5) & 5.991(9) & 9.57(9) & 8.40(11) & 1.621(6)\\
QB2ASM2 & 1.3916(21) & 4.221(5) & 5.874(12) & 9.74(14) & 8.81(8) & 1.701(8)\\
QB2ASM3 & 1.4754(27) & 3.983(5) & 5.877(14) & 10.3(4) & 9.24(8) &1.770(8)\\
QB2ASM4 & 1.587(4) & 3.697(5) & 5.869(16) & 10.53(15) & 9.27(15) & 1.815(10)\\
QB2ASM5 & 1.742(8) & 3.325(5) & 5.793(25) & 10.3(3) & 9.91(17) & 1.823(16)\\
QB2ASM6 &  & 2.803(7) &  &  &  & \\
\hline
QB3ASM1 & 1.382(3) & 4.379(6) & 6.053(15) & 9.81(13) & 8.91(8) & 1.652(10)\\
QB3ASM2 & 1.463(6) & 4.073(7) & 5.959(27) & 10.09(15) & 9.05(7) & 1.657(21)\\
QB3ASM3 & 1.636(7) & 3.704(8) & 6.061(28) & 10.56(15) & 9.48(10) & 1.801(16)\\
QB3ASM4 & 1.891(15) & 3.182(8) & 6.02(5) & 10.86(21) &  & 1.843(29)\\
\hline
QB4ASM1 & 1.1884(13) & 5.399(9) & 6.417(13) & 9.35(8) & 8.40(5) & 1.482(7)\\
QB4ASM2 & 1.328(4) & 4.644(8) & 6.168(20) & 9.63(14) & 8.67(6) & 1.563(15)\\
QB4ASM3 & 1.389(4) & 4.418(9) & 6.139(21) & 10.12(7) & 8.83(10) & 1.611(12)\\
QB4ASM4 & 1.504(8) & 4.027(8) & 6.06(3) & 10.10(15) & 8.86(14) & 1.647(23)\\
QB4ASM5 & 1.729(12) & 3.516(9) & 6.08(5) & 10.35(16) & 9.93(12) & 1.743(28)\\
QB4ASM6 & 1.879(19) & 3.170(8) & 5.96(6) & 11.4(4) &  & 1.70(3)\\
\hline
QB5ASM1 & 1.482(6) & 4.165(8) & 6.170(29) & 10.35(13) & 8.98(9) & 1.605(16)\\
QB5ASM2 & 1.805(28) & 3.435(9) & 6.20(10) & 10.29(20) &  & 1.73(5)\\
\hline\hline
\end{tabular}
\caption{%
\label{tab:meson_ratio_AS}%
Some useful ratios of (quenched) masses and decay constants of mesons made of Dirac fermions
transforming in the antisymmetric representation.
In parentheses we report the statistical uncertainties.
}
\end{center}
\end{table}

\begin{table}[t]
\begin{center}
\begin{tabular}{|c|c|c|c|c|}
\hline\hline
Measurement & $a\,m_{\rm t}$ & $m_{\rm t}/f_{\rm ps}$ & 
$a\,m_{\rm at}$ & $m_{\rm at}/f_{\rm ps}$  \\
\hline
QB1ASM1 & 0.7454(17) & 5.970(24) & 1.210(15) & 9.69(12) \cr
QB1ASM2 & 0.6850(18) & 5.876(23) & 1.132(19) & 9.71(17) \cr
QB1ASM3 & 0.6452(22) & 5.828(29) & 1.127(25) & 10.18(23) \cr
QB1ASM4 & 0.601(4) & 5.80(4) & 0.96(5) & 9.3(4) \cr
QB1ASM5 & 0.568(6) & 5.69(6) & 1.05(4) & 10.6(4) \cr
QB1ASM6 & 0.546(7) & 5.70(7) & 1.067(26) & 11.13(28) \cr
\hline
QB2ASM1 & 0.6375(7) & 5.989(10) & 1.027(12) & 9.64(12) \cr
QB2ASM2 & 0.5658(16) & 5.852(18) & 0.957(23) & 9.90(24) \cr
QB2ASM3 & 0.5488(15) & 5.901(18) & 0.986(17) & 10.60(20) \cr
QB2ASM4 & 0.5228(19) & 5.876(24) & 0.988(11) & 11.11(12) \cr
QB2ASM5 & 0.5022(29) & 5.91(3) & 0.94(3) & 11.0(4) \cr
QB2ASM6 &  &  &  &  \cr
\hline
QB3ASM1 & 0.4884(25) & 5.99(3) & 0.781(24) & 9.58(29) \cr
QB3ASM2 & 0.4682(19) & 6.016(25) & 0.829(8) & 10.65(10) \cr
QB3ASM3 & 0.4447(24) & 6.04(4) & 0.799(21) & 10.85((29) \cr
QB3ASM4 & 0.423(4) & 6.10(6) & 0.77(3) & 11.2(5) \cr
\hline
QB4ASM1 & 0.5285(6) & 6.415(13) & 0.782(9) & 9.49(12) \cr
QB4ASM2 & 0.4402(26) & 6.14(4) & 0.712(10) & 9.93(15) \cr
QB4ASM3 & 0.4238(20) & 6.12(3) & 0.719(9) & 10.38(14) \cr
QB4ASM4 & 0.4028(22) & 6.16(4) & 0.675(12) & 10.32(18) \cr
QB4ASM5 & 0.349(7) & 5.74(12) & 0.59(4) & 9.7(6) \cr
QB4ASM6 & 0.350(6) & 5.99(11) &  & \cr
\hline
QB5ASM1 & 0.3314(27) & 6.15(5) & 0.585(9) & 10.84(18) \cr
QB5ASM2 & 0.307(3) & 6.32(7) & 0.548(11) & 11.30(24) \cr
\hline\hline
\end{tabular}
\caption{%
\label{tab:meson_tensor_AS}%
Masses of t and at states, in units of $a$ and $f_{\rm ps}$, for
each of the ensembles and bare masses $a m_0$. 
The Dirac fermions transform in the 2-index antisymmetric representation. 
In parentheses we report the statistical uncertainties.
}
\end{center}
\end{table}

In this Appendix we present the numerical details and results of lattice measurements 
for meson masses and decay constants. 
In \Tab{meson_measurement_F} we first list the name of each measurement QB$i$FM$j$ 
at given values of $\beta$ and $m_0$ for fundamental (quenched) fermions 
and present the time intervals corresponding to the asymptotic region for the cases of PS, V, AV and S mesons 
along with the values of $\chi^2/N_{\rm d.o.f}$, which reflect the quality of the numerical fits using Eqs.~(\ref{eq:corr_M}) and (\ref{eq:corr_Pi}). 
Similar results for T and AT mesons are shown in \Tab{tensor_measurement_F}.
Analogously, in the case of antisymmetric (quenched) fermions, labeled as QB$i$ASM$j$, 
we presented the results in \Tab{meson_measurement_AS} for the cases of ps, v, av and s mesons, 
and in \Tab{tensor_measurement_AS} for t and at mesons.

In \Tab{meson_spec_spin0_F} we present the extracted values of PS masses and decay constants, as well as 
S masses. 
In the table we also present the results of $m_{\rm PS} \,L$ and $f_{\rm PS} \,L$. 
Similarly, in \Tab{meson_spec_spin1_F} we show the results of the masses and decay constants 
of V and AV mesons. 
For presentation purposes, we find it useful to show also the meson masses 
in units of the PS decay constant, 
as well as the ratios of $\hat{f}_{\rm V}/\hat{f}_{\rm PS}$, in \Tab{meson_ratio_F}. 
In Table~\ref{tab:meson_tensor_F} we report the masses of T and AT states,
both in units of the lattice spacing $a$ and of the PS
decay constant $\hat{f}_{\rm PS}$. 
Analogous results for the masses and decay constants of ps, s, v, av, t, and at mesons composed of antisymmetric fermions 
are presented in Tables~\ref{tab:meson_spec_spin0_AS}-- \ref{tab:meson_tensor_AS}.


\end{document}